\documentclass[preprint,11pt]{aastex}

\shortauthors{Burgasser et al.}
\shorttitle{T Spectral Classification}

\begin{document}

\title{The Spectra of T Dwarfs I: Near-Infrared Data and 
Spectral Classification}

\author{
Adam J.\ Burgasser\altaffilmark{1},
J.\ Davy Kirkpatrick\altaffilmark{2},
Michael E.\ Brown\altaffilmark{3},
I. Neill Reid\altaffilmark{4},
Adam Burrows\altaffilmark{5},
James Liebert\altaffilmark{5},
Keith Matthews\altaffilmark{1},
John E.\ Gizis\altaffilmark{2},
Conard C.\ Dahn\altaffilmark{6},
David G.\ Monet\altaffilmark{6},
Roc M.\ Cutri\altaffilmark{2},
and Michael F.\ Skrutskie\altaffilmark{7}
}
 
\altaffiltext{1}{Division of Physics, Mathematics, and Astronomy, M/S 103-33, 
California Institute of Technology, Pasadena, CA 91125; 
diver@its.caltech.edu, kym@caltech.edu}
\altaffiltext{2}{Infrared Processing and Analysis Center, M/S 100-22, 
California Institute of Technology, Pasadena, CA 91125; davy@ipac.caltech.edu,
gizis@ipac.caltech.edu, roc@ipac.caltech.edu}
\altaffiltext{3}{Division of Geological and Planetary Sciences, M/S 105-21, 
California Institute of Technology, Pasadena, California 91125;
mbrown@gps.caltech.edu}
\altaffiltext{4}{Space Telescope Science Institute, 3700 San Martin Drive,
Baltimore, MD 21218; inr@stsci.edu; also
Dept.\ of Physics \& Astronomy, University of
Pennsylvania, 209 S.\ 33rd Street, Philadelphia, PA 19104-6396}
\altaffiltext{5}{Steward Observatory, University of Arizona,
Tucson, AZ 85721; burrows@jupiter.as.arizona.edu, liebert@as.arizona.edu}
\altaffiltext{6}{U.S. Naval Observatory, P.O. Box 1149, 
Flagstaff, AZ 86002; dahn@nofs.navy.mil, dgm@nofs.navy.mil}
\altaffiltext{7}{Five College Astronomy Department, Department of Physics
and Astronomy, University of Massachusetts, Amherst, MA 01003;
skrutski@north.astro.umass.edu}

\begin{abstract}
We present near-infrared
spectra for a sample of T dwarfs, including eleven new 
discoveries made using the Two Micron
All Sky Survey.  
These objects are distinguished from warmer (L-type) brown dwarfs by the
presence of methane absorption bands in the 1--2.5 $\micron$ spectral region.
A first attempt at a near-infrared classification scheme for T dwarfs is made,
based on the 
strengths of CH$_4$ and H$_2$O bands and the shapes of the 1.25, 1.6, and
2.1 $\micron$ flux peaks.  
Subtypes
T1 V through T8 V are defined, and spectral indices useful for classification
are presented.  
The subclasses appear to follow a decreasing T$_{eff}$ scale,
based on the evolution of CH$_4$ and H$_2$O bands and the properties
of L and T dwarfs with known distances.  However, 
we speculate that this scale is not linear with spectral type
for cool dwarfs, due to 
the settling of dust layers below the photosphere
and subsequent rapid evolution of spectral morphology around 
T$_{eff}$ $\sim$ 1300--1500 K.
Similarities in near-infrared colors and
continuity of spectral features suggest that the gap between the latest
L dwarfs and earliest T dwarfs has been nearly bridged.  This argument is 
strengthened by the possible role of CH$_4$ as a minor
absorber shaping the K-band spectra of the latest L dwarfs.
Finally, we discuss one peculiar T dwarf, 2MASS 0937+2931, which has 
very blue near-infrared colors (J-K$_s$ = $-$0.89$\pm$0.24) due to
suppression of the 2.1 $\micron$ 
peak.  The feature is likely caused by enhanced
collision-induced H$_2$ absorption in a
high pressure or low metallicity photosphere.  
\end{abstract}


\section{Introduction}

Classification is an important first step in the characterization of any 
astronomical population.  It enables one to study the global properties 
of a group of similar objects and
generalize to larger, and perhaps undetected, members.  
A fundamental attribute of astronomical classification
is that it is based on observations; i.e., the way objects appear
morphologically, photometrically, or spectroscopically.  Stellar spectral
classification has been in practice for nearly 150 years, and 
calibration of spectral classes to temperature and luminosity scales
has provided crucial insight into the internal physics of
stars, their history and evolution, and the properties of the Galaxy and
extragalactic systems.

The stellar sequence of \citet{mor43},
the most widely accepted classification scheme, originates from the 
temperature-based
sequence first used in the Henry Draper catalog \citep{pic1890}, and 
luminosity
discriminants initially identified by \citet{mau1897}.  It extends the main
sequence from hot O-type stars to cool M dwarfs,
the latter of which until recently were the coolest and
faintest stars known.  With the advent of more sensitive optical and
near-infrared detectors and large-scale surveys, including the Two Micron
All Sky Survey \citep[hereafter 2MASS]{skr97}, 
the Sloan Digital Sky Survey \citep[hereafter SDSS]{yor00}, and
the Deep Near-Infrared Survey of the Southern Sky 
\citep[hereafter DENIS]{epc97},
cooler stars and brown dwarfs have now been identified.  
As a result, two new spectral classes have been introduced, the L \citep{kir99,mrt99} and
T \citep{kir99} classes, which are extensions of the main stellar sequence
into the brown dwarf regime.  

The L spectral class is comprised of a mix of stars and brown dwarfs,  
characterized in the red optical (6300--10100 {\AA})
by weakening bands of TiO and VO (dominant features in M dwarfs);
strengthening bands of FeH, CrH, and H$_2$O; and strengthening 
lines of the alkali metals Na I, K I, Cs I,
and Rb I \citep{kir99,mrt99}.  
The near-infrared spectra of these objects are similar to those
of M dwarfs, dominated by
H$_2$O and CO bands, but are redder
(1.3 $\lesssim$ J-K$_s$ $\lesssim$ 2.1), a characteristic attributed 
to warm photospheric dust \citep{tsu96a}.  L dwarfs 
cover an effective temperature (T$_{eff}$)
range from 2000--2200 K down to $\sim$ 1300--1700 K 
\citep{kir99,kir00,bas00,leg01}, corresponding to 
luminosities L $\sim$ 4{$\times$}10$^{-4}$ to 3{$\times$}10$^{-5}$ L$_{\sun}$ 
\citep{bur97}.  
Well over one hundred of these objects have now been
identified in the field \citep{del97,rui97,wol98,kir99,kir00,mrt99,rei00,fan00,giz00}, 
in stellar clusters \citep{mrt98,zap99,zap00}, and as companions to nearby 
stars \citep{bec88,reb98,gld99,kir01a,giz01a,wil01b}.  Classification
schemes for L dwarfs in the red optical have been defined 
by \citet{kir99} and \citet{mrt99}.

The T spectral class is comprised of brown dwarfs that
exhibit CH$_4$ absorption bands between 1.0 and 2.2
$\micron$.  The presence of these bands, broad H$_2$O features,
and collision-induced (CIA) H$_2$ absorption radically alter the
spectral energy distributions of T dwarfs,
and near-infrared colors become increasingly blue
(J-K$_s$ $\sim$ 0) as compared to L dwarfs.  
These distinctions
led \citet{kir99} 
to propose a second class, which at that time was solely occupied
by the first unequivocal brown dwarf, Gliese 229B \citep{nak95,opp95}.
Recently, a number of objects similar to Gliese 229B have been
discovered by SDSS \citep{str99,tsv00,leg00,geb01b}, 
2MASS \citep{me99,me00a,me00b}, and the
New Technology Telescope Deep Field \citep{cub99}.  These objects exhibit the
same CH$_4$ and H$_2$O bands seen in Gliese 229B, but of 
differing strengths, suggesting that near-infrared spectra can be 
used to derive a T dwarf classification scheme \citep{me99,leg00}.

In this paper, we make a first attempt to define such a
scheme based on the strengths of near-infrared
CH$_4$ and H$_2$O absorption bands and colors, using spectral
data obtained for newly discovered and previously known T dwarfs.
In $\S$2, we summarize our search for T dwarfs using the 2MASS database.
In $\S$3,
we discuss near-infrared spectroscopy acquired for candidate and known 
T dwarfs, and describe features observed in these
data.  Eleven new T dwarfs identified
in the 2MASS search are presented in $\S$4, and a rough determination of 
the T dwarf space density from this search 
are compared to estimates from other samples.  In $\S$5, we use the 
near-infrared spectral data, 
along with data obtained from the literature, to establish
a near-infrared classification scheme
for T dwarfs.  We also introduce a suite of spectral indices useful
for subtyping.  The properties of these T subclasses are discussed in 
$\S$6, in which we speculate on the effective temperature scale
of these objects based on their atmospheric evolution.  
The convergence of the L and T classes is addressed in $\S$7, based in part
on the increasing importance of CH$_4$ absorption in shaping
the spectra of late-type L dwarfs, including the possible presence of
weak CH$_4$ features in the K-band spectrum of the L7 V double
DENIS 0205-1159AB \citep{del97}. 
In $\S$8, we analyze our only peculiar
T dwarf, 2MASS 0937+2931, whose K-band flux is highly suppressed, possibly due
to increased surface gravity or 
diminished metallicity.  We summarize our
results in $\S$9.

\section{The 2MASS T Dwarf Search}

Since late 1998, we have been searching for T dwarfs using 
2MASS point source data.
This all-sky survey employs two automated 1.3m telescopes,
at Mt.\ Hopkins (USA) and
Cerro Tololo (Chile), each equipped with a three-channel camera
capable of simultaneous observations in the J (1.25 $\micron$), H
(1.65 $\micron$), and K$_s$ (2.17 $\micron$) near-infrared bands, 
down to nominal survey completeness limits of  
15.8, 15.1, and 14.3, respectively.   
Survey operations, begun in 1997 June at
Mt.\ Hopkins and 1998 March at Cerro Tololo, have recently been completed, and
final catalogs are expected to be released in 2002.  
Additional information about 2MASS can be found in Cutri
et al. 
(2001)\footnote{http://www.ipac.caltech.edu/2mass/releases/second/doc/explsup.html.}.

\subsection{Color Selection}

Molecular absorption and cool effective temperatures
cause the spectral energy distributions of T dwarfs to
peak at J-band (1.25 $\micron$), 
and near-infrared surveys such as 2MASS are particularly tuned to
detecting these objects.
Unfortunately, the near-infrared colors of T dwarfs (J-K$_s$ $\sim$ 0)
make them difficult to distinguish
from the overwhelming background of main sequence stars.
Figure 1 illustrates this problem in a near-infrared color-color diagram.
A typical sample of 2MASS point sources at moderate galactic latitude
are plotted as small points;
these objects were selected in a
one degree radius around 18$^h$ RA and +40$\arcdeg$ decl.\
(${\mid}b^{II}{\mid}$ = 26$\arcdeg$),
with J $<$ 15.8, H $<$ 15.1, and 
K$_s$ $<$ 14.3.  Their
distribution is well traced by the giant and dwarf stellar
tracks (solid lines) of \citet{bes88}.  
Extending from the red branch of the dwarf 
track, comprised of late M dwarfs \citep{leg98}, 
are L dwarfs (triangles) identified in 2MASS data
\citep{kir99,kir00}.  
At the opposite end of the main
sequence track lie Gliese 229B 
\citep[square]{leg99} and
the majority of T dwarfs (circles) 
identified by 2MASS and SDSS.
These objects lie just below the densest region of 
near-infrared color space,
occupied by early-type (O-G) stars.  ``Early'' T dwarfs (also known as 
L/T transition objects; Leggett et al. 2000) lie at colors intermediate
between the L dwarfs and Gliese 229B (J$-$K$_s$ $\sim$ 1--1.5) along 
the giant track.  Objects with even bluer near-infrared colors
(J$-$K$_s$ $\sim$ 0.5--1) are likely to be indistinguishable
from the vast majority of background sources.
 
Nonetheless, it is possible to distinguish T dwarfs using 
optical/near-infrared colors.
Low effective temperatures,
broadened K I absorption \citep{bur00,lie00},
and the possible presence of dust opacity \citep{tsu99},
make T dwarfs extremely red in the optical, with R$-$J $\gtrsim$ 9
\citep{mat96,gol98}.  Thus, 
T dwarfs similar to Gliese 229B with J $\gtrsim$ 10--12
would have been easily missed in earlier optical sky surveys
(e.g., POSS-II; Reid et al. 1991), which have typical limiting
magnitudes of R $\lesssim$ 19--21.

With these photometric properties in mind, we have constrained our 
search to the identification of the
coolest T dwarfs which have sufficient H$_2$O and 
CH$_4$ absorption to push their
near-infrared colors blueward of the 
majority of background stars (J$-$K$_s$ $\lesssim$ 0.6). 
We impose an additional 
optical/near-infrared color cut by rejecting candidate
objects with optical counterparts (from the USNO-A2.0 catalog; 
Monet et al.\ 1998) 
within 5$\arcsec$ of their 2MASS coordinates. 
We note that objects in the USNO-A2.0 catalog are
required to be detected on both the R- and B-band plates, so that faint red
objects visible only on the R-band plates 
(such as background M stars) generally remain on
our candidate lists.

\subsection{Search Samples}

We have drawn three samples from the 2MASS point source catalogues
in order to identify T dwarf candidates.  The properties of these
samples are summarized in Table 1.  Initial candidates were required 
to have ${\mid}b^{II}{\mid}$ $>$ 15$\arcdeg$ in order to reduce source
confusion near the Galactic plane, and high source density
regions around the Magellenic Clouds
and 47 Tuc were excluded.  Our main search sample, wdb0699, was drawn
from the 2MASS working database in 1999 June, at a time when 2MASS sky coverage 
was approximately 45\% complete.  Figure 2 displays 
a spatial map of the 35280
initial candidates (points) from this sample,
drawn from 16620 sq.\ deg.\ of 2MASS data.
Grey boxes highlight areas surveyed by 2MASS at the time of candidate selection;
data from 23510 scans were used in this sample, 12556 from the northern
hemisphere and 10954 from the southern hemisphere.
Note the increased source density close to the Galactic plane (dashed line),
due to a higher concentration of (typically) faint background stars.
Initial candidates
were further required to have J- and H-band detections with
J $<$ 16, J-H $<$ 0.3, and 
H-K$_s$ $<$ 0.3
(dashed lines in Figure 1).  Five T dwarfs
have already been published
from this sample 
\citep{me99,me00a,me00c}\footnote{A sixth T dwarf,
2MASS 1237+6526 \citep{me99} does
not fall into this sample as it is too faint at J (16.03$\pm$0.09).}.

The other search samples were drawn 
from the 2MASS Second Incremental Data Release
(2MASS IDR2), and
were used to investigate biases inherent to the  
color criteria of the primary wdb0699 sample.  The rdb0400 sample
reverses the H-K$_s$ color cut in order to identify faint sources
without K$_s$ detections, and thus artificially red H-K$_s$ colors.  The 
rdb0600 sample extends to J-H $\leq$ 0.4, with detections in all three
bands and J $\leq$ 15.  This sample was used to identify T dwarfs with weaker
CH$_4$ absorption, and was 
limited to the sky observable from the Southern Hemisphere
during the months of June and July.  Note that despite the stricter 
magnitude limit and smaller search area of this sample, a large number
of initial candidates was identified, emphasizing the 
difficulties caused by the high background source density 
in this region of color space.

In order to refine our candidate selection, we improved upon our 
optical/near-infrared color cut by visually examining optical images of
each candidate field.  R-band imaging data 
2$\farcm$5$\times$2$\farcm$5 around each candidate coordinate, taken
by the POSS-I, POSS-II, ESO/SERC, and AAO SES \citep{mor92} surveys,
were obtained from the Canadian Astronomy Data Centre's Digitized Sky
Survey image server\footnote{http://cadcwww.dao.nrc.ca/cadcbin/getdss.}.
Multiple epochs of each field were examined whenever possible.
Candidates with obvious optical counterparts at their 2MASS coordinates
were summarily rejected,
with the majority of these contaminants being close optical doubles
blended in the USNO-A2.0 catalog.  Objects identified as
proper motion stars based on 
multiple-epoch optical data were also removed from candidate lists.
Note that our criterion of no detection on the optical plates eliminates
many of the biases inherent to visual selection, including variable plate
quality and masking by bright star halos; these biases will tend to increase
the number of false candidates in our candidate list,
rather than eliminate bona fide T dwarfs.  
Nonetheless, 98.8\% of the initial 2MASS-selected
sources were eliminated in this manner, resulting in a substantially
reduced candidate pool (Cut \#2, Table 1, col.\ 6).

\subsection{Follow-up Imaging Observations}

Despite our optical and near-infrared color constraints,  
additional contaminants
remain.  Of primary concern are minor planets, which have 
near-infrared colors similar to T dwarfs \citep{syk00}, 
as indicated by the hashed boxes
in Figure 1.  These objects are also
absent in DSS images due to their 
motion.  Known minor planets are flagged in the 2MASS database and subsequently 
eliminated from search samples; however, uncatalogued asteroids remain.
To remove these objects from our candidate pool, we have conducted a 
near-infrared reimaging campaign, using the Near-Infrared Camera
\citep[hereafter IRCam]{mur95} mounted on the Palomar 60'' Telescope, 
and the Cerro Tololo Inter-American
Observatory (CTIO) Infrared Imager (CIRIM) 
and Ohio State InfraRed Imager/Spectrometer
\citep[hereafter OSIRIS]{dep93}
mounted on the CTIO 1.5m 
Telescope\footnote{Some additional observations were made using 
the Keck 10m Near Infrared Camera \citep{mat94} and Hale 5m D78
near-infrared camera.}.   
Observations of T candidates from our primary samples
are summarized in Table 2.  
Second epoch imaging has 
eliminated nearly 80\% of our Cut \#2 T dwarf candidates
with J-K$_s$ $<$ 0.5 (Table 1, col.\ 7).  To date, 99\%, 76\%, and 54\%
of our wdb0699, rdb0400, and rdb0600 Cut \#2 candidates have been reimaged,
respectively.

Table 3 lists all of the candidates from the three samples
present in the 2MASS IDR2 
that were not flagged as minor planets but were
absent in near-infrared reimaging.
Using ephemerides generated by D.\ Tholen for the 2MASS project, we
reexamined these objects for minor planet associations out to 25$\arcsec$.
A number of widely separated associations were made (Col.\ 8),
including the major asteroids Pallas, Irene, and Niobe.  The majority of these
apparently missed associations occurred between 1998 September and 1998
November, and are likely due to slight errors
in the ephemeris generation (D.\ Tholen, priv.\ comm.).  This problem is
currently being addressed and will be eliminated in the final release 
catalog; however, users of the 2MASS IDR2 should be aware of these
identifications. 

Determining the nature of our unconfirmed candidates can be done 
by investigating their spatial distribution.  Figure 3 shows 
the ecliptic latitude distribution of 144 unconfirmed targets from
our wdb0699 sample present in the 2MASS IDR2 and not
associated with a known asteroid in Table 3. 
We compared this to the distribution
of 288 flagged minor planets
in the 2MASS IDR2 having the same color and magnitude properties
as the wdb0699 sample, plus 25 missed associations
with the same constraints.  Both populations
peak near the ecliptic plane, and it is likely that 
the vast majority of our unconfirmed
candidates are probably uncatalogued minor
planets.  The distribution of the unconfirmed candidates
is distinctly broader, however.  We do not rule out that a few of
these transient sources may be faint variable
stars, infrared counterparts to
bursting sources, or, in at least one confirmed case, 
supernovae\footnote{2MASSI J0101271-073636 is the near-infrared counterpart
to SN 1998DN \citep{cao98}.}.

Additional imaging observations were made using the Palomar 60'' CCD
Camera with a Gunn r broad band filter \citep{wad79}
over a series of runs extending from 1999 May through 2000
December.  The purpose of these observations was to eliminate 
proper motion stars and background objects close to bright optical
sources from our candidate pool.  A total of 49 confirmed sources were
rejected based on r-band detections.  The typical limiting
magnitudes of r $\sim$ 19--20 correspond to color limits of 
r-J $\lesssim$ 6, equivalent to spectral types earlier than 
M6 V to M8 V \citep{kir99}.

\section{Near-Infrared Spectroscopic Observations}

The final confirmation of a candidate object as a T dwarf was made through the
identification of CH$_4$ absorption bands in 
1--2.5 $\micron$ spectral data.  Observations made
using the Near Infrared Camera \citep[hereafter NIRC]{mat94}, 
mounted on the Keck I 10m, and a similar instrument,
D78, mounted on the Palomar 5m Hale Telescope, are discussed in $\S\S$3.1
and 3.2; observations
made using OSIRIS, mounted on the CTIO 4m Blanco Telescope, are 
discussed in $\S$3.3.  The features seen in these spectra are discussed in
$\S$3.4.  Finally, the accuracy of the NIRC and OSIRIS datasets is 
examined in $\S$3.5 by comparison
with each other and to previously published data.

\subsection{NIRC Spectra}

The NIRC instrument is comprised of a 256$\times$256 InSb array camera
mounted on the f/25 Forward Cassegrain focus of the Keck I 10m Telescope.
Filters and grisms permit low-resolution ($\lambda$/$\Delta\lambda$ 
$\sim$ 60--120) spectroscopy from 1--2.5 $\micron$ in two
settings: 1--1.6 $\micron$, using the 150 lines mm$^{-1}$ 
grating blazed at 1.7 $\micron$
and JH order blocking filter; and 1.4--2.5 $\micron$, using the
120 lines mm$^{-1}$ 
grating blazed at 2.1 $\micron$
and HK order blocking filter.  Spectral resolutions on the 
0$\farcs$15 pixel$^{-1}$ chip are 
49 {\AA} pixel$^{-1}$ and 59 {\AA} pixel$^{-1}$, respectively.

A log of observations of identified T dwarfs and known L dwarfs, taken
over four nights, is given in Table 4a.  
Conditions
on 1999 November 18 (UT) were clear with good seeing ($\sim$ 0$\farcs$4);
conditions on 2000 January 23-24 (UT) were also clear but windy, with
poor seeing ranging from 0$\farcs$7 to 1$\farcs$3; finally, 
conditions on 2000 July 22 (UT) were plagued by heavy to light cirrus,
although seeing was
0$\farcs$4.  Targets were initially acquired in imaging mode and
placed into a 0$\farcs$53 slit.  Multiple 
sets of three exposures dithered 5$\arcsec$
along the slit were made for each object at the JH and HK settings.  F and
G dwarf standards close to the target objects, chosen for their weak 
Hydrogen lines, were observed using the same 
instrumental setup.  Finally, 
spectral dome flats were observed in order
to calibrate pixel response.  

Science images were initially divided by a median-combined,
dark-subtracted, normalized dome flat image to correct for pixel-to-pixel 
spectral response variations, and bad pixels were removed by
interpolation using a mask constructed from dark and flat field
frames.  These processed images were pairwise subtracted to remove dark
current and sky background, and curvature of the
dispersion along the chip was corrected by tracing OH sky lines.  
Spectra were then optimally extracted using a 
smoothed weighting function.   
At each dispersion position, object and standard spectra were ratioed and then 
multiplied by a blackbody appropriate to the standard star
\citep{tok00}.  The resulting flux-calibrated spectra for each
grism setting were then separately combined
using a clipped average (rejecting 3$\sigma$ outliers at each wavelength).  
Wavelength calibration was computed 
using coefficients listed in the NIRC instrument
manual\footnote{http://www2.keck.hawaii.edu:3636/realpublic/inst/nirc/manual/Manual.html.}.
 
Spectra from 1--1.6 $\micron$ and 1.4--2.5 $\micron$ were
then combined and flux calibrated.
First, we scaled the spectra by a multiplicative factor
to match the overlap region from 1.4--1.6 $\micron$.  Then, using 
spectral data of Vega 
\citep{ber95} and 2MASS photometry, 
we calculated flux corrections by integrating 
the 2MASS filter response curves\footnote{For 2MASS filter response functions, 
see http://www.ipac.caltech.edu/2mass/releases/second/doc/sec3\_1b1.html.}
over calibrator and object spectra, using the relation:
\begin{equation}
F_{obj}^{(corr)}({\lambda}) = F_{obj}^{(uncorr)}({\lambda}) {\times} 10^{-0.4m_{b}} {\times} 
\frac{\int F_{Vega}({\lambda}') T_b({\lambda}') d{\lambda}'}
{\int F_{obj}^{(uncorr)}({\lambda}') T_b({\lambda}') d{\lambda}'},
\end{equation}
where $F_{obj}^{(corr)}({\lambda})$ and $F_{obj}^{(uncorr)}({\lambda})$ are the
corrected and uncorrected flux densities of the object spectrum, 
$F_{Vega}({\lambda})$ is the flux density of Vega, $m_{b}$ is the magnitude
of the object through filter $b$, and $T_b({\lambda})$ is the
transmission function.  For this calibration, we chose to use 
$b$ = J-band magnitudes, as T dwarfs tend to be
brightest at this band, reducing photometric errors.  

Final NIRC spectra for confirmed T dwarfs are shown in Figure 4, along
with data for the L6.5 V 2MASS 0920+3517 and the L7.5 V 2MASS 0825+2115 
\citep{kir00}.  Spectra are normalized at their
J-band peaks, offset (dotted line), and ordered by
increasing 1.1 and 1.6 $\micron$ absorption.  Major absorption bands of
H$_2$O, CH$_4$, CO, FeH (1.19, 1.21, and 1.237 $\micron$), and CIA H$_2$ are
indicated, as are atomic lines of K I (1.169, 1.177, 1.243, and 1.252
$\micron$).  

To test our flux calibration, we measured spectrophotometric colors for all
of the calibrated spectra in the 2MASS near-infrared bands, and compared 
these to 2MASS photometric colors.  Results are given in Table 5.  
Residuals ($\delta$ $\equiv$ photometry minus spectrophotometry)
are typically on the order of 0.1 to 0.3 mag, slightly larger than the
2MASS photometric errors, with some colors discrepant by more than
0.6 mag.  Nonetheless, the mean of these residuals, 
${\langle}{\delta}_{J-H}{\rangle}$ = $-$0.04$\pm$0.17,
${\langle}{\delta}_{H-K_s}{\rangle}$ = $-$0.03$\pm$0.29, and
${\langle}{\delta}_{J-K_s}{\rangle}$ = $-$0.06$\pm$0.30, 
imply overall consistency in the relative calibration.  Significant color
differences may be due to the effects of cirrus during our 2000 July 22
observations (affecting data for
2MASS 1553+1532, 2MASS 2254+3123, and 2MASS 2339+1352), or possibly low
signal-to-noise photometry of objects barely detected by 2MASS
at K$_s$ (e.g., 2MASS 0937+2931).  Regardless, the flux calibration appears to 
be adequate for the characterization of gross spectral morphology. 

\subsection{D78 Spectra}

Additional low-resolution observations were made on 
2001 January 4-5 (UT) using the
D78 near-infrared camera, mounted at the f/70 Cassegrain
focus of the Palomar 5m Hale Telescope.  
This instrument is similar in construction
to NIRC, providing grism spectroscopy from 1--2.5 $\micron$ in three
orders (3$^{rd}$, 4$^{th}$, and 5$^{th}$) with $\lambda$/$\Delta\lambda$ 
$\sim$ 100.  Resolution on the 0$\farcs$125 pixel$^{-1}$ chip ranges from 
23 to 37 {\AA} pixel$^{-1}$.

Weather was generally clear on 2001 January 4 with good seeing
(0$\farcs$7); however conditions on 2001 January 5 were poor, with
clouds frequently obscuring observations.  
Only one T dwarf was identified in these observations,
2MASS 0755+2212, observed on January 5.  This object 
was acquired in imaging mode and placed into a 0$\farcs$71
slit.  Three exposures dithered 10$\arcsec$ along the slit were
obtained at each setting.  Pixel response and flux calibrations at J- and 
H-bands were made
using observations of the G0 V standard HD 4307 taken on 
January 4, which was 
smeared along the slit by driving the tip-tilt secondary with 
a 20 Hz, 40$\arcsec$ (peak-to-peak amplitude)
triangle wave.  Corresponding sky frames 
were also obtained.  Observations made at K-band were calibrated using 
observations of the G7 V HD 67767 (= $\psi$ Cnc) obtained on January 5
in a similar fashion. 

Science images
at each grism setting were initially 
divided by the sky-subtracted standard observations, 
and then pairwise differenced to remove dark current and sky background.
The resulting images were then shifted and coadded to boost signal-to-noise,
after correcting for dispersion curvature along the slit.  Spectra were then 
extracted from the combined images.
The wavelength
scale was computed using known dispersion coefficients
calibrated to observations 
of the 1.2818 Pa${\beta}$ emission line in NGC 2392 \citep{rud92}, obtained 
on January 5.  Finally, spectral orders were multiplied by
the appropriate blackbody and combined; no 
attempt was made to scale the individual orders or 
compute flux calibrations.

Figure 5 shows the reduced spectrum of 2MASS 0755+2212,
normalized at the J-band peak; the poor
observing conditions are evident in the significant amount of
noise present in the data.  
Nonetheless, H$_2$O absorption is seen at 1.2, 1.35, and 1.9
$\micron$, and CH$_4$ bandheads at 1.6 and 2.2 $\micron$ can be readily 
identified.
2MASS 0755+2212 was originally selected as a 2MASS T candidate in early
1999, but was initially rejected from the candidate list due to the presence of
a faint and extended
optical counterpart.  Based on Keck imaging observations
made with the Low Resolution Imaging Spectrograph \citep{oke95}, 
it appears that this 
counterpart is in fact a background galaxy coincident with the T dwarf.

\subsection{OSIRIS Spectra}

The OSIRIS instrument is a near-infrared imager/spectrometer mounted on
the f/14.5 Cassegrain focus of the CTIO 4m Blanco Telescope.  It is 
equipped with a 1024$\times$1024 HgCdTe HAWAII array with 18.5 $\micron$
pixels, sampling the entire illuminated field.
Use of the single diffraction grating blazed at 6.6 $\micron$ with the 
120 lines mm$^{-1}$ grating and f/2.8 camera provides
simultaneous, cross-dispersed, moderate resolution 
($\lambda$/$\Delta\lambda$ $\sim$ 1200) spectroscopy from
1.2--2.35 $\micron$ in four orders: J (3$^{rd}$ and 4$^{th}$), 
H (5$^{th}$), and K (6$^{th}$).  Resolution on the 0$\farcs$403 pixel$^{-1}$
chip ranges from 4.4 to 8.8
{\AA} pixel$^{-1}$.

Table 4b summarizes our OSIRIS observations.  Data obtained during 1999 
December 20-22 (UT) were taken in clear to hazy conditions, 
with seeing ranging from
0$\farcs$6 to 1$\farcs$0; observations made in 2000 July 15-19 (UT) were 
taken in conditions ranging from clear to cloudy, and seeing ranged from
0$\farcs$6 to 1$\farcs$8.  Additional observations were also made
during 1999 July 26-29 (UT), with preliminary
results reported in \citet{me00a}; however,
targets from this run were reobserved at later dates to
higher signal-to-noise, and the early
data are not included here.  A total of 14 
T dwarfs were observed using OSIRIS, along 
with a number of M and L dwarfs for comparison.  
Targets were acquired in imaging mode and placed into a 1$\farcs$2 $\times$
30$\arcsec$ slit.  Observations were made
in sets of 5 exposures dithered 4--5$\arcsec$ along the slit, with 
individual integrations ranging from 15 to 250 sec per exposure.  A-type
stars were observed near the target objects for flux calibration
and telluric corrections.
Spectral lamps reflected off of the 4m dome spot were observed
each night for pixel response calibration.

Science images were initially trimmed to eliminate vignetted
columns, divided by a median-combined, dark-subtracted flat field, and 
corrected for bad pixels by interpolation,
using a mask created from flat-field and dark
exposures.  Images were then pairwise subtracted to eliminate sky
background and dark current.  Curvature of the dispersion lines was
determined using standard star observations, 
and both standard and object spectra
were optimally extracted using a smoothed weighting function.  Spectra
from each order were then scaled by a multiplicative factor
and combined using a clipped average.
Wavelength calibration was done using OH line identifications from 
\citet{oli92}.
Telluric
H$_2$O absorption was corrected using spectra of the standard star, and a flux 
correction was calculated by interpolating over hydrogen 
Paschen and Brackett lines
and multiplying by the appropriate blackbody. Finally, 
the resulting flux-calibrated spectral
orders were combined by first scaling each
order by a multiplicative factor to match overlap regions
(typically 1.29--1.31, 1.53--1.57, and 1.96--1.98 $\micron$), and 
then correcting the combined spectra
to 2MASS H-band magnitudes (not J, due to the short
wavelength cutoff) using Equation 1.  Note that the overlap between the H- and
K-band orders falls within the 1.9 $\micron$ H$_2$O band; because of this,
we applied no scaling corrections between these two orders. 

Resulting spectra for confirmed T dwarfs are shown in Figure 6,
along with data for LHS 511AB \citep[M4.5 V]{luy79},
HB 2124-4228 \citep[M8.5 V]{haw88}, 
2MASS 2224-0158 \citep[L4.5 V]{kir00}, and DENIS 0205-1159AB
\citep[L7 V]{del97}.  Spectra are normalized, offset, and ordered as in 
Figure 4.  Bands of H$_2$O, CH$_4$, CO, FeH, and CIA H$_2$ are indicated, 
as are lines of K I (1.25 and 1.52 $\micron$ doublets), 
Na I (2.21 $\micron$ doublet), and Ca I 
(1.314, 1.98, and 2.26 $\micron$ triplets).  
The 1.52 $\micron$ K I doublet and Ca I lines, prominent in M dwarfs
\citep{jon94} are only detectable in 
the M4.5 V LHS 511AB.  Regions of increased noise are due to 
telluric H$_2$O opacity; note, however, that higher-energy H$_2$O transitions
originating from the target objects extend well into the J, H, and K
telluric windows.  

\subsection{Near-infrared Spectral Features}

The spectra in Figures 4 and 6 show that H$_2$O and ultimately CH$_4$
absorption are the dominant features shaping the spectra of objects cooler
than late M dwarfs.  The H$_2$O bands centered at 1.15, 1.4, and 1.9
$\micron$ segregate the spectral energy distributions into distinct
z-, J-, H-, and K-band peaks as early as L6.5 V, while the addition of CH$_4$
bands beginning at 1.05, 1.3, 1.6, and 2.2 $\micron$ further confine the
emitted flux into narrow peaks centered at 1.08, 1.27, 1.59, and 2.07
$\micron$.  Note that CH$_4$ is necessary to produce the strong absorption at
1.15 $\micron$, a feature also seen in the spectra of giant planets 
\citep{fin79}.  In addition, the gradually developing slope between the 
1.27 $\micron$ peak and the 1.4 $\micron$ H$_2$O band is also caused by 
CH$_4$ at 1.3 $\micron$.  

In addition to H$_2$O and CH$_4$ absorptions, 
the K-band spectral region is shaped by 2.3 $\micron$ CO 
in M and L dwarfs, a feature that has 
been detected in the early T dwarfs as well \citep{leg00}.
The distinct bandheads seen in LHS 511AB strengthen through
L4.5 V, but appear to weaken beyond this, as they are barely
detectable in our spectrum of SDSS 1254-0122 (an early T dwarf).  
None of the 2MASS T dwarfs show
any indications of CO absorption.
The other important molecular absorber at K-band is CIA H$_2$, which has
no distinct bandhead but likely suppresses flux throughout the 2--2.5
$\micron$ region starting in the latest L dwarfs (see $\S$7). 
With the presence of H$_2$O, CH$_4$, CO,
and CIA H$_2$ absorption features, the K-band peak undergoes a 
dramatic evolution.  
Starting from a flat slope with CO bandheads in mid-M dwarfs, 
the K-band becomes a plateau girded by CO and H$_2$O absorptions 
in the L dwarfs; it then develops
from a double plateau with the dual presence of CH$_4$ and CO in the
early SDSS T dwarfs to an increasingly suppressed and
rounded hump centered at 2.11 $\micron$ in the earliest
2MASS T dwarfs; finally, we see a smooth triangular peak centered at 2.07 
$\micron$ in the latest T dwarfs (those exhibiting the strongest 
CH$_4$ and H$_2$O absorption).

Finer features of FeH and K I
have been identified in the J-band spectra of M and L dwarfs
by \citet{mcl00}, and these features are also indicated in 
Figures 4 and 6. The resolution
of the NIRC data is generally insufficient to resolve these lines,
although the 1.25 $\micron$ K I
doublet can be seen as a notch in the blue wing of the J-band peak of the T
dwarf spectra.  OSIRIS data can resolve this doublet, however.  Figure
7 shows a close-up of the 1.19--1.34 $\micron$ spectral
region for the OSIRIS data; spectra are normalized and offset as in Figure 6.
The K I lines are present as early as M4.5 V, remain strong from the latest
M dwarfs through most of the T dwarfs, but are weak in Gliese 570D.  This 
behavior is mimicked in the evolution of the 
notch feature in the NIRC spectra.  Note that the
1.2432 $\micron$ line is blended with an FeH feature at 1.237 $\micron$ in 
HB 2124-4228 and 2MASS 2224-0158.  The 1.2522 $\micron$ line appears
to be relatively stronger in the T dwarfs, an
asymmetry noted by \citet{leg01} in the L7 V DENIS 0205-1159AB.  Lines
of Na I (1.138, 1.141 $\micron$) and Rb I (1.323 $\micron$) noted by 
\citet{mcl00} are likely present in the L dwarf spectra,
but are weakened by overlying H$_2$O opacity; 
we do not see these features in our data.

Finally, we note that several weak absorption 
features are seen in the H-band peaks
of our M and L dwarf spectra, near 1.58, 1.61, 1.63, and 1.67 $\micron$.
\citet{rei01} have also detected these features in
their spectral data, suspecting that they may be due
to an unidentified molecular absorber.  The features are
fairly close to both Brackett Hydrogen lines (1.5885, 1.6114,
1.6412, and 1.6811 $\micron$), which are strong in the
spectra of our A-type standards, and weak telluric
absorptions.  We believe that these features may be artifacts
produced during our flux calibration due to imperfect
interpolation over the Brackett lines (see $\S$3.3), as they
are not obviously present in the data prior to flux
calibration.  Recalibrating the data using later-type
standards may help to elucidate the authenticity and origin
of these weak features; however, their presence does not
effect the analysis presented here.

\subsection{Comparison of Data Sets}

Data from NIRC and OSIRIS observations comprise our primary spectral 
sample.  Because the instrumental setup, resolution, standards, and
reduction procedures were slightly different for these two datasets,
we have observed several objects with both instruments to
identify any inherent biases.  Figure 8 plots OSIRIS (light gray) and 
NIRC (black) data for 2MASS 2254+3123, 2MASS 0559-1404, 2MASS 2356-1553,
2MASS 0937+2931, 2MASS 1553+1532,
and Gliese 570D.  
In general,
wavelength and overall flux levels agree well, but there are some 
differences seen in the individual flux peaks.  This is most striking at H- and
K-bands, where the NIRC data are generally brighter, although differences
can also be seen at J-band at the 10--15\% level as well.  
Spectral data for 2MASS 0559-1404 are the most
discrepant, with NIRC observations as much as 
35\% brighter at K; there is also a
difference in slope seen in the 1.6 $\micron$ CH$_4$ band.  The 
discrepancies at K may be due to the
difficulty in determining overlap between the OSIRIS H- and K-band spectral
orders (see $\S$3.3).  A similar problem may be present in the
spectrum of 2MASS 1553+1532.
However, 
spectral differences generally do not exceed 10--15\%, similar to the 
differences seen between the 2MASS photometry and
NIRC spectrophotometry ($\S$3.1).

Figure 9 shows OSIRIS and NIRC data (light gray) compared to data from the
literature for DENIS 0205-1159AB \citep{leg01},
SDSS 1254-0122 \citep{leg00}, 2MASS 0559-1404 \citep{me00c},
SDSS 1346-0031 \citep{tsv00}, and SDSS 1624+0029 \citep{str99}.
Again, overall wavelength calibration and flux levels appear to be 
consistent, and band morphologies are generally coincident.
Most discrepant are data for 2MASS 0559-1404, which is roughly $\sim$ 50\% 
brighter at K-band in the NIRC spectrum (left panel).  
Data from \citet{me00c} were  
obtained with the Palomar 60'' 
Cornell-Massachusetts Slit Spectrograph
\citep{wil01a}; this instrument has
no overlap between the H- and K-bands, and it is
possible that the K-band segment of these data 
was improperly scaled, particularly
given the overall agreement from 1--2 $\micron$. 
OSIRIS data for 2MASS 0559-1404 match the \citet{me00c} spectrum better
at K (right panel), but not as well at H-band.
The spectra of the other
objects agree quite well with published results, and
small offsets 
may simply be due to differences in resolution or signal-to-noise.  
In general, we find that typical discrepancies do not exceed 
10--15\%, quite adequate for the
task of spectral classification.
 
\section{New T Dwarfs}

Based on our spectroscopic results, we have identified
eleven new T dwarfs in our 2MASS search samples.  
The photometric properties of these
objects are summarized in Table 6, and 5$\arcmin$$\times$5$\arcmin$
images of each field in the optical
(R-band) and near-infrared (J- and K$_s$-bands) are given in Figures 10a--d.
Images are oriented with North up and East to the left.  
These discoveries span a much broader range of colors than previous 
detections, with $-$0.89 $\leq$ J-K$_s$ $\leq$ 0.45, reflecting their significant
spectral variations.  Note, however, that none of these objects are as red as the 
\citet{leg00} early T dwarfs, a consequence of our near-infrared color
constraints.

Considerable follow-up remains for the rdb0400 and rdb0600 samples,
but a large percentage of the wdb0699 candidates have been characterized
through follow-up imaging and spectroscopy.  In addition to 34 confirmed
objects from this sample observed at r-band, 33 objects have had near-infrared
spectroscopic data taken.  The majority of these followed-up candidates
(Table 1, Col.\ 8) appear to be faint background stars or 
unidentified proper motion stars,
based on their r-J colors or spectral morphology.  The remaining 26 candidates
are mostly faint objects close to bright optical sources, and are likely to 
be background stars with compromised photometry.  Follow-up of
our best candidates from the wdb0699 sample is therefore nearly
complete, and we can use this sample to make a rough estimate
of the space density of 
T dwarfs with J-K$_s$ $\lesssim$ 0.6.  With a total of
14 identified T dwarfs (Table 1, Col.\ 9), the wdb0699 sample yields an areal
density of 8.4$\times$10$^{-4}$ T dwarfs deg$^{-2}$, or roughly
one T dwarf for every 1200 deg$^2$ down to J $\sim$ 16.  
This is nearly one-third the original estimate from
\citet{me99}, and implies only 35 T dwarfs with J-K$_s$ $\lesssim$ 0.6
detectable by 2MASS over the whole sky.
If we adopt a distance limit based on the absolute brightness of Gliese 229B
\citep[M$_J$ = 15.51$\pm$0.09]{leg99} and our J-band magnitude limit, we derive a space
density of 4.2$\times$10$^{-3}$ pc$^{-3}$.  

Our estimate can be compared to other T dwarf search samples with multiple
detections.
\citet{tsv00} calculate a density of 5$\times$10$^{-2}$ pc$^{-3}$ 
based on two detections in
130 deg$^2$ of SDSS data, with z$^*$ $\lesssim$ 19.8.  The three early T dwarfs
identified by \citet{leg00} in 225 deg$^2$ of SDSS data yield a space density 
between 1.9$\times$10$^{-2}$ and 1.8$\times$10$^{-3}$
pc$^{-3}$, the upper limit based on scaling the J-band magnitude of their 
faintest object (SDSS 0837-0000; J$_{UFTI}$ = 16.90$\pm$0.05) to Gliese 229B
(which is likely to be intrinsically fainter at J-band), while 
the lower limit scales the 
faintest K-band detection to the L8 V Gliese 584C 
\citep[M$_{K_s}$ = 12.89$\pm$0.09]{kir01a}.  
Our estimates are generally lower than those of SDSS, but without
quantification of biases inherent to our sample (e.g., color selection,
photometric completeness), 
we emphasize that our results are likely a lower
limit to the true T dwarf space density.
As a comparison, the L dwarf space density ranges from 
2$\times$10$^{-3}$ to 8$\times$10$^{-3}$ pc$^{-3}$, based on
2MASS \citep{kir99} and SDSS \citep{fan00} search samples, respectively, 
scaled to  
absolute magnitudes from \citet{kir00}.  
The low-mass stellar (0.1--1.0 M$_{\sun}$) density, on the other hand,
is roughly 
2$\times$10$^{-2}$ pc$^{-3}$ \citep{rei99}.  Our T dwarf
space density estimate applied to the mass function
simulations of \citet{rei99} predicts a 
substellar mass function
that goes as $\frac{dN}{dM} \propto M^{-1}$, consistent with results from
various cluster surveys \citep{luh00}.

Table 7 lists all of the 24 currently known T dwarfs\footnote{Additional 
T dwarfs identified by \citet[this volume]{geb01b} in SDSS data are not 
included here.}.
2MASS photometry is
listed where available.  Many more T dwarfs are
expected to be found with the anticipated completion of the 
2MASS and SDSS surveys.

\section{Spectral Typing}

The evolution of molecular features seen in Figures 4 and 6
suggest the 
makings of a classification scheme for T dwarfs.  To do this, we have
followed the philosophy of the MK system, basing our classification criteria
exclusively on the observed spectral features,
without presumption of the precise physical properties of these objects
(i.e., T$_{eff}$, gravity, or metallicity).  This is an important point,
as models of cool brown dwarfs, although greatly improved over
earlier work, do not yet adequately predict all of the features seen in
T dwarf spectra \citep{leg01}.  It is, in our opinion, better to classify 
an object on its actual 
appearance, rather than on possibly incorrect physical interpretations.

\subsection{Near-Infrared or Optical Classification?}

A classification scheme for T dwarfs
should be able to distinguish these objects from the warmer L dwarfs.
One might expect to do this at red optical wavelengths,
where both the \citet{kir99} 
and \citet{mrt99} L dwarf classification schemes are defined.
\citet{me01a} and \citet{kir01b} have shown that T dwarfs can be 
segregated from L dwarfs in this spectral regime, and
features useful for classification have been identified.
However, T dwarfs are exceedingly faint shortward of 1 $\micron$, 
(i$^*$-J $\gtrsim$ 7.5; Leggett et al.\ 2000), and data are 
difficult to obtain using even the largest ground-based telescopes.

In the near-infrared, however, T dwarfs 
are significantly
brighter, and large samples can be
observed using only moderate-sized (4m class) facilities.
Because of this utility, we advocate a T classification
scheme based on 1--2.5 $\micron$ spectroscopy.  Corresponding classification 
schemes for L dwarfs in this spectral regime have only recently been proposed,
however.
Preliminary efforts have been made by \citet{tok99} using indices
that measure H$_2$O and H$_2$ features at K-band.
\citet{rei01} have mapped the \citet{kir99} L dwarf scheme onto near-infrared
spectra using the strengths of the 1.4 and 1.9 $\micron$ H$_2$O bands.
\citet{tes01} find similar correlations using their own H$_2$O and color
indices. 
Other near-infrared investigations are currently underway by \citet{mcl01}, 
\citet{giz01b}, and \citet{wil01c}.

In order to gauge how well our classification system segregates L and T dwarfs
in the near-infrared, we have augmented our spectral sample with NIRC and
OSIRIS observations of known M and L dwarfs.  We have also included data for
M, L, and T dwarfs obtained from the literature 
\citep[hereafter CGS4 data]{geb96,str99,tsv00,leg00,leg01,rei01}.
Note that 
no attempt is made to define a near-infrared L classification scheme;
our purpose is simply to examine the behavior of our T classification
(in particular, spectral indices, $\S$5.3) over a broader sample.  We examine
how L dwarf near-infrared schemes may carry into the T dwarf regime in
$\S$5.6.

\subsection{Defining the Subtypes}

The first step toward classification is to determine groupings, or
subtypes.  Numerous techniques for identifying spectral subtypes
in the MK system have been proposed \citep{mor50,bai01}.  
We have chosen to define T subclasses by the following procedure: 
First, we visually compared the spectra,
normalized at their J-band peaks, with OSIRIS and CGS4 data
smoothed with a Gaussian filter
to match the resolution
of the NIRC data.  These spectra were then sorted into 
a morphological order based on the strengths of the 1.6 and 2.2
$\micron$ CH$_4$ bands, the 1.15 $\micron$ H$_2$O/CH$_4$ band, and the
overall appearance of the J, H, and K-band peaks, similar to the
ordering used in Figures 4 and 6. 
The CH$_4$ absorption bands are expected to strengthen with decreasing
T$_{eff}$ \citep{bur97}, so that a classification scheme based on these
features should adequately represent an underlying temperature
sequence. 
The spectra were then grouped into subtypes of similar morphology,
with the number of subtypes chosen to evenly represent the spectral
variations seen.  This process was repeated until a consistent system
developed, resulting in seven groups we have labeled 
T1 V, T2 V, T3 V, T5 V, T6 V, T7 V, and T8 V.
Our two subtype omissions, T0 V and T4 V, are based on apparent gaps in the 
photometric colors and spectral morphologies
of the currently known T dwarf population.  
The earliest T dwarf so far identified, SDSS 0837-0000, is considerably
bluer in J-K$_s$ color that the latest L8 dwarfs, suggesting the presence
of an additional (earlier) subtype with weaker methane
features.  We have also omitted a class between the latest of the Sloan 
transition dwarfs, SDSS 1021-0304, and the reddest 
2MASS T dwarf, 2MASS 2254+3123, again based on the
significant near-infrared color differences, 
as well substantial relative suppression 
of the H- and K-band peaks
in the spectrum of the latter object.  Note that all of the 2MASS discoveries
fall exclusively in the last 
four subtypes, due to the color constraints
discussed in $\S$2.

Representative bright standards for each subtype were then chosen. 
Low resolution spectra of these standards are shown in Figure 11, along
with data for the L7 V DENIS 0205-1159AB \citep{leg01}.  Spectra are
normalized at their J-band peaks with zero point offsets indicated by
dashed lines.  Data from \citet{leg00,leg01} have been
degraded to NIRC resolution using a Gaussian filter.  
The spectral properties of these
subtypes are summarized in Table 8.  The 
progressive strengthening of the H$_2$O
and CH$_4$ bands throughout the sequence is readily apparent.  
Early types are distinguishable from L dwarfs by the
presence of weak CH$_4$ absorption at 1.6 and 2.2 $\micron$, the latter
of which appears in conjunction with 
CO at 2.3 $\micron$ in types T1 V, T2 V, and T3 V.
Other important features
include the strengthening 1.15 $\micron$ H$_2$O/CH$_4$ band,
weakening 1.25 $\micron$ K I doublets, suppression of
H- and K-band flux relative to J (significant between T3 V and T5 V), and
narrowing of the 1.08, 1.27, 1.59, and 2.07 $\micron$ peaks through T8 V.
The K-band peak of the later subtypes evolves from a 
rounded hump centered at 2.11 $\micron$
(T5 V), to a sloped, asymmetric peak notched at 2.17 $\micron$ (T7 V),
and finally into a sharper, symmetric peak centered at 2.07 $\micron$ (T8 V).

\subsection{T Dwarf Spectral Indices}

T dwarfs can be classified, to first order, by a simple visual
comparison against the standard sequence plotted in Figure 11.  A more
quantitative approach is to use spectral indices, ratios of fluxes in
spectral regions which measure specific morphological features,
such as the CH$_4$ and H$_2$O 
bands and the changing appearance of the K-band peak.
Based on the subtype properties listed in Table 8, we have defined a suite
of spectral indices, listed in Table 9.
For each index, flux regions were
chosen to sample the feature of interest and a nearby pseudo-continuum
point (typically the flux peaks at J, H, or K), and the mean of the data
in these regions were ratioed.  The H$_2$O, CH$_4$, and CO indices
measure the relative depths of their respective bands, while the color indices
H/J, K/J, and K/H are approximate measures of color around the 1.25, 
1.6, and 2.1 $\micron$ flux peaks.  Both the 2.11/2.07
and K shape indices measure the change in the K-band peak, influenced by
the molecular absorbers H$_2$O, CH$_4$, CO, and CIA H$_2$.

The behavior of these indices was examined for the T dwarf standards,
and a subset of indices showing obvious trends 
were chosen for use in spectral typing.  These include the 
H$_2$O-A, H$_2$O-B, CH$_4$-A, CH$_4$-B, and CH$_4$-C band indices; the
H/J and K/J color indices; and the 2.11/2.07 K-band ratio.  Values of these
indices measured for late-L \citep{kir99} and T dwarf
standards are listed in Table 10.  NIRC data
were used to compute the standard
ratios for types T5 V through T8 V, while CGS4 data 
were used for types L5 V through T3 V. 

Using the standard values as benchmarks, we then computed
spectral types for the remaining T dwarfs by directly comparing the
individual indices.  Tables 11a--c list the index values and corresponding
subtypes derived from 
NIRC and D78, OSIRIS, and literature data, respectively.  
Only measurable indices 
are listed in each set; for example,
the spectral limits of the OSIRIS and D78 data, and significant K-band noise,
restrict the use of the H$_2$O-A, CH$_4$-C,
and 2.11/2.07 indices.  We note that the CH$_4$-C, H/J, and K/J indices
appear to saturate in the latest subtype standards, obviating
their use in distinguishing between types T7 V and T8 V; in addition,
the 2.07/2.11
index shows a consistent trend only for types T3 V through T8 V.

Individual subtypes were then averaged into a decimal classification
after rejecting the single highest and lowest values; these are 
listed in the last column of each table.  Uncertainties 
were determined from the scatter of the index subtypes,
which in no case exceeded $\pm$0.9 subclasses.  Finally, the
decimal classifications
were rounded off to the nearest 0.5 subclass to produce the  
final subtypes listed in Table 7 (col.\ 2).  In general, preference
was given to classifications derived from NIRC and CGS4 data, as these
samples were used to derive the standard ratios.  Nonetheless, differences
between the various datasets for common objects did not exceed 
$\pm$0.9 subclasses.   
Objects observed on multiple occasions by the same instrument showed no
differences in the their final spectral types,
indicating that the classifications are robust (i.e., not changed by 
intrinsic variability in the source).
Note that final subtypes for 2MASS 0755+2212
and NTTDF 1205-0744 are assigned tentative (uncertain) values due to 
the poor quality
of their spectral data.  Ignoring these objects, comparison between
the datasets indicates that the
derived subtypes have typical uncertainties of $\pm$0.5 subclasses.

\subsection{Spectral Index Relations}

The behavior of our derived spectral indices as compared to spectral
type allows us to ascertain  
the usefulness of these indices for classification, while also probing
the evolution of spectral features.
Figures 12a--d
plot these ratios for objects with types M4.5 V through T8 V, measured from
the NIRC (diamonds), OSIRIS (triangles), and CGS4 (squares) datasets.
Filled circles indicate spectral standard values;  
T spectral types are those listed in Table 7.
Note that we have ended the L sequence in this (and subsequent) diagrams
at L8 V; 
motivation for this choice is discussed in $\S$7.

The H$_2$O indices (Figure 12a) are seen to decrease monotonically
over the entire sequence, except for the H$_2$O-C index, which increases
from T2 V to T8 V.  Decreasing indices imply strengthening bands, which can be 
easily seen in the spectral data shown in Figures 4 and 6.  The slopes
are greater
in the T dwarfs, reflecting either increased band contrast with the loss
of higher-energy (and hence higher temperature) wing transitions; increased
H$_2$O photospheric abundance with the formation of CH$_4$, via:
\begin{equation}
CO + 3H_2 \rightarrow CH_4 + H_2O;
\end{equation}
or increased transparency due to the settling of dust cloud layers 
\citep{ack01,bur01}.  The sharper increase in slope for the
H$_2$O-A index is due to the additional contribution of CH$_4$
absorption in this feature.  Note that the H$_2$O-C index has a great
deal of scatter in the T dwarf regime, and its behavior is likely
affected by CH$_4$ and CIA H$_2$ absorption at K-band.

The CH$_4$ indices (Figure 12b) are generally flat for most
of the M and L dwarfs earlier than type L5 V, but rapidly decrease starting
around L5 V to T1 V. 
Note that this behavior makes the CH$_4$ indices 
particularly useful in distinguishing
T dwarfs from other late-type objects, as expected.  
The CH$_4$-A and CH$_4$-B
indices start to decrease around L8 V to T1 V, while CH$_4$-C decreases
as early as L3 V to L5 V.  The latter downturn is likely due to
absorption by CIA H$_2$ in the late-L and T dwarfs, although we cannot rule
out the early influence of CH$_4$ as a minor absorber.  

The color indices (Figure 12c) reflect the well-known reddening seen in 
near-infrared photometry of M and L dwarfs, but shows that there is a
smooth transition to bluer colors in the T dwarf regime.  All three
of these indices peak around L5 V, at which point colors become steadily 
bluer, with late T dwarf values below those of mid-M dwarfs.  It is interesting
to note that both H/J and K/J turn over at the same spectral type, since
CIA H$_2$ likely affects only the latter index.  
\citet{kir99} see a similar downturn in FeH and CrH bandstrengths 
in the red optical
around L4 V to L5 V, and it is possible that these trends
are in some way correlated, possibly through the precipitation of
photospheric dust (see $\S$6.3.2).
Both the H/J and K/J 
indices appear to saturate and possibly reverse their downward trends
between types T7 V and T8 V.  The K/H index
shows more scatter and is less useful than the other color indices.

The remaining indices (Figure 12d) are generally not useful for 
spectral classification over the full range of spectral types shown.
The CO index shows a gradual decline from mid-M to early-T
(implying increasing band strength), but is predominantly scatter for
most of the T dwarfs due to 2.2 $\micron$ CH$_4$ absorption.
The 2.11/2.07 index is generally constant from mid-M to early-T, but shows
a downturn beyond T3 V.  Despite the significant scatter, this index appears
to separate mid- and late-T dwarfs from warmer objects, probably reflecting
changes in the relative strengths of absorbers at K-band beyond T5 V.  
Finally, the K-shape index appears to 
show no real trends over the entire range of subtypes.

Overlaid on these plots are linear fits to select indices over discrete
subtype ranges.  Fits to T dwarfs are indicated by long-dashed lines, 
while fits to M and L dwarfs are indicated by short-dashed lines.  
Coefficients for
these linear fits are given in Tables 12a and 12b, with values for
the root mean square (RMS) deviations over the subtype ranges used.  
Most of these relations yield accurate spectral types to within $\pm$1.0
subclasses for the T dwarfs.

\subsection{A Recipe for Spectral Classification}

Based on the discussion above, we outline a procedure for determining the
spectral type of T dwarfs using near-infrared data:
\begin{enumerate}
\item Measure the H$_2$O-A, H$_2$O-B, CH$_4$-A, CH$_4$-B, CH$_4$-C,
H/J, K/J, and 2.11/2.07 indices as defined in Table 9.  A subset of these
indices can be used when spectral data do not extend into the required
spectral regions.
\item Compare each index with standard values, either from Table 10 or 
(preferably) from measurements using the same 
instrumental setup.  Compute spectral types for each index.
\item Average the derived subtypes, rejecting the single high and low values,
to determine a mean decimal spectral type.  This decimal value 
rounded off to the nearest 0.5 subclass yields the final discrete subtype.
\end{enumerate}

The linear relations given in Table 12a can be used to derive individual 
spectral types in step 2, although trends in the indices are not necessarily
linear and direct comparison to standard values is generally
preferable.  Spectral types
should be computed using as many of the indices listed in step 1 as possible,
in order to reduce any biases due to differences in instrumentation
(resolution or response characteristics) or
observing conditions.  Nonetheless, we
find that this procedure yields consistent results for data with low to
moderate resolutions to better than $\pm$1.0 subclasses.

\subsection{Comparison to L Dwarf Classification Indices}

A number of spectral indices have been defined to classify
M and L dwarfs in the near-infrared, and it is useful to examine the
behavior of these indices in the T dwarf regime.  
\citet{tok99} defined two indices at K-band
to measure the strengths of the 1.9 $\micron$ H$_2$O and CIA H$_2$:
\begin{equation}
K1 = \frac{{\langle}F_{2.10-2.18}{\rangle}-{\langle}F_{1.96-2.04}{\rangle}}{0.5({\langle}F_{2.10-2.18}{\rangle}+{\langle}F_{1.96-2.04}{\rangle})}
\end{equation}
\begin{equation}
K2 = \frac{{\langle}F_{2.20-2.28}{\rangle}-{\langle}F_{2.10-2.18}{\rangle}}{0.5({\langle}F_{2.20-2.28}{\rangle}+{\langle}F_{2.10-2.18}{\rangle})},
\end{equation}
the latter of which is also sensitive to CH$_4$ in the T dwarfs.  Figure 13 is an 
extension of Figure 4 from \citet{tok99}, comparing these indices for 
types M (triangles), L (squares), and T (circles) from the NIRC (open symbols) 
and CGS4 (filled symbols) datasets.  T dwarf standards and representative
M and L dwarfs are labeled by their spectral types.
The K1 index appears to peak in the early T dwarfs,
then decreases toward the later T dwarfs, similar to what is seen in our  
H$_2$O-C index.  K2 decreases from late L through late T, 
reflecting increased CH$_4$ and CIA H$_2$ toward cooler
temperatures.  Despite
significant scatter in the T dwarf regime, 
these indices, defined for objects with 2600 $\lesssim$
T$_{eff}$ $\lesssim$ 1500 K, do show trends in objects as 
cool as $\sim$ 800 K.

\citet{rei01} have defined two indices for L dwarfs that measure
the blue and red wing of the 1.4 $\micron$ H$_2$O band; respectively,
\begin{equation}
H_2O^A = \frac{{\langle}F_{1.33-1.35}{\rangle}}{{\langle}F_{1.28-1.30}{\rangle}}
\end{equation}
\begin{equation}
H_2O^B = \frac{{\langle}F_{1.47-1.49}{\rangle}}{{\langle}F_{1.59-1.61}{\rangle}}.
\end{equation}
We plot these indices versus spectral type in Figure 14, using NIRC (diamonds)
and CGS4 (squares) data; filled circles denote values for L and T dwarf spectral
standards.  The index-spectral type relations
from \citet{rei01}, defined over the range M8 V to L8 V, 
are plotted as short-dashed lines.  These relations are extended into the
T dwarf region for comparison.
The H$_2$O$^A$ index does not track linearly into the T regime,
due to contamination by the 1.3 $\micron$ CH$_4$ band.  The 
H$_2$O$^B$ index is remarkably linear, however,  
over the entire subtype range shown.  A fit to this index over
types M5 V through T8 V yields (long-dashed line):
\begin{equation}
SpT = (12.6{\pm}0.9) - (26.7{\pm}0.6){\times}[H_2O^B],
\end{equation}
with an RMS scatter of 1.2 subtypes, where SpT(T0 V) = 0, SpT(T5 V) = 5, 
SpT(L5 V) = -4, etc.
This relation is consistent with that derived by \citet{rei01}, and it appears
that the H$_2$O$^B$ index is useful for  
classification over a broad range of late-type dwarfs.

\section{Properties of T Dwarf Subtypes}

Having established a spectral classification scheme that consistently 
characterizes changes in near-infrared spectral morphology,
we now examine some of the general properties of the individual subtypes.
From the behaviors of spectral features and colors, it is possible to
probe some of the fundamental properties of these cool brown dwarfs.

\subsection{1.25 $\micron$ K I doublet}

The 1.2432 and 1.2522 $\micron$ K I transitions are particularly useful
diagnostics of photospheric temperature.  These higher order lines
originate in the 4p $^2$P$_0$ level, 1.6 eV above the 
4s $^2$S ground state.  For local thermodynamic equilibrium (LTE), 
the relative number density of the excited state goes as T$^{12.5}$ at a
photospheric temperature of $\sim$ 1500 K\footnote{The 
0.77 $\micron$ resonance doublet, which
populates the 4p $^2$P$_0$ level, has a similar
transition rate \citep{wie66} 
but is a much stronger feature, due to the greater number of atoms in
the ground state.}.  We might therefore expect to see 
the 1.25 $\micron$ lines weaken sharply as the photospheric temperature 
at J-band
decreases.  

We have measured the pseudo-equivalent widths (PEWs) of the K I lines by
integrating over their line profiles: 
\begin{equation}
PEW = \frac{{\int}[C({\lambda})-f({\lambda})]d{\lambda}}{C({\lambda}_c)},
\end{equation}
where $C({\lambda})$ is the neighboring pseudo-continuum and ${\lambda}_c$
the transition wavelength.  Results for OSIRIS data and
literature CGS4 data are given in
Tables 13a and 13b, respectively, which list the 
central wavelength and PEW for each transition.  
Errors were determined
from the noise in the neighboring pseudo-continuum.
Figure 15 plots these values versus spectral type, with OSIRIS data
shown as triangles and CGS4 data as squares.
There is an apparent strengthening of the 1.2432 $\micron$ line
from the mid-M dwarfs to the mid-L dwarfs,
which is present but more subtle at 1.2522 $\micron$.  The growth of the
1.2432 $\micron$ line is likely influenced by contamination of FeH at
1.237 $\micron$; however, the gradual strengthening of both
lines mimic that of the 0.77 $\micron$ K I resonance doublet
in red optical data \citep{kir99}.  There is a significant drop in the
PEWs of the 1.2432 $\micron$ line
between L5 V and L7 V, likely due to the disappearance of the FeH lines
\citep{mcl00}, although a subtle dip may be present in the 1.2522 $\micron$
line strengths.  Both lines appear to peak up slightly in the mid-T dwarfs,
and then drop toward T8 V (Gliese 570D).  The mid-T peak is unexpected
assuming LTE, both in terms
of the presumably cooler photospheres of these objects and the presence of
overlying H$_2$O and CH$_4$
absorption centered at 1.15 $\micron$.  Relative opacity and depth
effects are clearly important in the formation of the K I line, 
and their re-strengthening may be the result of 
greater transparency at J-band as
dust layers settle deeper into the atmosphere or begin to rain out 
\citep{bur99,ack01}.  Dropping temperatures and/or H$_2$O and CH$_4$ opacity
likely leads to the disappearance of these lines in the latest
T dwarfs.
The complicated behavior of the K I lines reflects the competing
influences of decreasing temperature and increasing transparency
as dust settles out of the photosphere, but contributions by
neighboring molecular features must be considered as well.

\subsection{Near-infrared Colors}

The 2MASS near-infrared colors of T dwarfs 
versus spectral type are shown in 
Figure 16.
Colors for individual objects are indicated by open diamonds, 
with arrows indicating
upper limits for objects not detected in both bands.
Solid points give the weighted (by photometric uncertainty)
mean colors for each half subtype, computed only from those objects
with definite color measurements.
These values are listed in Table 14.  
L dwarf average colors from \citet{kir00} are also shown in Figure 16 for
comparison.
The reddening of near-infrared colors through the L dwarfs, and subsequent
shift to bluer colors in the T dwarfs, closely resemble the trends seen in the
H/J, K/J, and K/H color indices in Figure 12c, although the L5 V peak
is not as readily apparent.  Color breaks between L8 V and T1 V, and between
T3 V and T5 V, are evident in the J-H and J-K$_s$ colors.  We note, however,
that the T3 V object, SDSS 1021-0304, appears to be 
slightly redder than the T2 V SDSS 1254-0122 in all three colors.
Beyond T5 V, colors appear to saturate
around J-H $\sim$ 0.0, H-K$_s$ $\sim$ 0.1, and J-K$_s$ $\sim$ 0.1, 
although any trends (such as the slight rise in the H/J and K/J color
indices between T7 V and T8 V) may
be hidden by photometric uncertainties of 0.1--0.3 mag.  
The mean H-K$_s$ and J-K$_s$ colors for
T6 V objects are biased by the very blue colors of the T6 Vp 2MASS 0937+2931
(see $\S$8); excluding this object, colors for this subtype generally
match those
for T5.5 V through T8 V.  
The largest color changes therefore occur between L8 V and T5 V, with later
objects showing little variation in near-infrared colors, despite apparent
differences in spectral morphology.

\subsection{Effective Temperatures}

Spectral types in the MK system generally map onto an underlying
temperature (or more accurately, ionization) scale,
while separate luminosity classes yield radius/surface gravity information.
The temperature and luminosity of a brown dwarf are directly related,
however, as substellar radii remain constant to within 35\%
for effective temperatures
less than 2200 K and ages greater than of 0.01 Gyr \citep{bur97},
with little variation due to metallicity \citep{cha00a}.
Thus, for most field L and T dwarfs:
\begin{equation}
T_{eff}(K) = (\frac{L}{4{\pi}R^2{\sigma}})^{0.25} 
\sim 555 (\frac{L}{10^{-6} L_{\sun}})^{0.25}
\end{equation}
to within 20\%, assuming R = 7.5$\times$10$^9$ cm = 
1.06 R$_{Jup}$ = 0.11 R$_{\sun}$. 
This simple relation is important when we consider the 
difficulty in estimating
effective temperature from spectral data.
These objects are clearly not blackbodies, and brightness temperatures
derived from spectral data can range over 1000 K in the 1--2.5 $\micron$ regime
\citep{sau00}.  However, a reasonable estimate of T$_{eff}$ can be obtained
if the total luminosity of a degenerate brown
dwarf is known.  This approach only requires assumptions
based on the well-understood internal physics of brown dwarfs, rather than
relying on extensive atmospheric modeling.

\subsubsection{The Temperatures of L dwarfs}

Luminosities for a number of late-M and L dwarfs have recently been compiled by 
\citet{leg01} and \citet{rei01}.  These authors calculate bolometric corrections
by extrapolating near-infrared
spectral and photometric data to shorter and longer  wavelengths.
Luminosities are then determined by:
\begin{equation}
log(\frac{L}{L_{\sun}}) = -0.4(m_b + BC_b +5log{\pi} + 5 - M_{bol,\sun})
= -0.4M_{bol} + 1.896
\end{equation}
where M$_{bol}$ = M$_b$ + BC$_b$ is the intrinsic bolometric magnitude 
(intrinsic brightness plus bolometric correction at band $b$), determined
from the parallax $\pi$ (in arcsec), 
and using M$_{bol,\sun}$ = 4.74 \citep{liv00}.  T$_{eff}$ can 
then be computed from Equation 9.

Figure 17 plots the effective temperatures 
for late-M and L dwarfs with known parallaxes
\citep{kir00,kir01a,wil01b}, using 2MASS photometry and
J-band bolometric corrections from \citet{rei01}:
\begin{equation}
BC_J = 1.904 - 0.034{\times}SpT
\end{equation}
(here, SpT(L0) = 0, SpT(L5) = 5, SpT(M5) = $-$4, etc.) and Equations 9 and 10 above.  
Based on uncertainties in the T$_{eff}$--luminosity relation and
scatter in Equation 11 (about 0.1 mag), we estimate T$_{eff}$ uncertainties 
at 15\%. 
Also plotted in this diagram
are T$_{eff}$ estimates from \citet{leg01}, based on K-band
bolometric corrections and structure models from \citet{cha00b}.
There is a good correlation between these two temperatures sets, with
differences for common objects never exceeding $\pm$100 K.
The data show a fairly monotonic decrease in T$_{eff}$ with 
spectral typefrom M9 V to L8 V; a linear fit yields
(dashed line):
\begin{equation}
T_{eff}(K) = 2190 - 113{\times}SpT,
\end{equation}
with an RMS deviation of 120 K.  We compare this relation 
to the \citet{kir99,kir00} and \citet{bas00}
L dwarf temperature scales (solid lines) in Figure 17, where the latter scale
has been adjusted to the \citet{kir99} L dwarf types 
using transformations given in
\citet{mrt99}.  These scales clearly bracket the effective temperatures 
calculated here and
from \citet{leg01} for most of the L dwarfs, although
the Kirkpatrick scale generally underestimates 
temperatures
for the early L dwarfs, while the Basri scale overestimates temperatures
for the late L dwarfs.  Additional discussion on the L dwarf temperature
scale is given in \citet{giz01b}.

\subsubsection{The Temperatures of T dwarfs}

Currently, only two T dwarfs have known distances and hence absolute brightness
determinations: Gliese 229B and Gliese 570D.  Because they are companions to 
well-studied bright stars, accurate estimates of effective temperature for
these objects have been 
derived using evolutionary models and the properties of their primaries
\citep{mar96,all96,tsu96b,me00a,geb01a};
these temperatures are plotted as solid circles in Figure 17.  

It is clear that an extrapolation of 
Equation 13 into the T dwarf regime
substantially underestimates the temperatures of these objects.
Indeed, as pointed out by \citet{kir00}, 
there appears to be a relatively small difference in T$_{eff}$ between
the coolest L dwarfs and Gliese 229B, only $\sim$ 350 K.
How do we resolve this situation with the observed changes in
spectromorphology and photometric colors, 
including the apparently continuous strengthening of H$_2$O
indices across the L/T boundary?
The solution to this problem may lie in the behavior of atmospheric dust.
Recent models by \citet{ack01} indicate that dust clouds, which
significantly modify the emergent spectra of L dwarfs, reside at the 
1500--1800 K temperature layer in brown dwarfs, and thus begin to settle
below the photosphere around the L/T transition.  \citet{tsu99}
find a similar critical temperature at T$_{cr}$ = 1550 K.  Loss of dust 
opacity likely causes substantial redistribution of emergent flux,
increasing transparency in some spectral regions, while also increasing
H$_2$O and (ultimately) CH$_4$ band strengths as the photosphere cools.
In addition, the cooling of the atmosphere above the dust layers will
drive the formation of CH$_4$ and H$_2$O, increasing their column abundances
and hence further increasing band strengths.  These events should result in a
significant strengthening of molecular bands over a narrow T$_{eff}$ range.
The dust loss scenario also
explains the strengthening of the 1.25 $\micron$ K I doublet 
from L8 V to T5 V, inducing greater transparency at J-band and a
corresponding increase in column abundance.
Another intriguing scenario suggested by \citet{ack01}
is the formation of holes in the dust clouds,
or ``hot spots'', similar to what is seen on Jupiter at 5 $\micron$ 
\citep{wes74}.  Small bright regions could potentially dominate over 
discrete spectral bands, modifying the emergent flux
with little change to the total luminosity
or effective temperature.

Based on these considerations, we speculate that the evolution of brown dwarfs
from spectral types L8 V to T5 V occurs over a narrow range in T$_{eff}$,
perhaps as little as 200 K (dot-dashed line).  
This temperature difference is similar to
that between Gliese 229B (T6.5 V) and Gliese 570D (T8 V).  Because
dust resides in a relatively narrow range of thermal layers,
its effect on the
emergent spectrum below T$_{eff}$
$\sim$ 1000 K will no longer be important.
Indeed, dust-free models by \citet{tsu96b} provided better fits
to the spectrum of Gliese 229B than corresponding dusty models.
The evolution of spectral features at this point
should once again be tied to effective temperature, with 
spectral types later than roughly 
T6 V sampling a broader range of T$_{eff}$.  
Parallax observations of T dwarfs 
are necessary to test this hypothesis, and
variability observations of the latest L dwarfs and earliest T dwarfs
could also provide clues to the behavior of atmospheric dust as it
settles below the photosphere.  
Note that a narrow T$_{eff}$ range for objects of type L8 V through T5 V would
imply a paucity of these brown dwarfs relative to later-type T dwarfs;
more accurate space densities are needed to test this prediction.
Nonetheless, we believe that the rapid depletion of dust
provides a natural explanation to the observed evolution of spectral 
features and estimated effective temperatures.

\section{DENIS 0205-1159AB and the L/T Transition}

The behavior of spectral indices and colors between the latest L 
dwarfs and early T dwarfs shows a small morphological gap between
L8 V and T1 V.  The recent detection of the fundamental band of CH$_4$ 
at 3.3 $\micron$ in objects as early as L5 V \citep{nol00} and the 
lack of objects later than L8 V in 2MASS L dwarf searches \citep{kir00} 
both suggest that the separation in temperature between these two classes 
is also not large.
To investigate the L/T transition,
we examined the 
near-infrared spectrum of the L7 V DENIS 0205-1159AB,
a bright, well-studied, equal-magnitude binary \citep{koe99,leg01}, 
17.5 pc from the Sun \citep{kir00}.
\citet{del97} discovered this object in DENIS survey data, and 
identified a feature
at 2.22 $\micron$ that they attributed to weak CH$_4$ absorption.  
\citet{tok99} rejected this hypothesis, however, favoring CIA H$_2$ absorption
based on the spectral models of \citet{tsu99}.  

Figure 18 plots spectral data from 2.01 to 2.35 $\micron$ 
for DENIS 0205-1159AB obtained from 
\citet{del97}, \citet{tok99}, \citet{leg01}, and \citet{rei01}.
Near-infrared spectral data 
for SDSS 0539-0059 (L5 V)
and SDSS 1254-0122 (T2 V) from \citet{leg00} are also shown for comparison.
Spectra are normalized at their K-band peaks (short-dashed lines) 
and offset for comparison. 
Above these data are opacity spectra for CIA H$_2$ (thick line), 
CH$_4$ (thin line), H$_2$O (dashed line), and CO (dot-dashed line)
at T = 1400 K and P = 1 bar
(see references in Burrows et al.\ 1997).  This combination of
temperature and pressure is expected to represent 
the typical photospheric conditions 
for a 1 Gyr, T$_{eff}$ = 1500 K brown dwarf \citep{bur97,ack01}.  Opacity
data for CH$_4$, H$_2$O, and CO have been scaled by their 
number densities relative to H$_2$ using
the chemical equilibrium models of \citet{bur99}.  

The opacity spectra indicate that the broad absorption
feature seen from 2.18 to 2.29 $\micron$
is a combination of CIA H$_2$, CH$_4$, and H$_2$O, 
with CIA H$_2$ being the strongest contributor beyond 2.1 $\micron$.
In addition, there is a weak absorption feature at
2.20 $\micron$ (arrows) present in all of the DENIS 0205-1159AB spectra
except that of \citet{leg01}
that closely corresponds to a sharp feature in the CH$_4$ opacity spectrum.  
This absorption is only slightly
offset from the 2.2056/2.2084 $\micron$ Na I doublet
(4s $^2$S $\rightarrow$ 4p $^2$P$_0$), indicated by vertical lines in Figure 18.  
The high-energy Na I lines form 
3.2 eV above the ground state and appear to weaken over types 
late-M to mid-L \citep{rei01}.  It is likely, therefore, that Na I is
not responsible for this feature in DENIS 0205-1159AB. 

Does the weak 2.20 $\micron$ feature
constitute a clear detection of CH$_4$ at K-band
in the latest L dwarfs?  In our opinion it does not, 
due to the weakness and 
occasional non-detection of the absorption.
Its apparent variability in the spectra of Figure 18 may be 
intrinsic to the source, as a number of L- and T-type brown dwarfs have
recently been shown to be variable, both photometrically \citep{bai99,bai01b} and
spectroscopically \citep{nak00,kir01a}, likely due to patchy clouds of
dust in the photosphere \citep{ack01}.  Differences in instrumental
resolution and observing conditions, however, make this interpretation
ambiguous, and the weakness of this feature may lend doubt to its reality.   
Higher resolution, higher signal-to-noise, and time-resolved data
are required to confirm its authenticity and possible variable nature.

Nonetheless, the opacity spectra in Figure 18 do suggest that CH$_4$
is an important minor absorber at these temperatures,
and its emergence as an identifiable feature likely occurs in
objects only slightly cooler than DENIS 0205-1159AB.
This suggests that the earliest T dwarfs (i.e., those with 
clear CH$_4$ bands) are probably not significantly cooler than the
latest L dwarfs currently known.
In that case, L8 V may be the last
subtype of the L spectral class, with cooler objects showing obvious
CH$_4$ absorption features
at K-band, classifying them as T dwarfs.  
In order to verify this hypothesis, it is necessary to identify
more of the so-called ``L/T transition'' objects and determine at what point
the detection of CH$_4$ becomes unambiguous at J-, H-, and K-bands.
Note that even if the 2.20 $\micron$ CH$_4$ feature
is confirmed in DENIS 0205-1159AB or in
other L7/L8 dwarfs, it does not necessitate a reclassification of these
objects, much as the 3.3 $\micron$ absorption seen in mid- and late-L dwarfs 
reclassify them as T dwarfs \citep{nol00}, or the presence of TiO bands in
bright K5 and K7 dwarfs \citep{rei95} reclassify them as M dwarfs.
The weak feature described here may indicate that the latest-type L dwarfs
known are close to the
L/T boundary, but not that they have traversed it.

\section{2MASS 0937+29: A Template for Measuring Gravity/Composition?}

One T dwarf that deserves special attention is the T6 Vp 2MASS 0937+2931,
designated peculiar because of its highly suppressed K-band peak.  Indeed,
this object's near-infrared colors are significantly bluer that any other T
dwarf so far identified.  As can be seen in Table 13a and Figures 7 and 15, 2MASS
0937+2931 also has very weak or absent 1.25 $\micron$ K I absorption, weaker
even than the T8 V Gliese 570D.

To identify the source of this anomalous absorption, we again examined the
molecular contributors to the photospheric opacity in this object.  Figure 19
plots 1.9--2.3 $\micron$ NIRC data for 2MASS 0937+2931 (solid line) along with
data for the T6 V standard 2MASS 0243-2453.  Both spectra are normalized at
their J-band peaks and match very well from 1--1.8 $\micron$; however, 2MASS
0937+2931 is nearly 50\% fainter at K-band.  Plotted above this are opacity
spectra for CIA H$_2$ (thick solid line), CH$_4$ (solid line), and H$_2$O
(dotted line) at T = 1000 K and P = 1 bar, with CH$_4$ and H$_2$O scaled to
their chemical equilibrium relative number densities \citep{bur99}.  H$_2$O
and CH$_4$ are clearly more important at cooler temperatures, as CIA H$_2$ at
1 bar is only a minor absorber outside of the 2.1 $\micron$ window.  However,
because H$_2$ features form via collisional processes, they strengthen with
increasing pressure, scaling roughly as $\sigma$ $\propto$ P.  This effect can
be seen in the increased opacity of CIA H$_2$ for P = 10 bar (thick dashed
line).  The corresponding opacities of H$_2$O and CH$_4$ are generally
unaffected by changes in pressure; and at or below T $\sim$ 1000 K, number
densities at equilibrium remain essentially unchanged \citep{bur99}.  Thus,
CIA H$_2$ is the only major molecular
absorber sensitive to local pressure in T dwarf
atmospheres.

Photospheric pressure is directly related to specific
gravity by hydrostatic equilibrium:
\begin{equation}
\frac{dP}{d{\tau}} \sim \frac{P}{\tau} = \frac{g}{{\kappa}_R} \propto \frac{M}{R^2{\kappa}_R}.
\end{equation}
Because brown dwarf radii are roughly constant, this implies the (simplistic) 
relation P$_{phot}$ $\propto$ M for constant ${\kappa}_R$.  
Thus, at a given T$_{eff}$ and composition, 
an older, more massive object will tend to have a 
higher photospheric pressure than a younger, less massive object, 
and hence stronger CIA H$_2$ absorption.  If we assume that most T dwarfs
identified to date have masses between 40 to 60 M$_{Jup}$
(assuming 800 $<$ T$_{eff}$ $<$ 1200 and 1 $< \tau <$ 5 Gyr), 
then CIA H$_2$ absorption should be 1.5 times stronger in 2MASS 0937+2931 
if it were a 75 
M$_{Jup}$ object.  The structure models of \citep{bur97} predict a
factor of 1.75 increase in specific gravity due to the increased contraction 
of the older, more massive object.  Thus, signficant differences in 
photospheric pressures are possible.  However, a 75 M$_{Jup}$ 
brown dwarf with T$_{eff}$ = 1000 K
would also have to be older than 11.5 Gyr \citep{bur97}. 

An alternate hypothesis for the strong K-band absorption in
2MASS 0937+2931 is decreased metallicity.  Zero-metallicity brown dwarf
spectral models are dominated by CIA H$_2$ opacity and peak shortward of 1
$\micron$; when molecular absorption is included, the suppression of flux at
shorter wavelengths increases the relative luminosity at K-band by roughly a
factor of 2 \citep{sau95}.  If the metallicity of a brown dwarf atmosphere
were decreased from Z = Z$_{\sun}$ to 0.1 Z$_{\sun}$, the number densities of
CH$_4$ and H$_2$O and (in an approximate sense) their relative opacities would
also decrease roughly by a factor of 10.  Decreased metallicity would not
significantly affect the amount of photospheric H$_2$ present, although it may
slightly alter the opacity spectrum.  Assuming photospheric pressure remained
constant, one might expect the same relative opacity contributions in this
scenario as in the Z = Z$_{\sun}$, P = 10 bar case.  Decreased metallicity
would also explain the deficit of K I absorption seen at J-band.  Note that
because of the reduced opacity and corresponding increased luminosity, brown
dwarfs with subsolar metallicity cool more rapidly \citep{cha97}.  Thus, a
metal-poor model for 2MASS 0937+2931 is either younger (same mass) or more
massive (same age) than its solar metallicity, T$_{eff}$ = 1000 K counterpart,
while a high-pressure model must be both older and more massive.

These arguments are based on very simple scaling laws of abundance and
opacity, and disentangling the effects of T$_{eff}$, specific gravity, and
metallicity on the emergent spectrum of 2MASS 0937+2931 requires more
extensive modeling.  It is clear, however, that CIA H$_2$ plays an important
role in shaping the spectra of both L and T dwarfs, and its sensitivity to the
ambient pressure makes it an excellent tracer of surface gravity and hence
mass.  Diagnostics that measure the relative contributions of CIA H$_2$,
CH$_4$, and H$_2$O may allow us to extract the physical characteristics of
these constantly evolving objects.

\section{Conclusions}

In 1995, the discovery of the uniquely cool brown dwarf Gliese 229B was
announced, an object exhibiting CH$_4$ absorption features reminiscent of the
reflectance spectrum of the giant planet Jupiter \citep{opp95}.  Six years
later, at least two dozen counterparts have been identified, the majority
found only in the past two years by the 2MASS and SDSS surveys.  We now know
that these objects, T dwarfs, form a distinct spectral class, defined by the
presence of CH$_4$ absorption bands in the near-infrared.

In this article, we have examined the morphologies of the CH$_4$ and H$_2$O
bands and J, H, and K flux peaks that characterize the 1--2.5 $\micron$
spectra of known and newly discovered T dwarfs.  Based on these features, we
have made a first attempt to define a classification scheme for T dwarfs 
which segregates the
currently known population into subtypes T1 V, T2 V, T3 V, T5 V, T6 V, T7 V,
and T8 V.  We have also introduced a prescription for near-infrared
classification using a suite of spectral indices that measure the strengths of
CH$_4$ and H$_2$O bands, near-infrared colors, and the evolution of the K-band
peak.  Our work parallels independent efforts reported by \citet{geb01b} in
this volume, who produce a nearly identical sequence based on CGS4 spectral
data (D.\ Golimowski, priv.\ comm.).  That this should be the case is not
surprising, since both schemes are based on the observed strengths of 
H$_2$O and CH$_4$ features.  Nevertheless, their independent convergence
intimates a natural order in the spectra of T dwarfs.

From the properties of spectral features, particularly H$_2$O bands, and the
progression of near-infrared colors, it appears that the latest L dwarfs are
not very different from the earliest T dwarfs.  The possible role of CH$_4$ as
a minor contributor in shaping the K-band peak of the latest L dwarfs, and the
recent detection of CH$_4$ at 3.3 $\micron$ in objects as early as L5 V
\citep{nol00}, lead us to believe that perhaps only one subclass exists
between the latest L subtype (L8 V) and the earliest T subtype (T1 V).  Thus,
the gap between these two separately defined spectral classes appears to be
effectively bridged.

The evolution of spectral features from late L to early T dwarfs is likely
dominated by the behavior of atmospheric dust, which greatly influences
the near-infrared spectra of L dwarfs but appears to be absent in 
T dwarfs such as Gliese 229B \citep{tsu96b,all96,mar96}.  The condensation
and gravitational settling of dust clouds leads to a substantial evolution
of spectral morphology, possibly over a narrow T$_{eff}$ range.  A rapid
depletion of observed dust opacity, perhaps stimulated by the formation
of holes in dust clouds extending to significantly hotter layers,
could explain both the small effective temperature difference between the
$\sim$ 1300 K L8 V Gliese 584C \citep{kir00} and the $\sim$ 950 K 
T6.5 V Gliese 229B \citep{sau00}, as well as the behavior of K I line
strengths at J-band.  One may argue that the spectral transition between
the L and T classes is driven more by the physics of dust condensation
than by the thermochemical formation of CH$_4$ bands (in addition,
the latter process is likely driven by the former).  Modeling the L/T
transition clearly requires a more accurate treatment of the 
condensation, coagulation, and settling of dust species \citep{tsu99,ack01}.
A similarly complex transition
may occur with the condensation of H$_2$O at T$_{eff}$ $\sim$ 400 K
\citep{bur97}, a possible endpoint to the T spectral class.

In any case, the near-infrared classification scheme derived in this 
article appears to adequately represent nearly all of the currently
known T dwarfs.  Atmospheric models, spectral indices, and the properties
of the two known T dwarf companions Gliese 229B and Gliese 570D imply
that this scheme maps directly onto a decreasing temperature sequence.  
Extensive modeling is required to derive the true physical properties
of the observed objects, and hence our
interpretations are best deemed preliminary.  However, we believe that 
the one notable anomaly, 2MASS 0937+2931,
which has a highly suppressed K-band peak likely caused by increased
opacity from CIA H$_2$, could be an extreme example of
an old, high-mass brown dwarf or one with diminished metallicity.
Determining the properties of this unusual T dwarf will enable us to 
understand how the physical characteristics of mass, age, and composition
separately shape the emergent energy distribution of cool brown dwarfs.
It is clear that the identification of new T dwarfs in the 
2MASS and SDSS surveys will help fill in the current gaps, both in the spectral
sequence (which must be developed to include such discoveries) 
and in our understanding of  
processes occurring in cool brown dwarf atmospheres.

\acknowledgements

A.\ J.\ B. would like to thank 
J.\ Cuby, X.\ Delfosse, T.\ Geballe, S.\ Leggett, M.\
Strauss, A.\ Tokunaga, and Z.\ Tsvetanov for providing electronic versions of
their published spectra, and acknowledges useful discussions with
R.\ Blum, D.\ Golimowski, M.\ Marley, G.\ Neugebauer, 
D.\ Saumon, and M.\ R. Zapatero Osorio in the preparation 
of the manuscript.  Observations described
in this article would not have been possible without the assistance of 
our knowledgable telescope operators and instrument specialists:
Mike Doyle, Karl Dunscombe,
Jean Mueller, Kevin Rykowski, Barrett ``Skip'' Staples, and Merle Sweet
at Palomar; 
Teresa Chelminiak, Bob Goodrich, Chuck Sorenson, and Meg Whittle at Keck;
and Maria Theresa Acevedo, Alberto Alvarez, Robert Blum,
Mauricio Fernandez, Angel Guerra, and Patricio Ugarte at CTIO. 
A.\ J.\ B., J.\ D.\ K., and J.\ E.\ G.\ acknowledge the support 
of the Jet Propulsion
Laboratory, California Institute of Technology, which is operated under
contract with the National Aeronautics and Space Administration.  
A.\ B.\ acknowledges support from NASA grants NAG 5-7073, NAG 5-7499, and
NAG 5-10629.  Portions
of the data presented herein were obtained at the W.\ M.\ Keck Observatory
which is operated as a scientific partnership among the California 
Institute of
Technology, the University of California, and the National Aeronautics 
and Space
Administration.  The Observatory was made possible by the generous financial 
support of the W.\ M.\ Keck Foundation.
The Digitized Sky Survey was produced at the Space
Telescope Science Institute under US Government grant NAG W-2166.
DSS images were obtained from the Canadian Astronomy Data Centre, 
which is operated by the Herzberg Institute of Astrophysics, 
National Research Council of Canada. 
This research has made use of the SIMBAD database,
operated at CDS, Strasbourg, France.
This publication makes use of data from the Two Micron
All Sky Survey, which is a joint project of the University of
Massachusetts and the Infrared Processing and Analysis Center,
funded by the National Aeronautics and Space Administration and
the National Science Foundation.

\clearpage

\figcaption[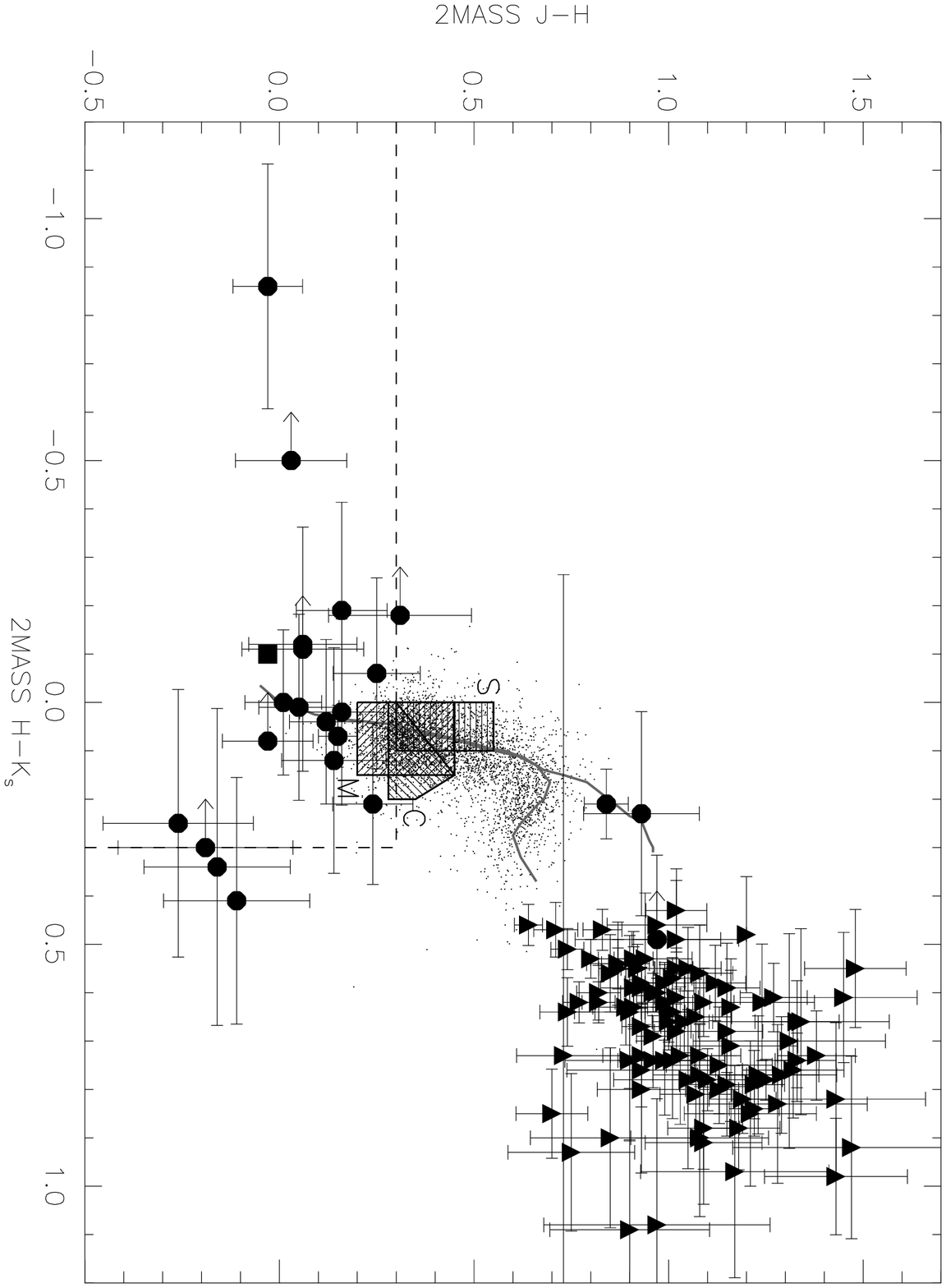]{Near-infrared color-color diagram of 
objects detected by 2MASS.  
Small points are unresolved sources with J $<$ 15.8,
H $<$ 15.1, and K$_s$ $<$ 14.3,
selected within a 1$\degr$ radius around 18$^h$ R.A., 40$\degr$ decl.
Solid lines trace out the \citet{bes88} dwarf and giant tracks.
L dwarf (triangles) and T dwarf (circles) colors
are shown with error bars, based on 2MASS photometry. 
Colors for Gliese 229B (square) 
are taken from \citet{leg99}.  
The locations of S-, C-, and M-type asteroids are
indicated by hash-marked boxes \citep{syk00}.  Dashed lines delineate the color
constraints of the wdb0699 sample.
\label{fig-1}}

\figcaption[f2.eps]{Aitoff projection map of sky area 
searched in wdb0699 sample.  Light gray boxes map out areas scanned by
2MASS at the time of sample selection, covering 16620 sq.\ deg., 
while individual points mark the locations
of the 35280 Cut \# 1 candidates selected in this sample.
\label{fig-2}}

\figcaption[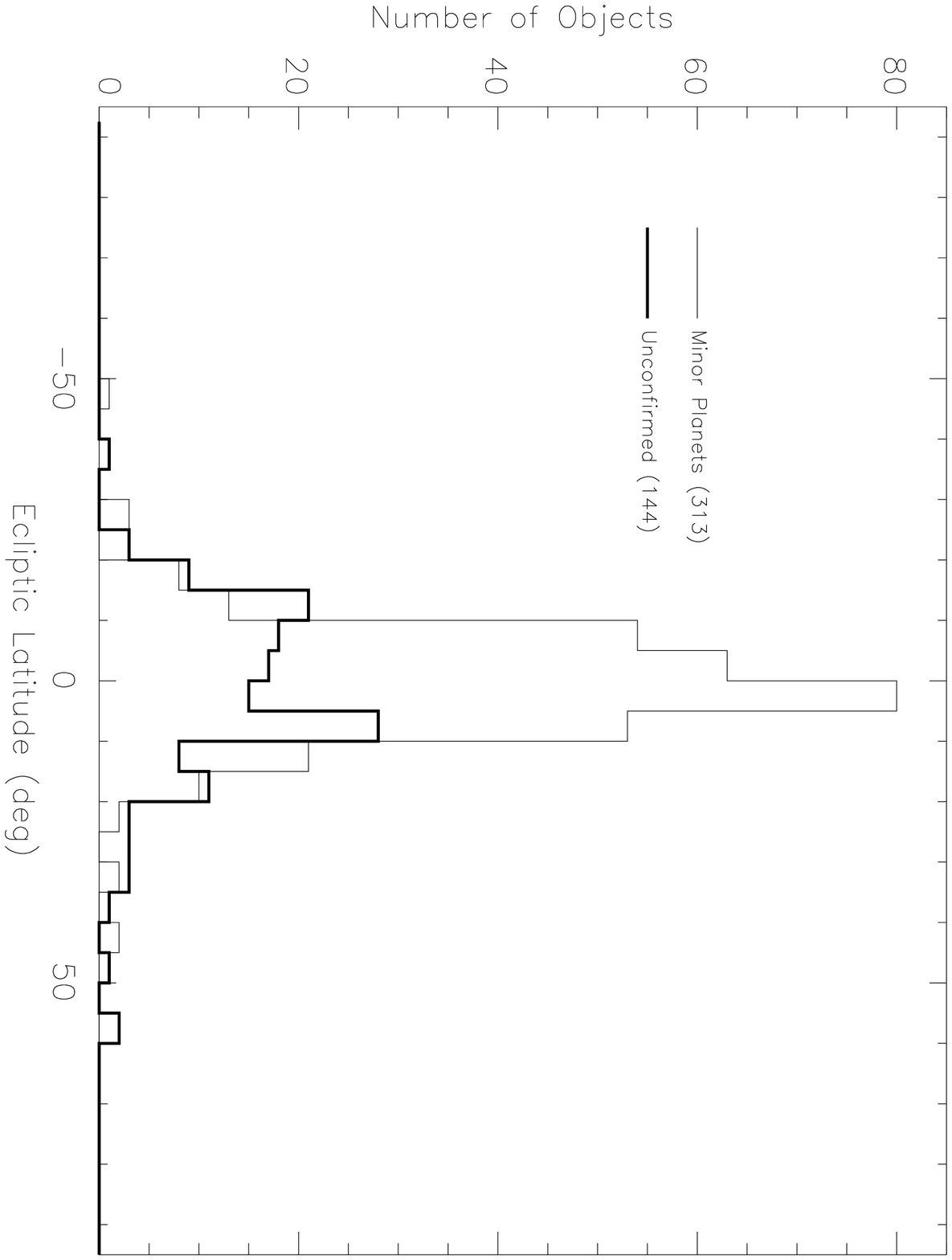]{Ecliptic latitude distribution of 144 objects
in the 2MASS IDR2,
unconfirmed in near-infrared reimaging observations (thick line).
These objects, with J $<$ 16, J-H $<$ 0.3, and H-K$_s$ $<$ 0.3,
are not flagged as minor planets and are not associated 
with any catalogued asteroid within 25$\arcsec$, based on ephemerides
provided by D.\ Tholen.  The distribution of 288 flagged minor planets
and 25 missed associations (see Table 3) in the 2MASS IDR2 with the
same magnitude and color constraints
is overlain for comparison (thin line).
\label{fig-3}}

\figcaption[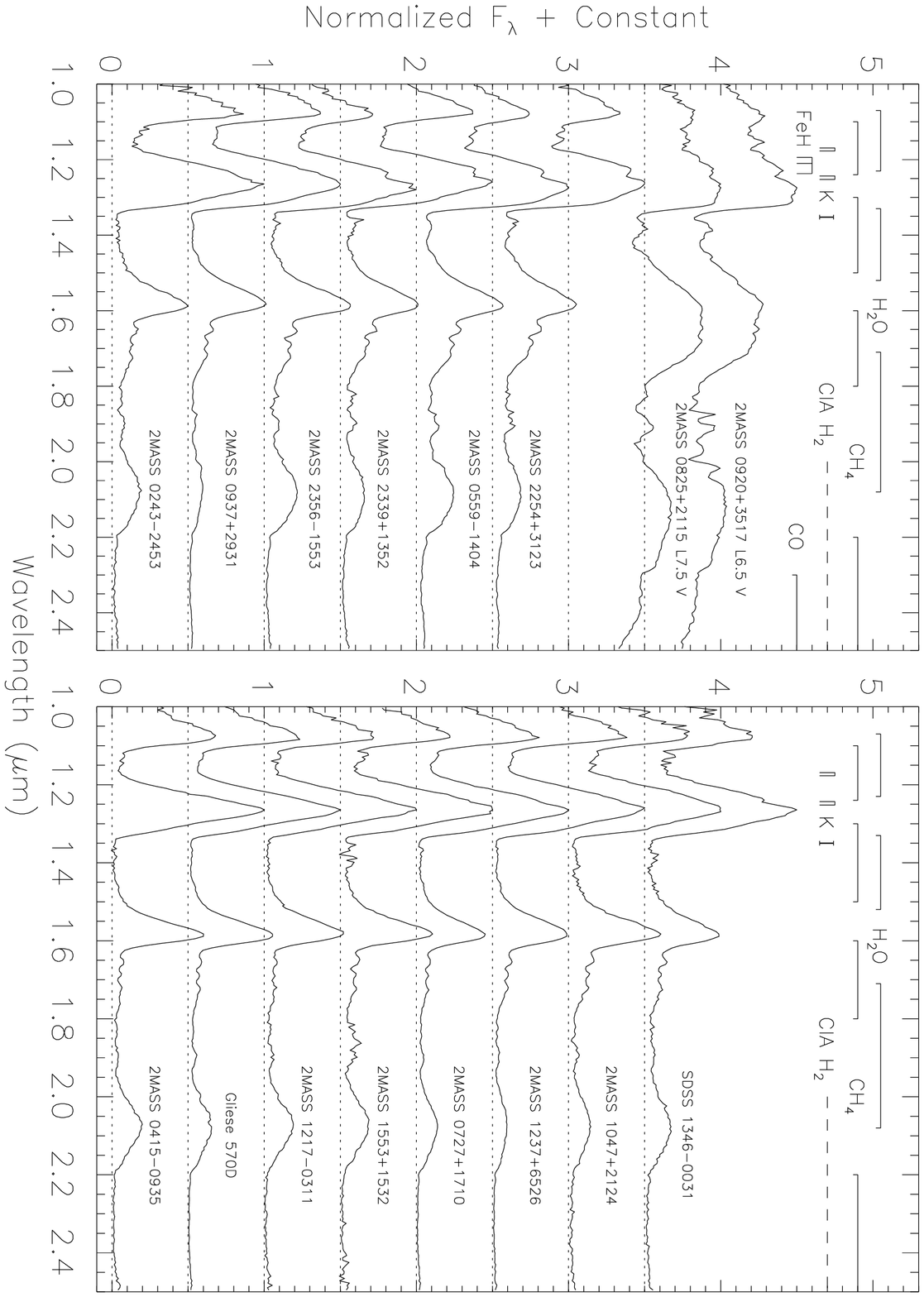]{1--2.5 $\micron$ spectra of T dwarfs observed using
the Keck I NIRC instrument.  Spectra are nomalized at their J-band
peaks, offset by a constant (dotted lines), and ordered by increasing
1.15 and 1.6 $\micron$ absorption.  Major H$_2$O, CH$_4$, CO, 
and FeH absorption
bands are indicated, as are lines of K I (1.169, 1.177, 1.243, and 1.252
$\micron$) and the region of strongest CIA H$_2$ absorption.  The L dwarfs
2MASS 0920+3517 (L6.5 V) and 2MASS 0825+2115 (L7.5 V) are shown for comparison. 
\label{fig-4}}

\figcaption[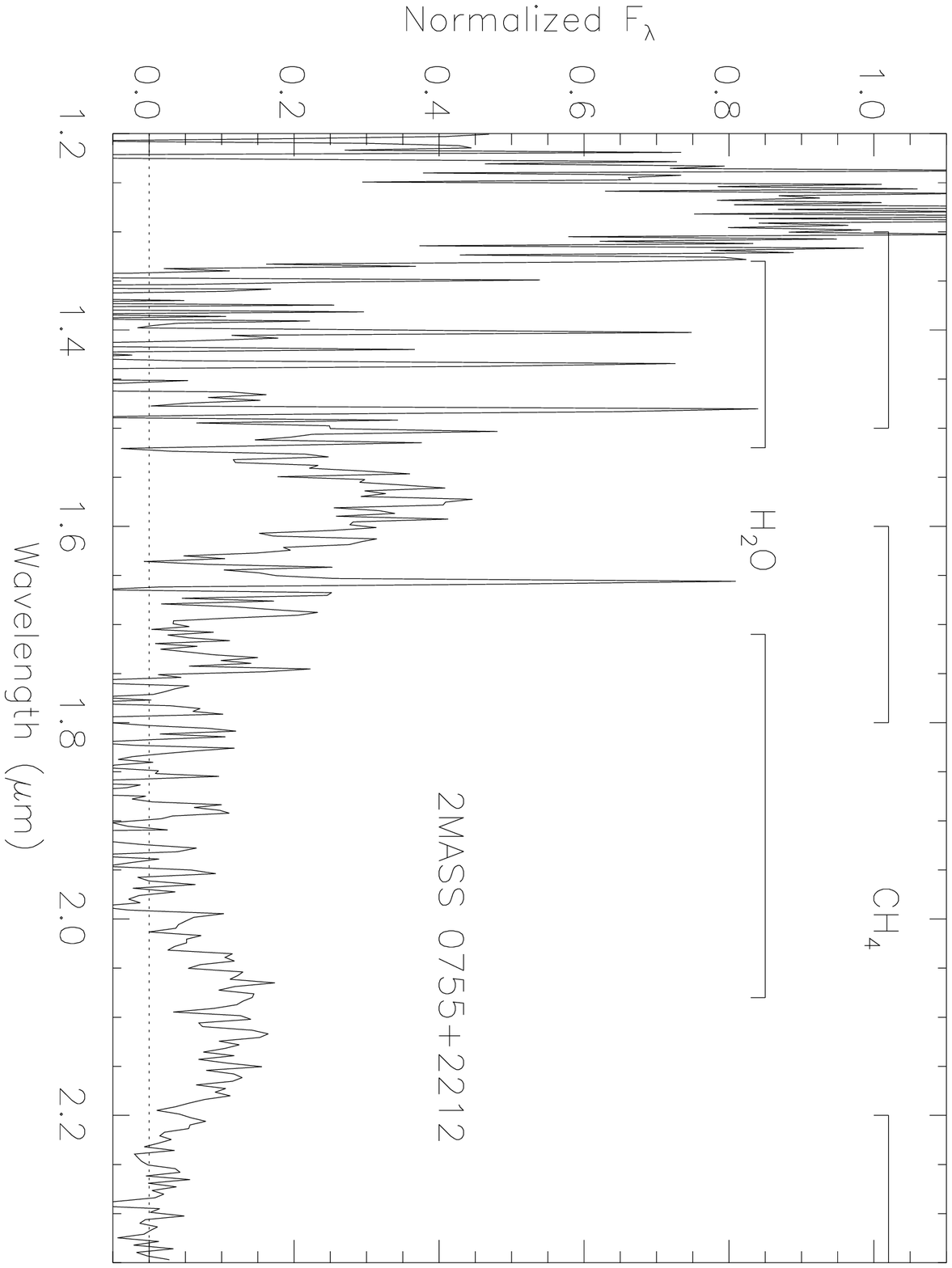]{1.2--2.3 $\micron$ spectrum of 2MASS 0755+2212 
obtained in poor observing conditions
using the Hale D78 near-infrared camera.
 Data are normalized at 1.27 $\micron$.  Identifying
H$_2$O and CH$_4$ absorption bands are indicated.
\label{fig-5}}

\figcaption[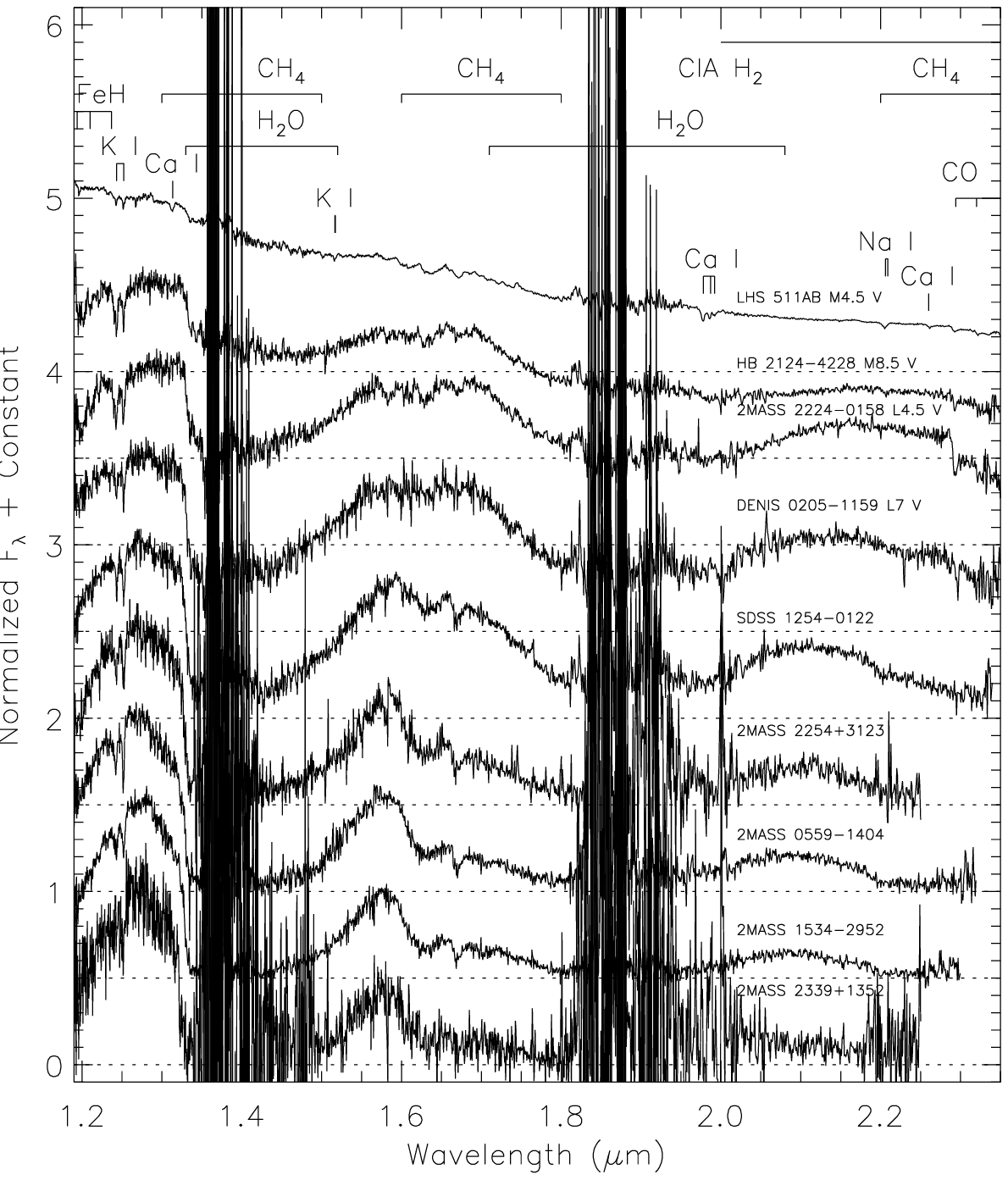,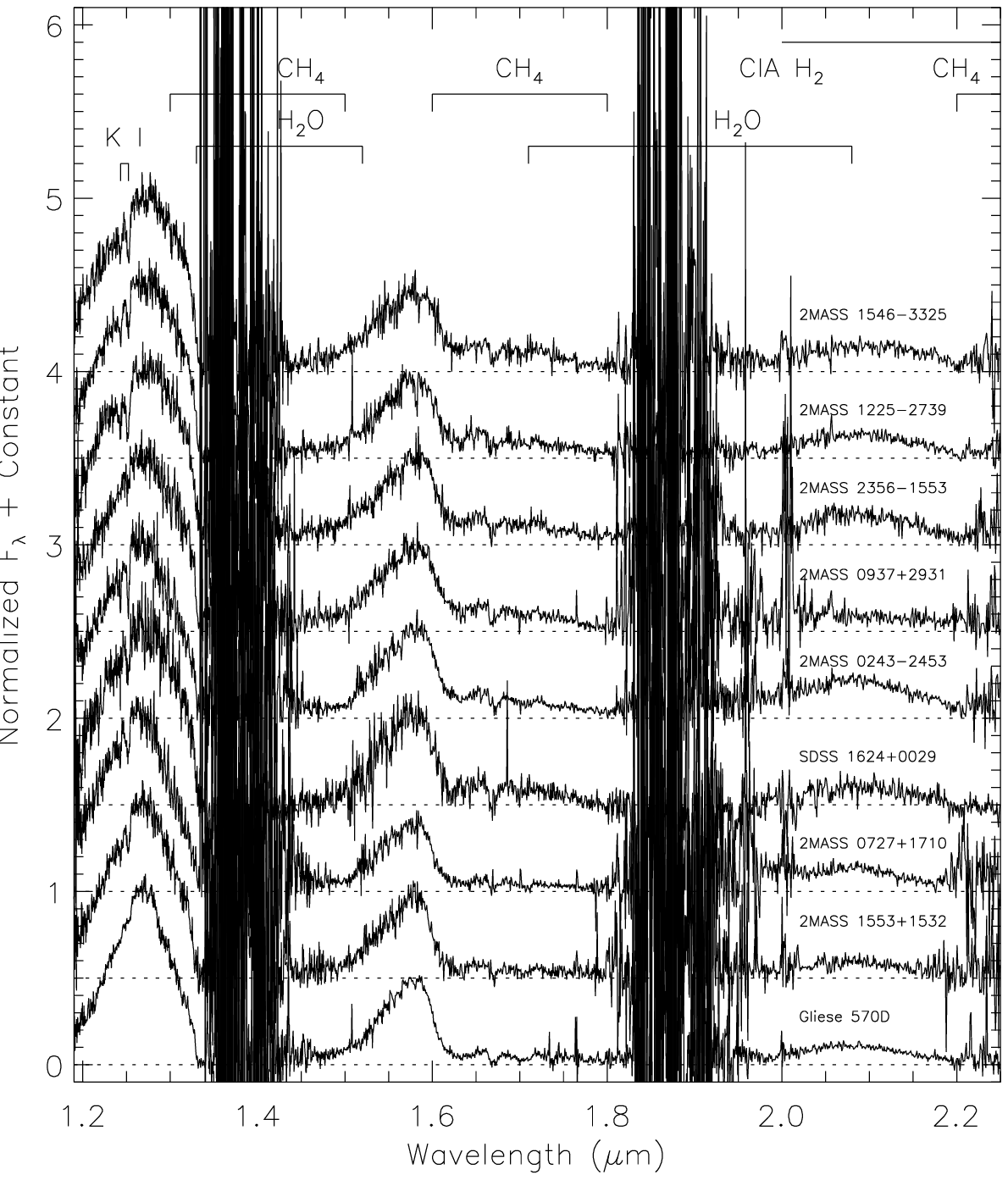]{1--2.3 $\micron$ spectra of objects observed using
the CTIO 4m OSIRIS instrument.  Data are normalized at 1.27 $\micron$, offset
by a constant (dotted lines), and sorted as in Figure 4.  
Bands of H$_2$O, CH$_4$, CO, FeH, and CIA H$_2$
are indicated, as are lines of K I (1.25 $\micron$ doublet) and 
Na I (2.2 $\micron$ doublet).  Ca I (1.314, 2.26, 
and 1.98 $\micron$ triplets) and the 1.52 $\micron$ K I doublet present in the 
spectrum of the M4.5 V LHS 511AB is also noted.  
Noisy regions around 1.4 and 1.9
$\micron$ are due to telluric H$_2$O absorption.
\label{fig-6}}

\figcaption[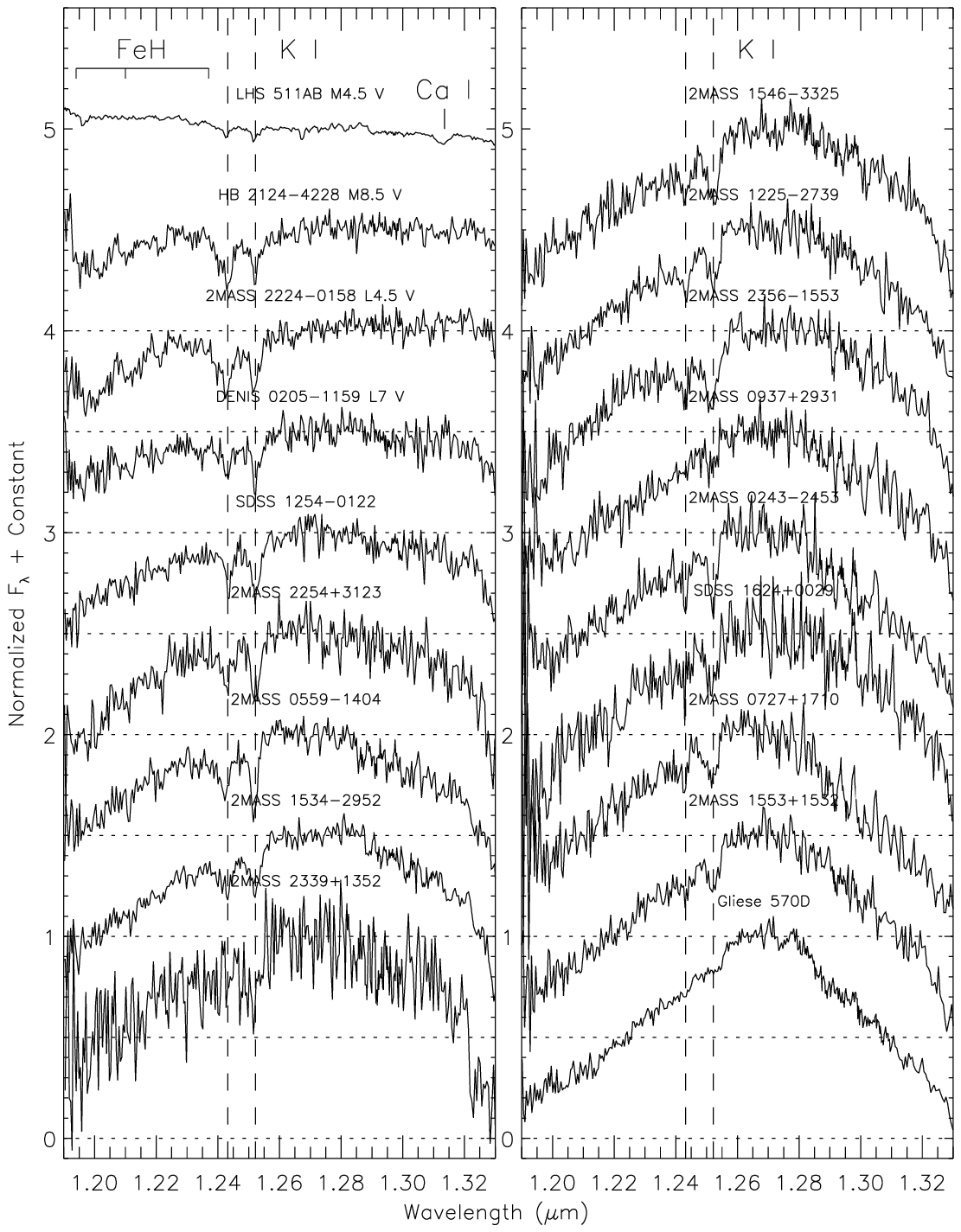]{OSIRIS data from 1.19 to 1.33 $\micron$.  Spectra are
scaled and offset as in Figure 6.  The 1.25 $\micron$ K I doublet is indicated
by dashed lines, and FeH features at 1.194, 1.21, and 1.237 $\micron$,
and the Ca I triplet at 1.314 $\micron$ (present in LHS 511AB) are
noted.
\label{fig-7}}

\figcaption[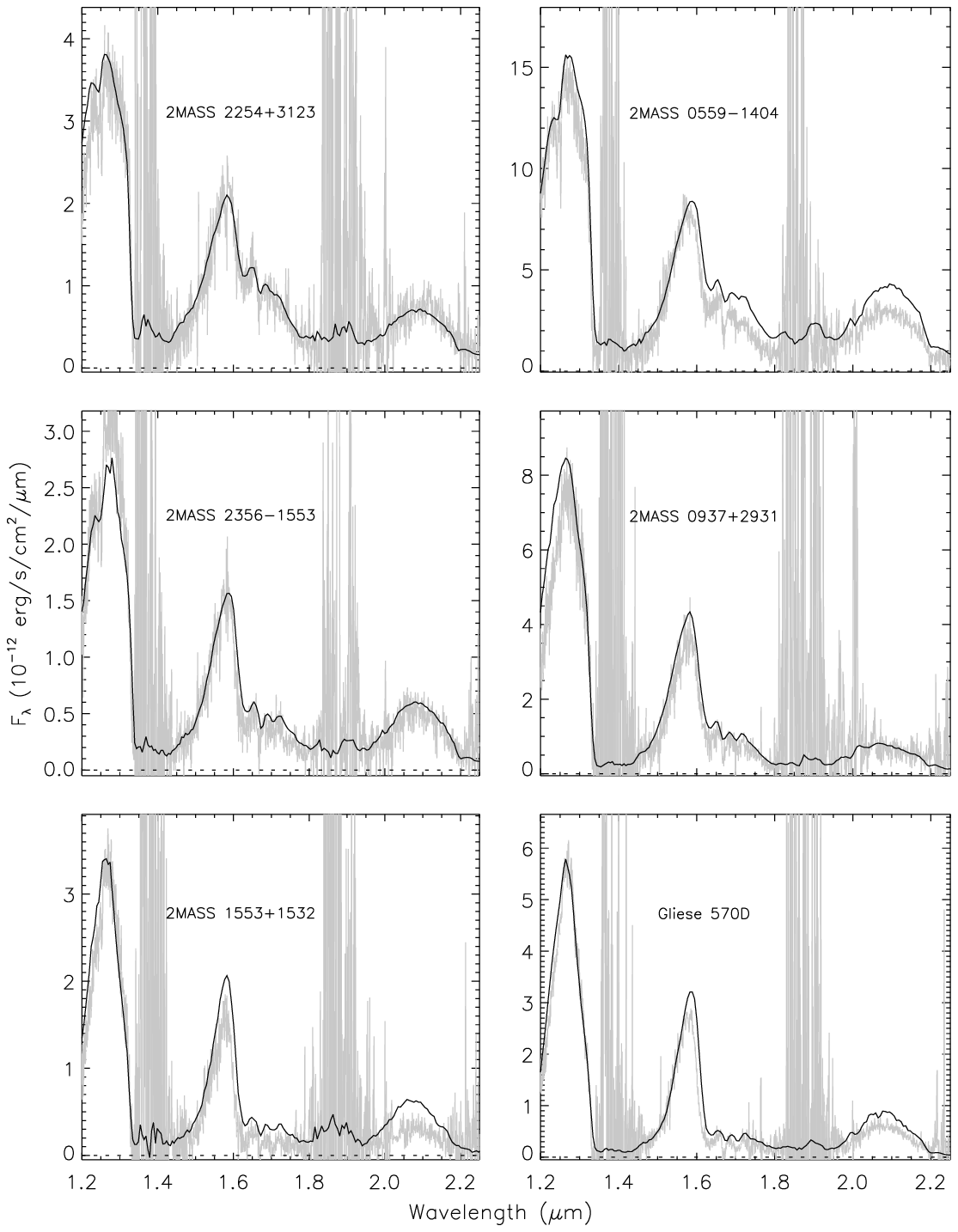]{Comparison between NIRC (black lines) and OSIRIS 
data (gray lines) for the T dwarfs 2MASS 2254+3123, 2MASS 0559-1404, 
2MASS 2356-1553, 2MASS 0937+2931, 2MASS 1553+1532, and Gliese 570D.
\label{fig-8}}

\figcaption[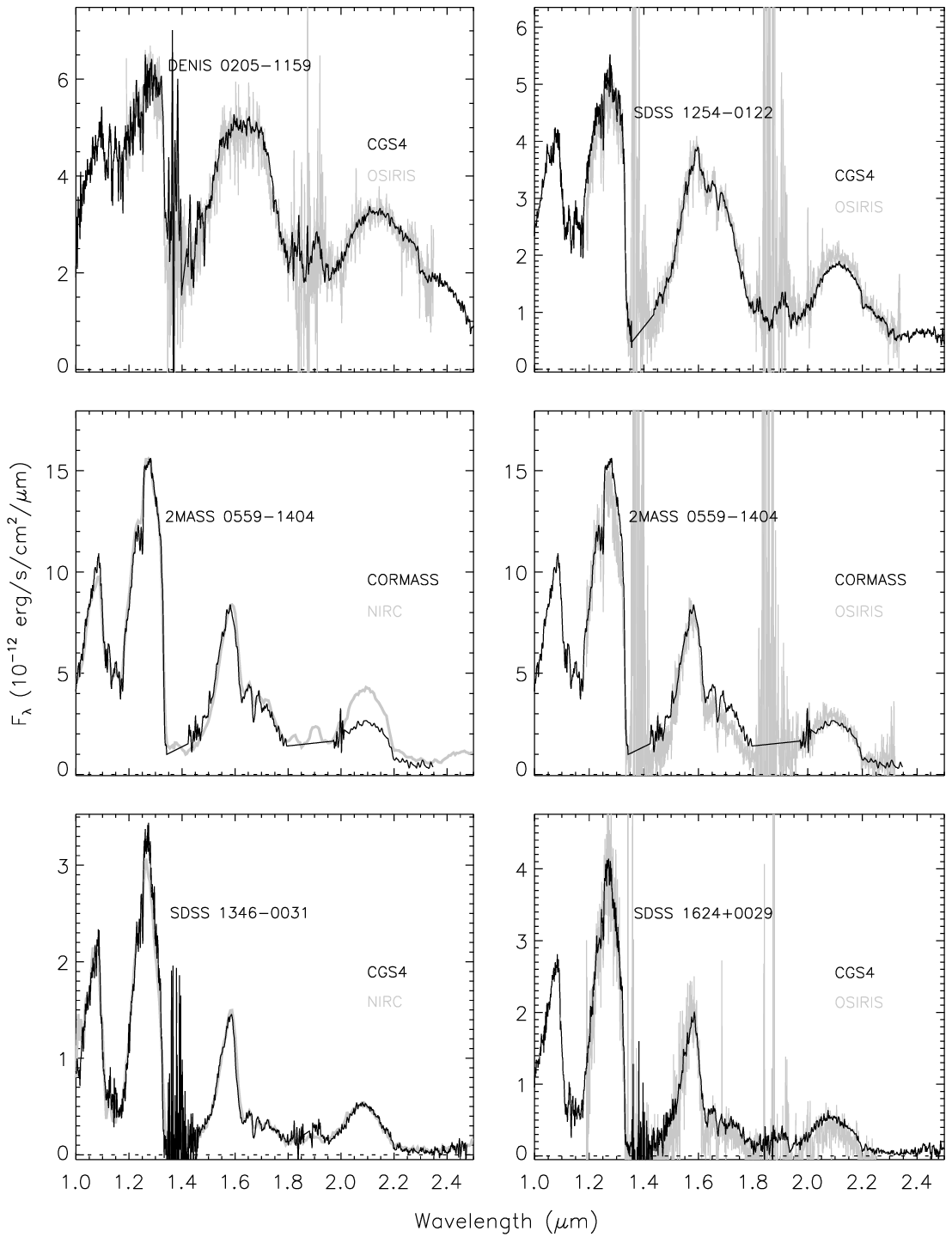]{Comparison between NIRC/OSIRIS data (gray lines)
and data from the literature (black lines) for DENIS 0205-1159AB (L7 V) and the
T dwarfs SDSS 1254-0122, 2MASS 0559-1404, SDSS 1346-0031, and SDSS 1624+0029.
\label{fig-9}}

\figcaption[f10a.eps,f10b.eps,f10c.eps,f10d.eps]{Finder images for 
newly discovered T dwarfs.  5$\arcmin$
$\times$ 5$\arcmin$ optical (R-band) and near-infrared (2MASS J- and K$_s$-band)
images are shown, oriented with North up and East to the left.  10$\arcsec$
$\times$ 10$\arcsec$ boxes in each image highlight the positions of the 
identified T dwarfs.
\label{fig-10}}

\figcaption[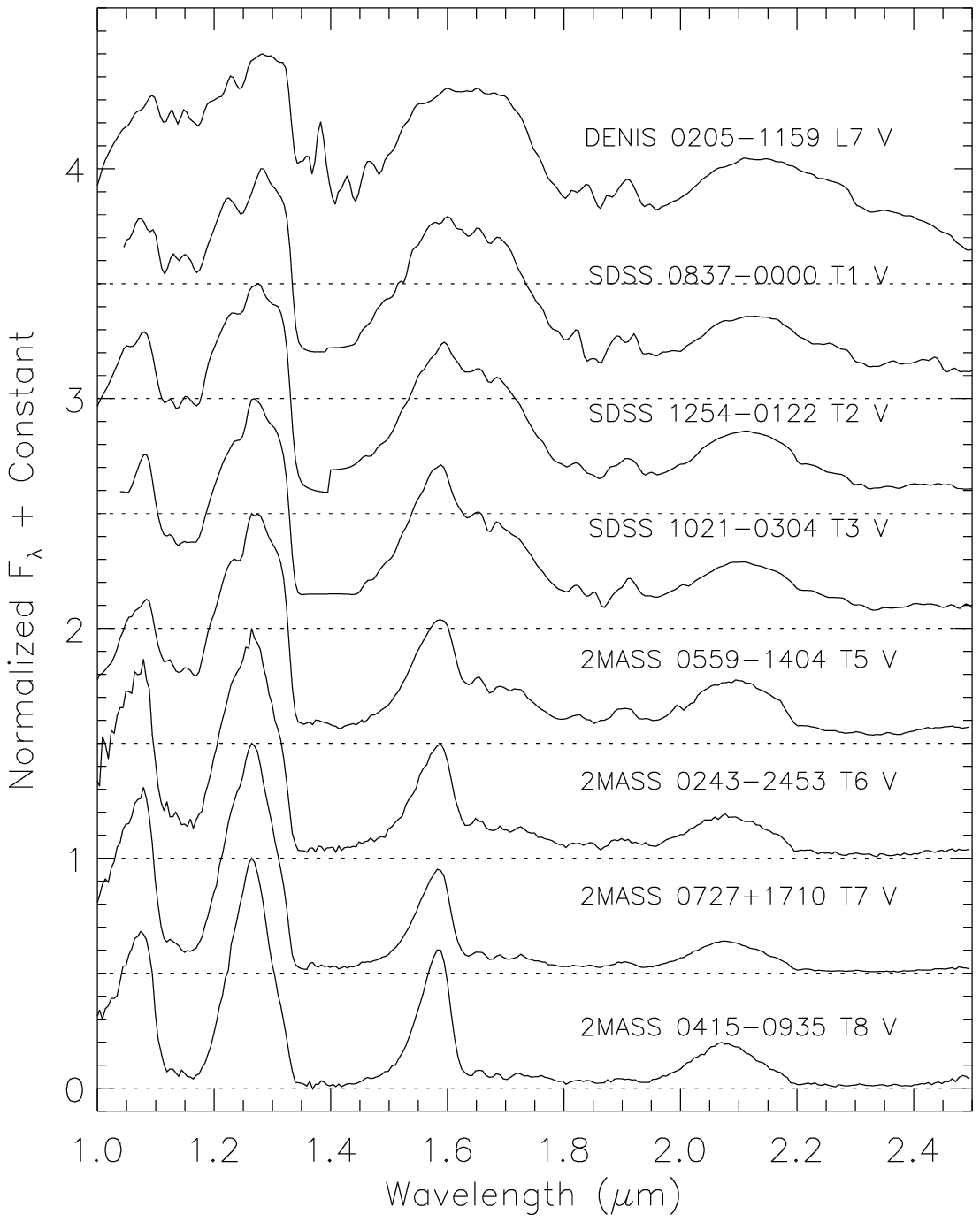]{Low resolution spectra of T dwarf standards and the
L7 V DENIS 0205-1159AB.  Data are
normalized at their J-band peaks, and zero point offsets are indicated by 
dotted lines.  Data for DENIS 0205-1159AB \citep{leg01}, SDSS 0837-0000,
SDSS 1254-0122, and SDSS 1021-0304 \citep{leg00} have been degraded
to the resolution of the NIRC spectra 
using a Gaussian filter.  
\label{fig-11}}

\figcaption[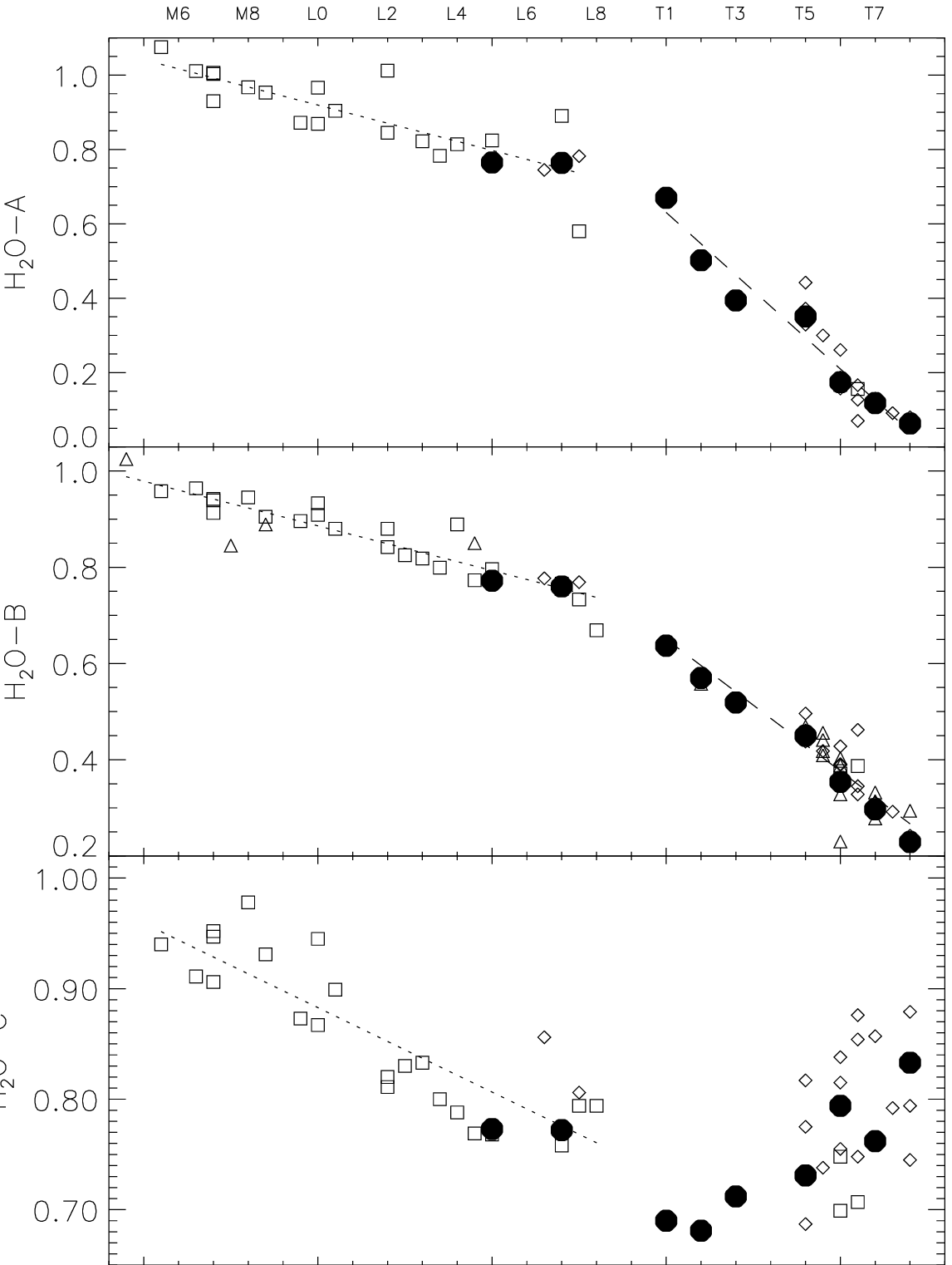,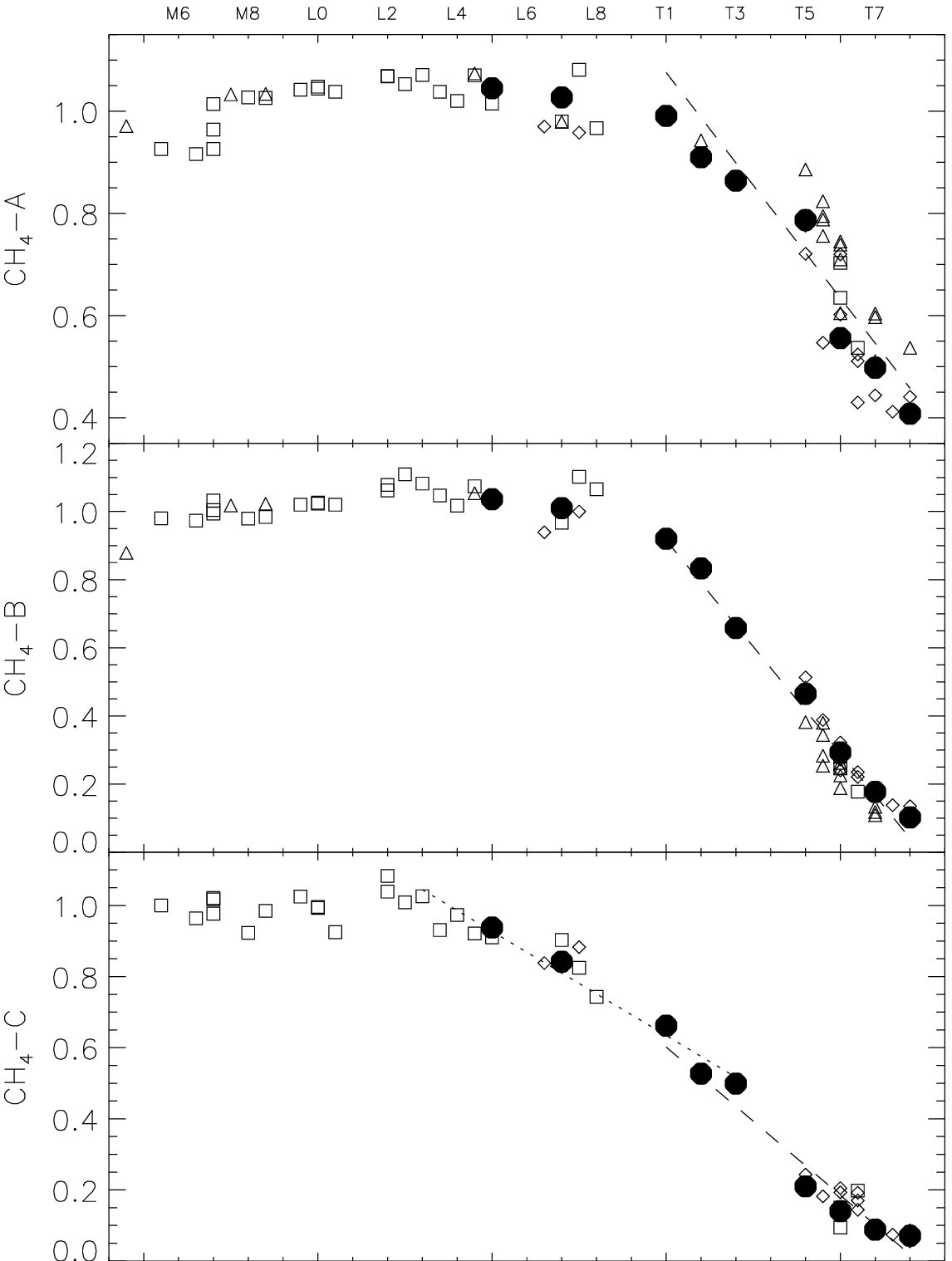]{Near-infrared
spectral indices for M, L, and T dwarfs versus spectral
type.  NIRC data are represented by diamonds,
OSIRIS data by triangles, and CGS4 data (from the literature) by squares.  
Filled circles indicate
measures for late-L and T dwarf standards.  Long-dashed and short-dashed
lines trace linear fits to indices in the T and late-M and L dwarf regimes,
respectively.  (a) H$_2$O indices; (b) CH$_4$ indices; (c) color indices; 
(d) CO, 2.11/2.07, and K shape indices.
\label{fig-12}}

\figcaption[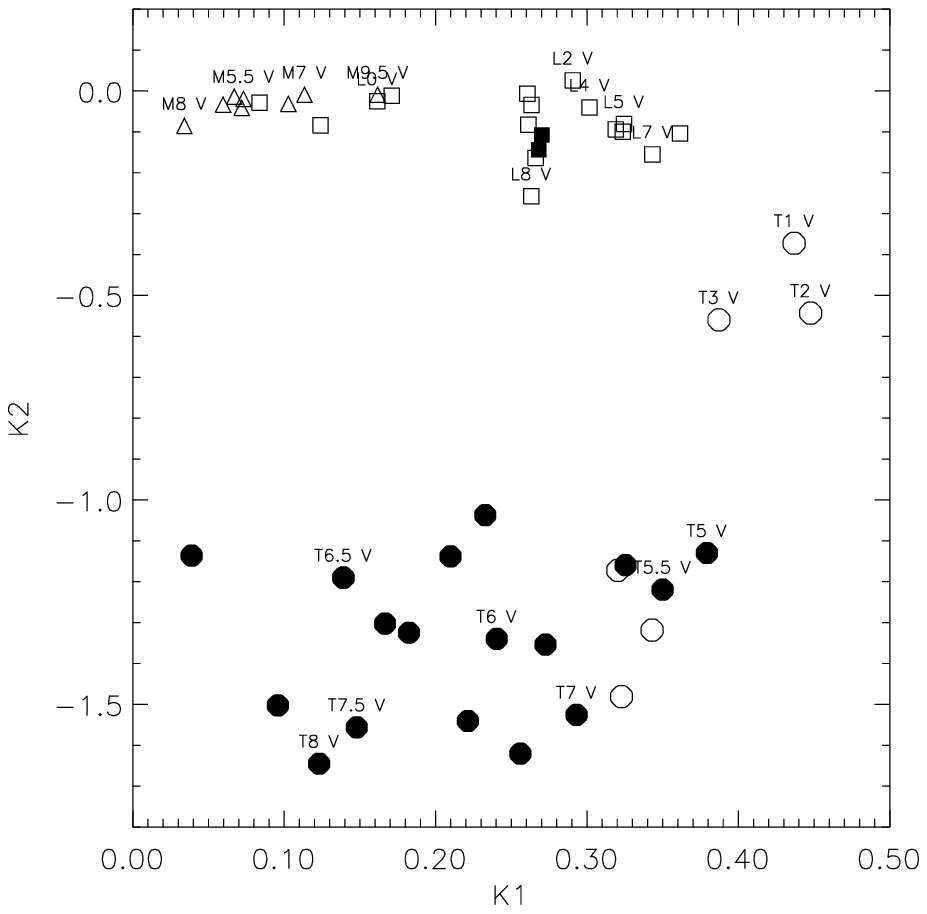]{Comparison of K1 and K2 indices from \citet{tok99} for 
M (triangles), L (squares), and T dwarfs (circles).  Filled symbols are data
taken from NIRC, open symbols are CGS4 data from the literature.
Representative M, L, and T dwarfs are labeled by their
subtypes.
\label{fig-13}}

\figcaption[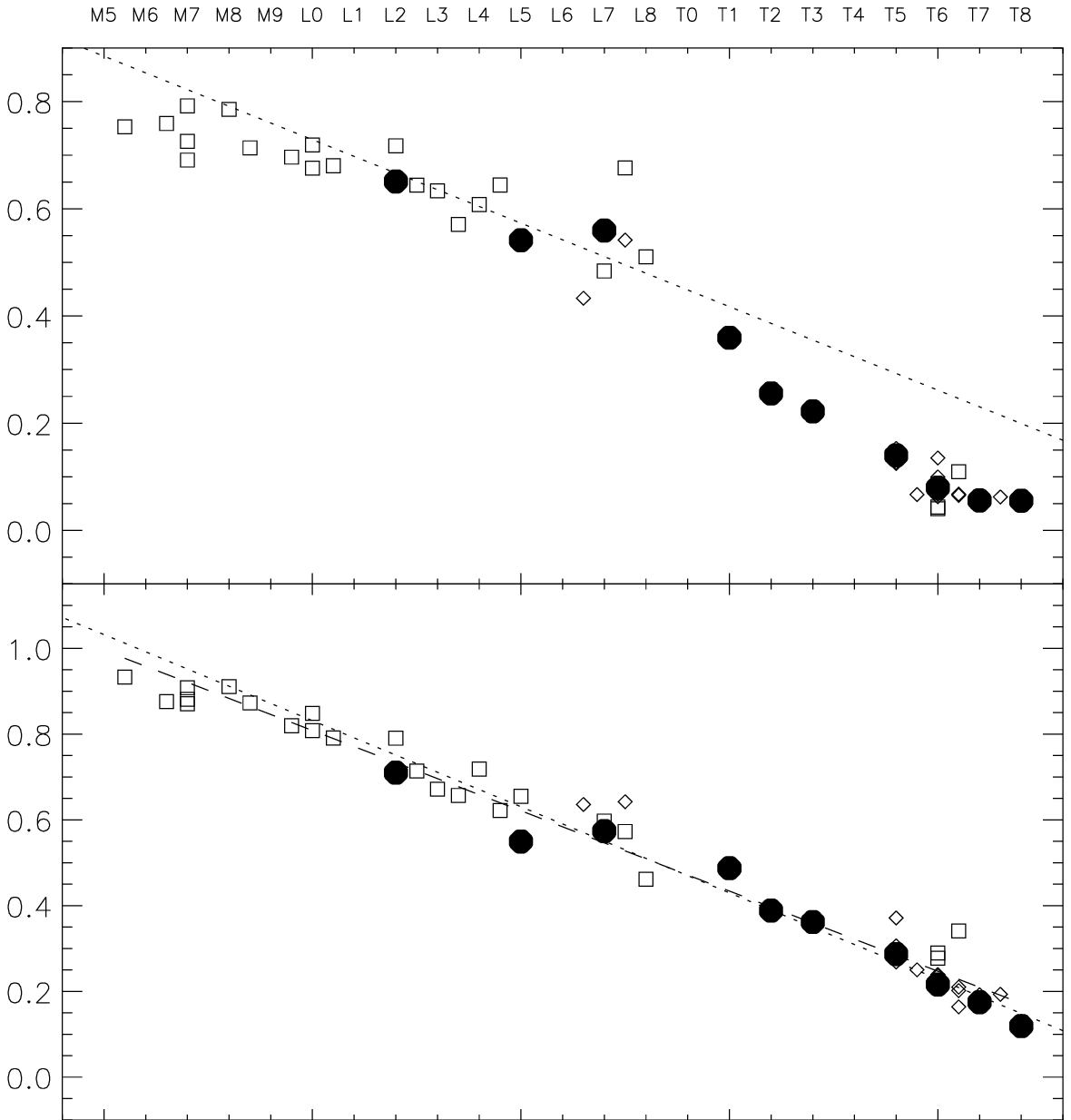]{Values for H$_2$O$^A$ and H$_2$O$^B$ indices from 
\citet{rei01} versus spectral type.  Symbols are those used in Figure 12.
Short-dashed 
lines are the index/spectral type relations determined by \citet{rei01}
for spectral types M8 V through L8 V, extended over the spectral
range shown.  The long-dashed line 
traces a linear fit to the H$_2$O$^B$ indices for types
M5 V through T8 V.
\label{fig-14}}

\figcaption[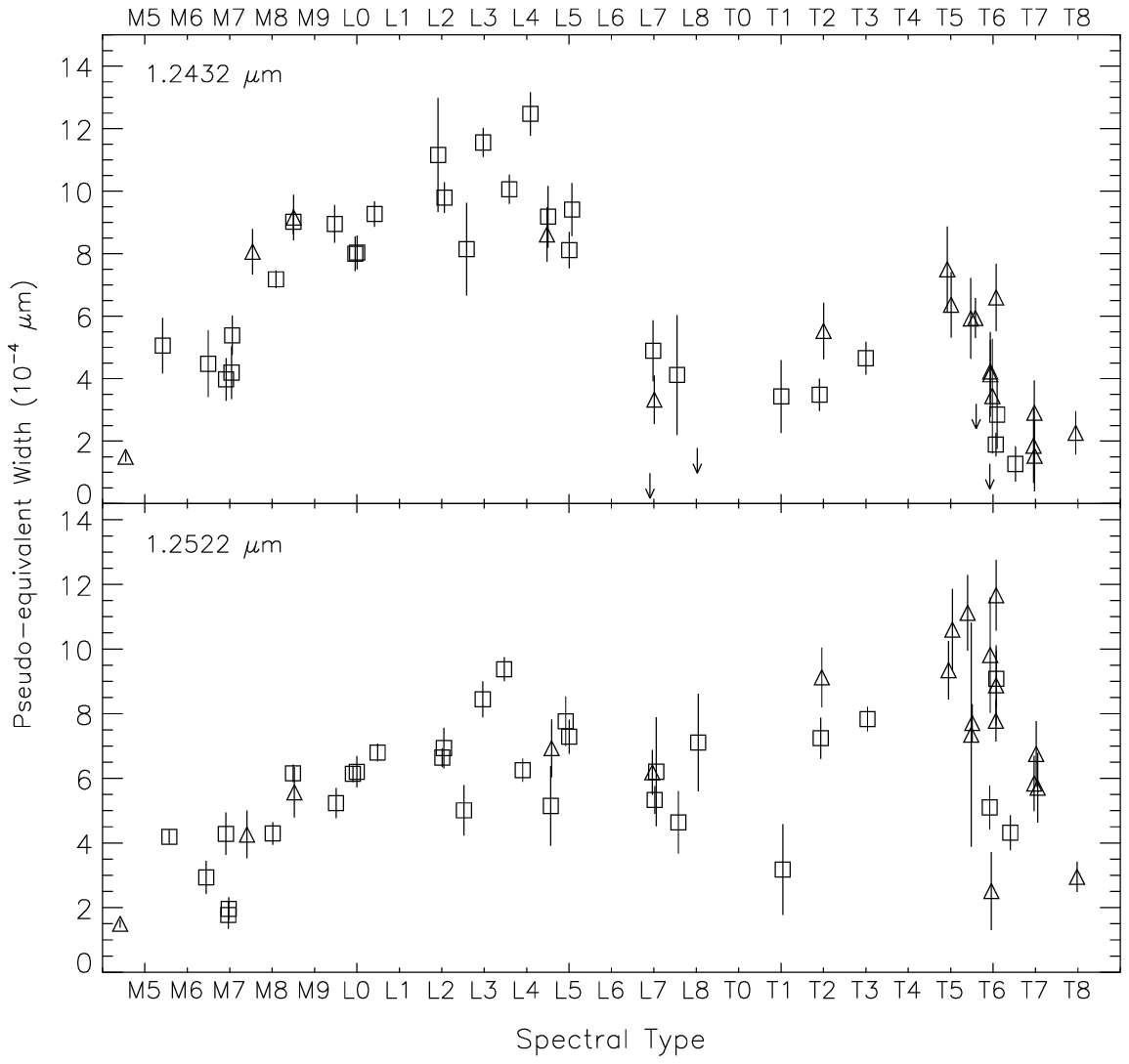]{1.2432 and 1.2522 $\micron$
K I psuedo-equivalent width (PEW) measurements for late-M, 
L, and T dwarfs.  OSIRIS data are indicated with triangles; CGS4
data from the literature are indicated by squares.
\label{fig-15}}

\figcaption[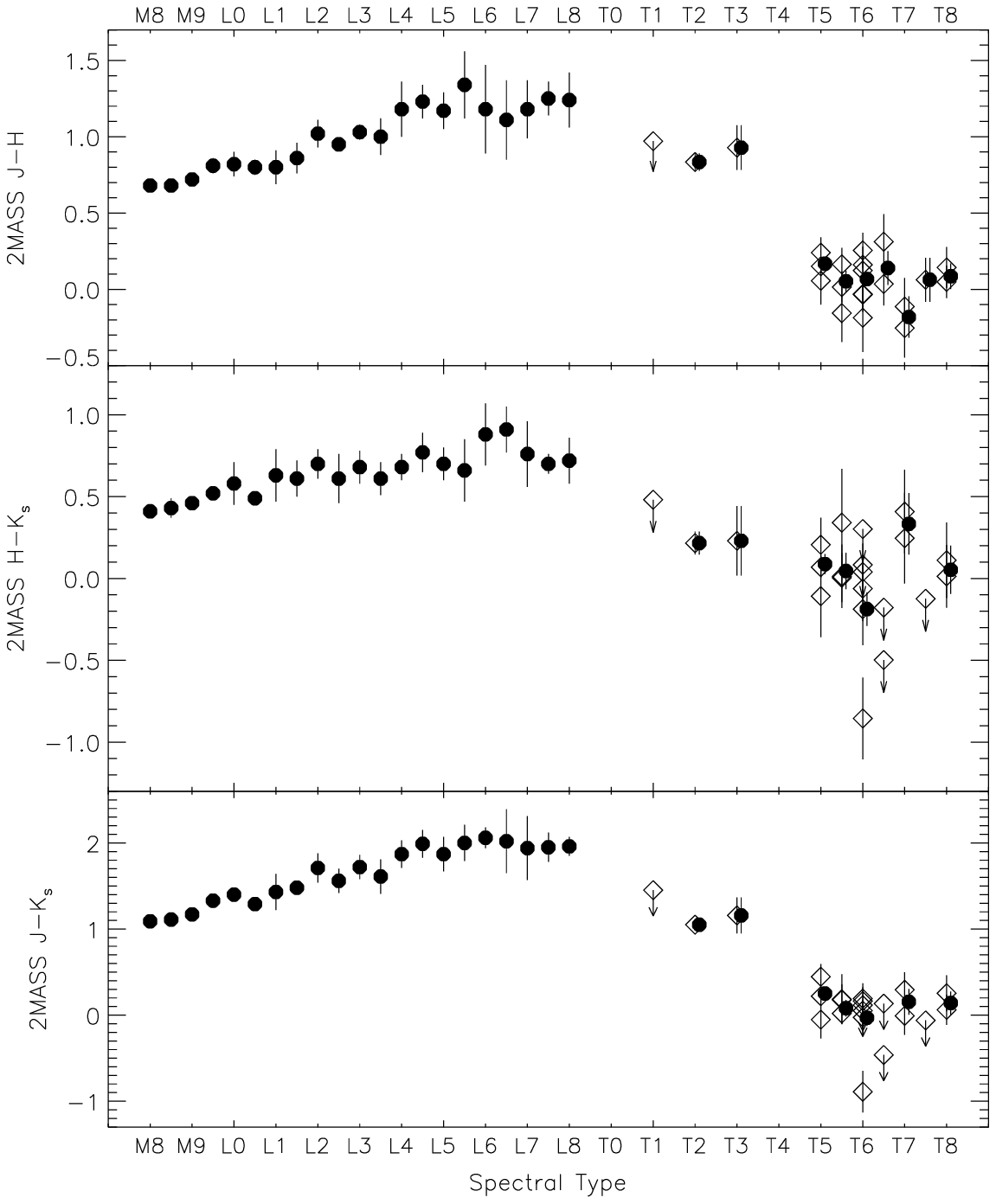]{2MASS colors for L and T dwarfs
versus spectral type.  Individual T dwarf colors with errors are indicated by 
open symbols, with upper limit detections indicated by downward arrows.
Weighted averages at each half-subtype are indicated by slightly offset
filled circles.  L dwarf mean colors (with one-sigma scatter) 
are taken from \citet{kir00}.
\label{fig-16}}

\figcaption[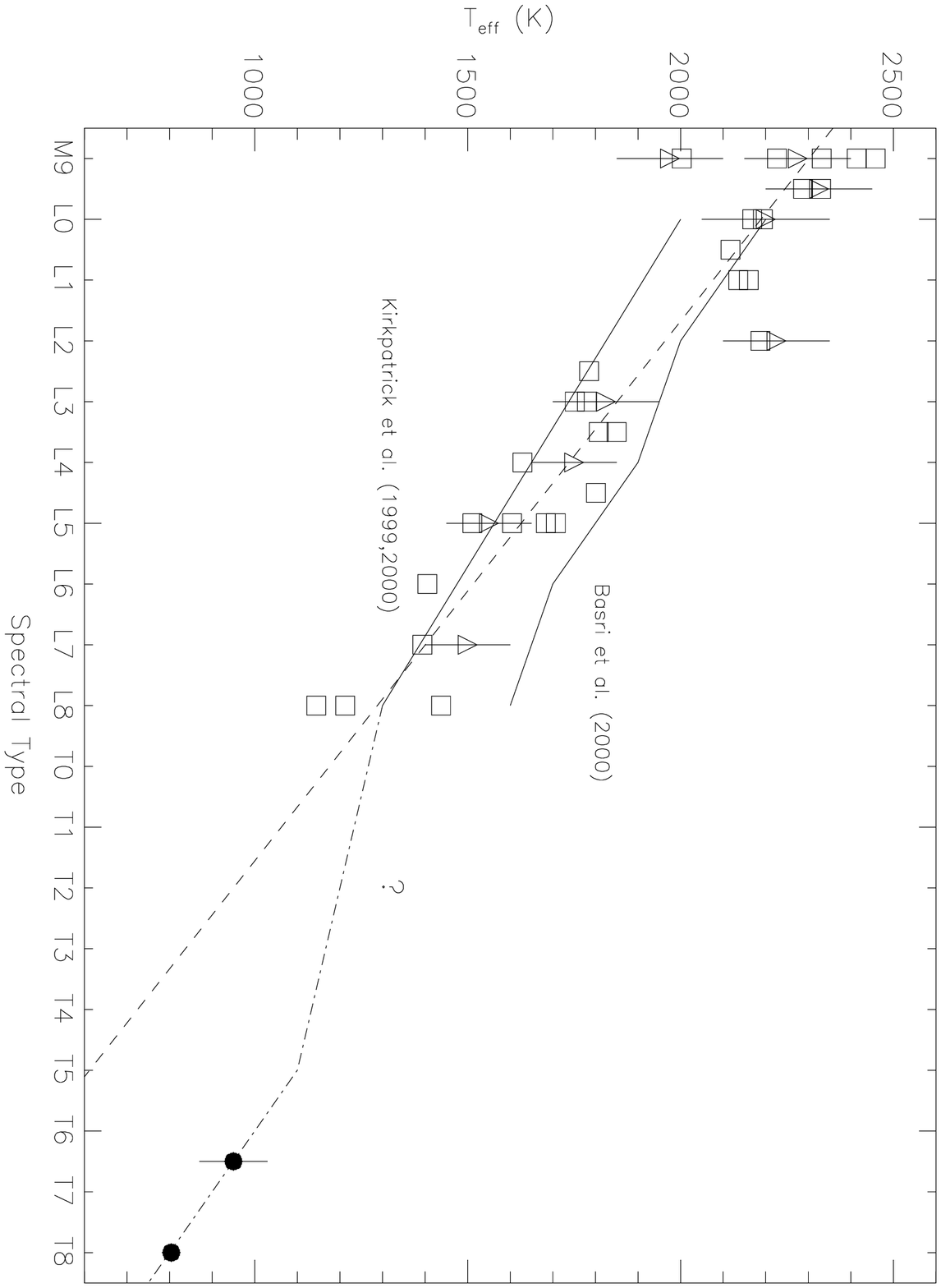]{Effective temperatures for late-type dwarfs.  
Temperatures for M and L dwarfs computed
from bolometric corrections from \citet{rei01} are shown as squares.
T$_{eff}$
estimates from \citet{leg01} are indicated by triangles.  
The two solid circles plot T$_{eff}$ estimates for
Gliese 229B \citep{sau00} and Gliese 570D \citep{me00a,geb01a}.  
Spectral type/effective temperature
scales from \citet{kir99,kir00} and \citet{bas00} are indicated,
with the latter
scale corrected to the \cite{kir99} spectral sequence.  The dashed line
shows a linear fit of T$_{eff}$ for objects of type M9 V through L8 V,
while the dot-dashed line traces a hypothetical T$_{eff}$ scale from
L8 V through T8 V. 
\label{fig-17}}

\figcaption[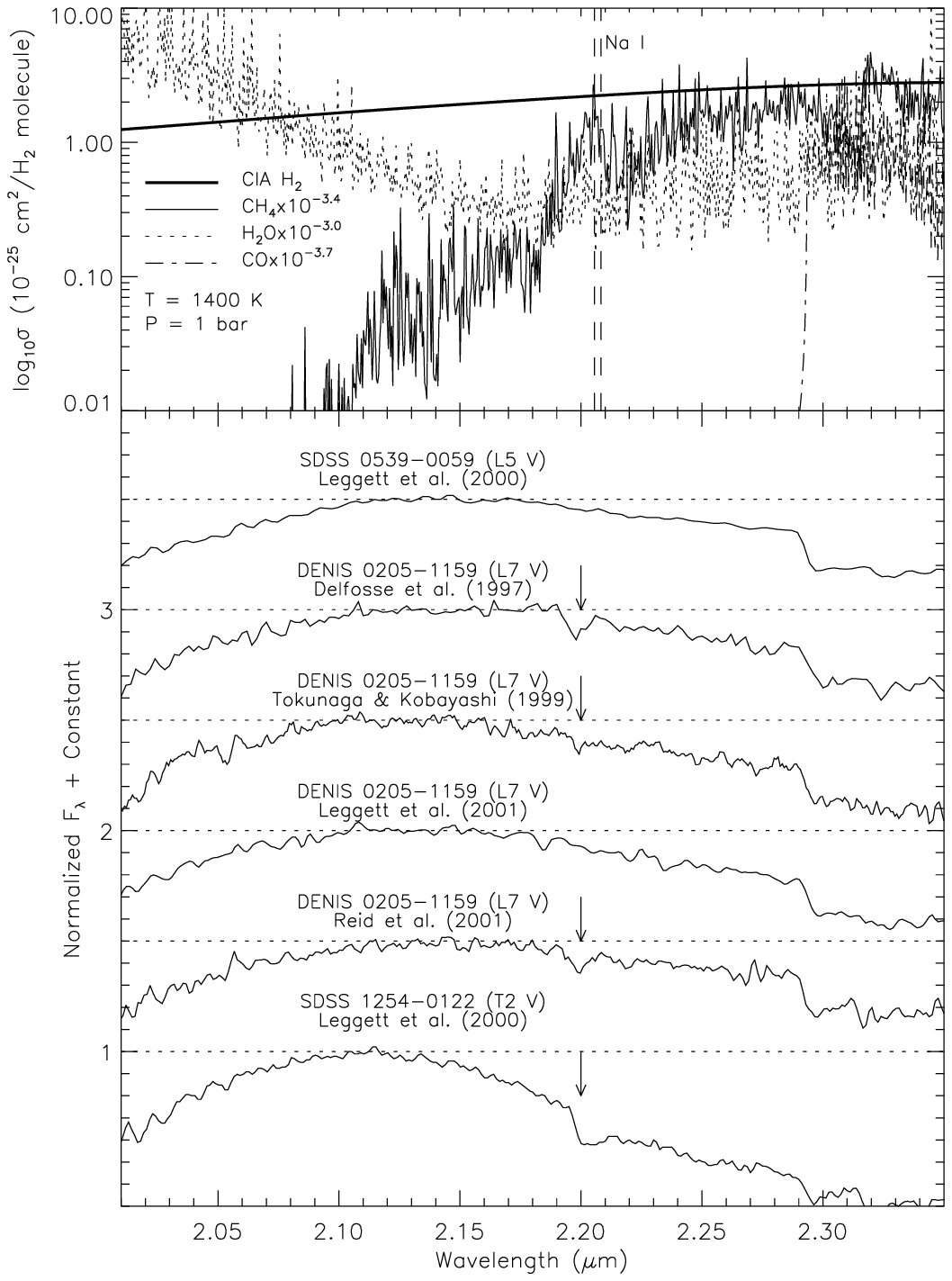]{The K-band spectrum of DENIS 0205-1159AB.  Top panel shows opacity
spectra of CH$_4$ (solid line), H$_2$O (dotted line), CO (dot-dashed line)
and CIA H$_2$ (thick solid line) at
T = 1400 K and P = 1 bar.  Data for CH$_4$, H$_2$O, 
and CO have been scaled to their
chemical equilibrium number densities at this temperature and pressure 
\citep{bur99}.
The location of the 2.2 $\micron$ Na I doublet is also indicated.
Bottom panel shows spectra of DENIS 0205-1159AB from \citet{del97};
\citet{tok99}; \citet{leg01}; and \citet{rei01}.
Data for SDSS 0539-0059 (L5 V) and SDSS 0837-0000 (T1 V)
from \citet{leg00} are also shown for comparison.  
Spectra are normalized at 2.15 $\micron$ (dotted lines)
and offset.  Arrows identify a possible weak CH$_4$ feature at 2.20 $\micron$.
\label{fig-18}}

\figcaption[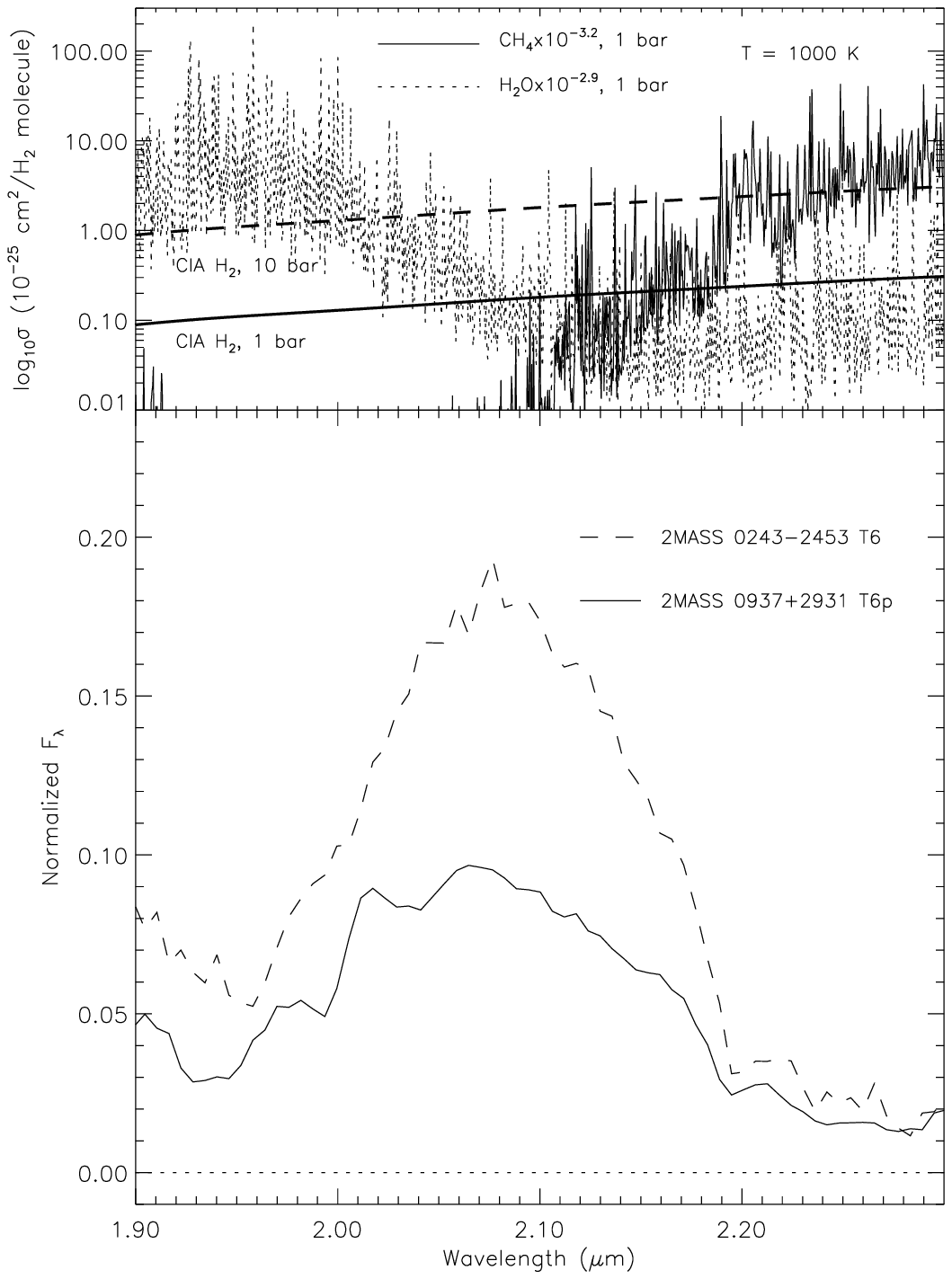]{The K-band spectrum of 2MASS 0937+2931.  
Top panel shows opacity
spectra of CH$_4$ (solid line) and H$_2$O (dotted line) at
T = 1000 K and P = 1 bar, and CIA H$_2$ opacity at T = 1000 K and P = 1 bar
(thick solid line) and 10 bar (thick dashed line).  CH$_4$ and H$_2$O data
have been scaled as in Figure 18.  
Bottom panel shows spectra of 2MASS 0937+2931 (solid line) 
and the T6 V standard 2MASS 0243-2453 (dashed line).
Both spectra have been normalized at their J-band peaks.
\label{fig-19}}

\clearpage

\begin{deluxetable}{ccccccccc}
\tabletypesize{\scriptsize}
\tablenum{1}
\tablewidth{0pt}
\tablecaption{2MASS T Dwarf Search Samples.}
\tablehead{
\colhead{} &
\colhead{} &
\colhead{} &
\colhead{Area} &
\colhead{Candidates} &
\colhead{Candidates} &
\colhead{} &
\colhead{} &
\colhead{Identified} \\
\colhead{Sample} &
\colhead{Mag.\ Cut} &
\colhead{Color Cuts} &
\colhead{(sq.\ deg.)} &
\colhead{(Cut \#1)\tablenotemark{a}} &
\colhead{(Cut \#2)\tablenotemark{b}} &
\colhead{Confirmed\tablenotemark{c}} &
\colhead{Followed up\tablenotemark{d}} &
\colhead{T Dwarfs} \\
\colhead{(1)} &
\colhead{(2)} &
\colhead{(3)} &
\colhead{(4)} &
\colhead{(5)} &
\colhead{(6)} &
\colhead{(7)} &
\colhead{(8)} &
\colhead{(9)} 
}
\startdata
wdb0699 & J $<$ 16 & J-H $<$ 0.3 & 16620 & 35280 & 404 & 93 & 67 & 14 \\
 & & H-K$_s$ $<$ 0.3 & & & &  \\
rdb0400\tablenotemark{e} & J $<$ 16 & J-H $<$ 0.3 & 15315 & 15558 & 272 & 137 & 33 & 2 \\
 & & H-K$_s$ $\geq$ 0.3 & & & &  \\
rdb0600\tablenotemark{e} & J $\leq$ 15 & J-H $\leq$ 0.4 & 4735 & 16560 & 138 & 9 & 5 & 1 \\
 & & H-K$_s$ $\leq$ 0.3 & & & &  
\tablenotetext{a}{Initial candidate pool from 2MASS catalogues.}
\tablenotetext{b}{Number of candidates remaining after visual inspection of DSS images.}
\tablenotetext{c}{Number of candidates remaining after near-infrared reimaging observations.}
\tablenotetext{d}{Number of confirmed candidates with follow-up 
Gunn r-band imaging or near-infrared spectroscopy.}
\tablenotetext{e}{Substantial follow-up remains for these samples.}
\enddata
\end{deluxetable}

\clearpage

\begin{deluxetable}{llccl}
\tabletypesize{\scriptsize}
\tablenum{2}
\tablewidth{0pt}
\tablecaption{Log of Confirmation Imaging Observations.}
\tablehead{
\colhead{} &
\colhead{} &
\colhead{\# Observed} &
\colhead{\% Confirmed} &
\colhead{} \\
\colhead{UT Date} &
\colhead{Instrument} &
\colhead{All / (J-K$_s$ $<$ 0.5)} &
\colhead{All / (J-K$_s$ $<$ 0.5)} &
\colhead{Conditions} \\
\colhead{(1)} &
\colhead{(2)} &
\colhead{(3)} &
\colhead{(4)} &
\colhead{(5)} 
}
\startdata
1998 Oct 7-9 & Palomar 60'' IRCam & 11 (10) & 0.0 (0.0) & thin cirrus to clear \\
1999 Feb 23-27 & Palomar 60'' IRCam & 6 (6) & 16.7 (16.7) & clear to scattered clouds \\
1999 May 3 & Palomar 60'' IRCam & 4 (2) & 50.0 (0.0){\phn} & cirrus and fog \\
1999 May 27 & Keck 10m NIRC & 4 (4) & 50.0 (50.0) & clear \\
1999 July 23-25 & CTIO 1.5m CIRIM & 194 (176) & 7.2 (6.3) & clear to hazy \\
1999 Sept 22-25 & Palomar 60'' IRCam & 83 (70) & 37.3 (37.1) & fog to hazy \\
1999 Nov 19-21 & Palomar 60'' IRCam & 4 (3) & 0.0 (0.0) & thin clouds to cloudy \\
1999 Dec 17-19 & CTIO 1.5m CIRIM & 28 (22) & 50.0 (36.4) & hazy to scattered clouds \\
2000 Jan 23-24 & Keck 10m NIRC & 5 (5) & 0.0 (0.0) & clear and windy \\
2000 Apr 14-15 & Palomar 60'' IRCam & 54 (45) & 55.6 (57.8) & clear to low clouds and wind \\
2000 May 18-20 & Palomar 60'' IRCam & 83 (8){\phn} & 80.7 (25.0) & clear to fog and haze \\
2000 June 21-23 & CTIO 1.5m OSIRIS & 104 (45){\phn} & 35.6 (15.6) & cloudy to clear \\
2000 Aug 18-20 & CTIO 1.5m OSIRIS & 26 (6){\phn} & 53.8 (16.7) & light cirrus \\
2000 Oct 11-13 & Palomar 60'' IRCam & 27 (5){\phn} & 63.0 (40.0) & clear, high humidity \\
2001 Jan 5 & Palomar 5m D78 & 1 (1) & 100.0 (100.0) & cloudy \\
\hline \hline 
TOTAL & & 634 (408) & 36.3 (21.3) & \\
\enddata
\end{deluxetable}

\clearpage

\begin{deluxetable}{lccccccl}
\rotate
\tabletypesize{\scriptsize}
\tablenum{3}
\tablewidth{0pt}
\tablecaption{2MASS IDR2 Sources Unconfirmed in Second Epoch Imaging (Partial).}
\tablehead{
\colhead{Object\tablenotemark{a}} &
\colhead{Date (UT)\tablenotemark{b}} &
\colhead{2MASS J} &
\colhead{2MASS H} &
\colhead{2MASS K$_s$} &
\colhead{Instrument\tablenotemark{c}} &
\colhead{Date (UT)} &
\colhead{Note\tablenotemark{d}} \\
\colhead{(1)} &
\colhead{(2)} &
\colhead{(3)} &
\colhead{(4)} &
\colhead{(5)} &
\colhead{(6)} &
\colhead{(7)} &
\colhead{(8)} 
}
\startdata
2MASSI J0004085-101237 & 1998 Oct 07 & 15.95${\pm}$0.07 & 15.79${\pm}$0.14 & 15.78${\pm}$0.25 & CTIO CIRIM & 1999 Jul 24 &  \\
2MASSI J0006171-072119 & 1998 Oct 07 & 15.98${\pm}$0.07 & 15.76${\pm}$0.13 &  $>$ 15.19 & CTIO OSIRIS & 2000 Aug 20 & 13$\farcs$0 from 1998 QP104 \\
2MASSI J0009296+163226 & 1997 Jul 12 & 15.84${\pm}$0.07 & 15.32${\pm}$0.09 & 15.52${\pm}$0.22 & CTIO CIRIM & 1999 Jul 24 &  \\
2MASSI J0009332-111547 & 1998 Oct 16 & 15.27${\pm}$0.05 & 15.02${\pm}$0.09 & 15.00${\pm}$0.14 & CTIO CIRIM & 1999 Jul 23 &  \\
2MASSI J0011144-043556 & 1998 Sep 17 & 15.24${\pm}$0.04 & 14.94${\pm}$0.06 & 14.69${\pm}$0.10 & CTIO CIRIM & 1999 Jul 24 &  \\
2MASSI J0012248-020023 & 1998 Sep 17 & 12.94${\pm}$0.03 & 12.59${\pm}$0.03 & 12.49${\pm}$0.03 & CTIO OSIRIS & 2000 Jun 21 & 10$\farcs$3 from (927) Ratisbona \\
2MASSI J0012529-062929 & 1998 Oct 16 & 14.88${\pm}$0.04 & 14.59${\pm}$0.06 & 14.49${\pm}$0.08 & CTIO CIRIM & 1999 Jul 24 & 11$\farcs$0 from (2321) Luznice 1980 DB1 \\
2MASSI J0014365-130605 & 1998 Jul 20 & 15.62${\pm}$0.07 & 15.36${\pm}$0.11 & 15.27${\pm}$0.18 & CTIO CIRIM & 1999 Jul 24 &  \\
2MASSI J0014407-080107 & 1998 Oct 16 & 14.06${\pm}$0.03 & 13.72${\pm}$0.03 & 13.50${\pm}$0.05 & CTIO OSIRIS & 2000 Jun 21 & 9$\farcs$6 from (2132) Zhukov 1975 TW3 \\
2MASSI J0015297+032103 & 1998 Oct 18 & 14.84${\pm}$0.04 & 14.70${\pm}$0.06 & 14.64${\pm}$0.10 & CTIO OSIRIS & 2000 Jun 21 & 6$\farcs$8 from (6046) 1991 RF14 \\
$\vdots$ & & & & & & & \\
\tablenotetext{a}{Source designations for 2MASS sources in the Incremental
Release Catalogs are given
as ``2MASSI Jhhmmss[.]s$\pm$ddmmss'.  The suffix conforms to IAU
nomenclature convention and is the sexigesimal R.A. and decl. at J2000 equinox.}
\tablenotetext{b}{UT Date observed by 2MASS.}
\tablenotetext{c}{Telescope shorthand as follows: P60 = Palomar 60'', CTIO = 
Cerro Tololo 1.5m, and KECK = Keck 10m.}
\tablenotetext{d}{Unconfirmed objects near asteroid positions are likely
missed associations due to errors in ephemerides.  See text for discussion.}  

\enddata
\end{deluxetable}

\clearpage

\begin{deluxetable}{lccccll}
\tabletypesize{\scriptsize}
\tablenum{4a}
\tablewidth{0pt}
\tablecaption{Log of NIRC Observations.}
\tablehead{
\colhead{Object} &
\colhead{Type} &
\colhead{UT Date} &
\colhead{Int.\ (s)} &
\colhead{Airmass} &
\colhead{Calibrator} &
\colhead{Type\tablenotemark{a}} \\
\colhead{(1)} &
\colhead{(2)} &
\colhead{(3)} &
\colhead{(4)} &
\colhead{(5)} &
\colhead{(6)} &
\colhead{(7)}  
}
\startdata
2MASSI J0243$-$2453 & T & 2000 Jan 24 & 1080 &  1.41 & HD 19378 & G0 V \\ 
2MASSI J0415$-$0935 & T & 1999 Nov 18 & 1080 &  1.21 & SAO 150589 & G5 V \\ 
2MASSI J0415$-$0935 & T & 2000 Jan 24 & 1080 &  1.35 & HD 22855 & G3 V \\ 
2MASSI J0559$-$1404 & T & 2000 Jan 23 & 1080 &  1.21 & HD 41083 & G2/3 V \\ 
2MASSI J0559$-$1404 & T & 2000 Jan 24 & 1080 &  1.26 & HD 41083 & G2/3 V \\ 
2MASSI J0727+1710 & T & 2000 Jan 23 & 1080 &  1.00 & SAO 96796 & G5 V \\ 
2MASSI J0825+2115\tablenotemark{b} & L7.5 V & 2000 Jan 23 & 1080 &  1.00 & SAO 96796 & G5 V \\ 
2MASSI J0920+3517\tablenotemark{b} & L6.5 V & 2000 Jan 24 & 1080 &  1.04 & HD 73617 & F5 V \\ 
2MASSI J0937+2931 & T & 2000 Jan 23 & 1080 &  1.01 & HD 76332 & G2 V \\ 
2MASSI J0937+2931 & T & 2000 Jan 24 & 1080 &  1.03 & HD 76332 & G2 V \\ 
2MASSI J1047+2124 & T & 2000 Jan 23 & 1080 &  1.00 & HD 98562 & G2 V \\ 
2MASSI J1047+2124 & T & 2000 Jan 24 & 1080 &  1.00 & HD 98562 & G2 V \\ 
2MASSI J1217$-$0311 & T & 2000 Jan 23 & 1080 &  1.09 & HD 108754 & G7 V \\ 
2MASSI J1237+6526 & T & 2000 Jan 24 & 1080 &  1.42 & HD 110276 & G0 Vw \\ 
SDSSp J1346$-$0031 & T & 2000 Jan 23 & 1080 &  1.07 & HD 127913 & G2 V \\ 
Gliese 570D & T & 2000 Jan 24 & 1080 &  1.39 & HD 126253 & G0 V \\ 
2MASSI J1553+1532 & T & 2000 Jul 22 & 900 &  1.15 & HD 146759 & G5 V \\ 
2MASSI J2254+3123 & T & 2000 Jul 22 & 900 &  1.08 & HD 219213 & G5 V \\ 
2MASSI J2339+1352 & T & 2000 Jul 22 & 900 &  1.08 & HD 219213 & G5 V \\ 
2MASSI J2356$-$1553 & T & 1999 Nov 18 & 1080 &  1.28 & HD 990 & F5 V \\ 
\tablenotetext{a}{Spectral types obtained from the SIMBAD database.}
\tablenotetext{b}{Previously identified L dwarfs observed for spectral comparision.}
\enddata
\end{deluxetable}

\clearpage

\begin{deluxetable}{lccccll}
\tabletypesize{\scriptsize}
\tablenum{4b}
\tablewidth{0pt}
\tablecaption{Log of OSIRIS Observations.}
\tablehead{
\colhead{Object} &
\colhead{Type} &
\colhead{UT Date} &
\colhead{Int.\ (s)} &
\colhead{Airmass} &
\colhead{Calibrator} &
\colhead{Type\tablenotemark{a}} \\
\colhead{(1)} &
\colhead{(2)} &
\colhead{(3)} &
\colhead{(4)} &
\colhead{(5)} &
\colhead{(6)} &
\colhead{(7)} 
}
\startdata
DENIS J0205$-$1159\tablenotemark{b} & L7 V & 2000 Jul 19 & 1250 & 1.07 & HD 13406 & A3 Vm \\ 
2MASSI J0243$-$2453 & T & 1999 Dec 20 & 3600 & 1.04--1.15 & HD 17168 & A1 V \\ 
2MASSI J0243$-$2453 & T & 1999 Dec 21 & 4200 & 1.01--1.09 & HD 17168 & A1 V \\ 
2MASSI J0559$-$1404 & T & 1999 Dec 20 & 2400 & 1.05--1.09 & HD 40972 & A0 V \\ 
2MASSI J0727+1710 & T & 1999 Dec 20 & 3600 & 1.52--1.72 & HD 60275 & A1 V \\ 
2MASSI J0937+2931 & T & 1999 Dec 21 & 3000 & 1.98--2.00 & HD 79248 & A2 V \\ 
2MASSI J1225$-$2739 & T & 2000 Jul 18 & 2500 & 1.16--1.24 & HD 109142 & A3 Vm \\ 
SDSSp J1254-0122 & T & 2000 Jul 16 & 4500 & 1.21--1.54 & HD 112846 & A3 V \\ 
Gliese 570D & T & 2000 Jul 16 & 4500 & 1.09--1.30 & HD 131992 & A2/3 V \\ 
SDSSp J1624+0029 & T & 2000 Jul 15 & 3450 & 1.49--2.07 & HD 148207 & A0 V \\ 
2MASSI J1534$-$2952 & T & 2000 Jul 18 & 2500 & 1.19--1.28 & HD 139202 & A1 Vm \\ 
2MASSI J1546$-$3325 & T & 2000 Jul 19 & 3750 & 1.04--1.13 & HD 140442 & A1 V \\ 
2MASSI J1553+1532 & T & 2000 Jul 17 & 3750 & 1.43--1.53 & HD 143936 & A0 V \\ 
2MASSI J1553+1532 & T & 2000 Jul 18 & 2500 & 1.46--1.51 & HD 143936 & A0 V \\ 
HB 2115$-$4518\tablenotemark{b} & M8.5 V & 2000 Jul 19 & 750 & 1.04 & HD 203725 & A0 III \\ 
HB 2124$-$4268\tablenotemark{b} & M7.5 V & 2000 Jul 19 & 750 & 1.02 & HD 203725 & A0 III \\ 
LHS 511AB\tablenotemark{b} & M4.5 V & 2000 Jul 19 & 75 & 1.08 & HD 205178 & A0 V \\ 
2MASSI J2224$-$0158\tablenotemark{b} & L4.5 V & 2000 Jul 19 & 1250 & 1.15 & HD 212417 & A3 V \\ 
2MASSI J2254+3123 & T & 2000 Jul 16 & 3750 & 2.10--2.20 & HD 216716 & A0 V \\ 
2MASSI J2339+1352 & T & 2000 Jul 16 & 2500 & 1.44--1.51 & HD 222250 & A3 V \\ 
2MASSI J2356$-$1553 & T & 1999 Dec 20 & 1800 & 1.27--1.37 & HD 4065 & A0 V \\ 
2MASSI J2356$-$1553 & T & 2000 Jul 17 & 5000 & 1.04--1.15 & HD 223785 & A1 V \\ 
\tablenotetext{a}{Spectral types obtained from the SIMBAD database.}
\tablenotetext{b}{Previously identified M and L dwarfs observed for 
spectral comparision.}
\enddata
\end{deluxetable}

\clearpage

\begin{deluxetable}{lccccccccc}
\tabletypesize{\scriptsize}
\tablenum{5}
\tablewidth{0pt}
\tablecaption{Comparison of NIRC Spectrophotometric Colors to 2MASS Photometry.}
\tablehead{
\colhead{} &
\multicolumn{3}{c}{2MASS J-H} &
\multicolumn{3}{c}{2MASS H-K$_s$} &
\multicolumn{3}{c}{2MASS J-K$_s$} \\
\colhead{Object} &
\colhead{Phot.} &
\colhead{Spec.} &
\colhead{$\delta$\tablenotemark{a}} &
\colhead{Phot.} &
\colhead{Spec.} &
\colhead{$\delta$\tablenotemark{a}} &
\colhead{Phot.} &
\colhead{Spec.} &
\colhead{$\delta$\tablenotemark{a}} \\
\colhead{(1)} &
\colhead{(2)} &
\colhead{(3)} &
\colhead{(4)} &
\colhead{(5)} &
\colhead{(6)} &
\colhead{(7)} &
\colhead{(8)} &
\colhead{(9)} &
\colhead{(10)} 
}
\startdata
2MASSI J0243$-$2453 &  {\phs}0.25$\pm$0.11 & {\phs}0.10 & {\phs}0.15 & $-$0.06$\pm$0.20 & {\phs}0.15 & $-$0.21 & {\phs}0.19$\pm$0.18 & {\phs}0.25 & $-$0.06 \\ 
2MASSI J0415$-$0935\tablenotemark{b} &  {\phs}0.14$\pm$0.13 & {\phs}0.12 & {\phs}0.02 & {\phs}0.12$\pm$0.23 & {\phs}0.21 & $-$0.09 & {\phs}0.26$\pm$0.21 & {\phs}0.33 & $-$0.07 \\ 
2MASSI J0415$-$0935\tablenotemark{b} &  {\phs}0.14$\pm$0.13 & {\phs}0.12 & {\phs}0.02 & {\phs}0.12$\pm$0.23 & $-$0.04 & {\phs}0.16 & {\phs}0.26$\pm$0.21 & {\phs}0.08 & {\phs}0.18 \\ 
2MASSI J0559$-$1404\tablenotemark{b} &  {\phs}0.14$\pm$0.05 & {\phs}0.25 & $-$0.11 & {\phs}0.07$\pm$0.06 & {\phs}0.18 & $-$0.11 & {\phs}0.21$\pm$0.06 & {\phs}0.43 & $-$0.22 \\ 
2MASSI J0559$-$1404\tablenotemark{b} &  {\phs}0.14$\pm$0.05 & {\phs}0.26 & $-$0.12 & {\phs}0.07$\pm$0.06 & {\phs}0.34 & $-$0.27 & {\phs}0.21$\pm$0.06 & {\phs}0.60 & $-$0.39 \\ 
2MASSI J0727+1710 &  $-$0.27$\pm$0.19 & $-$0.15 & $-$0.12 & {\phs}0.26$\pm$0.28 & {\phs}0.05 & {\phs}0.21 & $-$0.01$\pm$0.22 & $-$0.10 & {\phs}0.09 \\ 
2MASSI J0825+2115 &  {\phs}1.33$\pm$0.06 & {\phs}0.98 & {\phs}0.35 & {\phs}0.74$\pm$0.06 & {\phs}0.77 & $-$0.03 & {\phs}2.07$\pm$0.06 & {\phs}1.75 & {\phs}0.32 \\ 
2MASSI J0920+3517 &  {\phs}0.93$\pm$0.10 & {\phs}0.85 & {\phs}0.08 & {\phs}0.73$\pm$0.11 & {\phs}0.68 & {\phs}0.05 & {\phs}1.66$\pm$0.11 & {\phs}1.52 & {\phs}0.14 \\ 
2MASSI J0937+2931\tablenotemark{b} &  $-$0.03$\pm$0.08 & {\phs}0.15 & $-$0.18 & $-$0.86$\pm$0.25 & $-$0.30 & $-$0.56 & $-$0.89$\pm$0.24 & $-$0.16 & $-$0.73 \\ 
2MASSI J0937+2931\tablenotemark{b} &  $-$0.03$\pm$0.08 & {\phs}0.02 & $-$0.05 & $-$0.86$\pm$0.25 & $-$0.49 & $-$0.37 & $-$0.89$\pm$0.24 & $-$0.48 & $-$0.41 \\ 
2MASSI J1047+2124\tablenotemark{b} &  {\phs}0.03$\pm$0.13 & {\phs}0.15 & $-$0.12 & $<$ $-$0.50 & $-$0.22 & -- & $<$ $-$0.47 & $-$0.06 & -- \\ 
2MASSI J1047+2124\tablenotemark{b} &  {\phs}0.03$\pm$0.13 & {\phs}0.19 & $-$0.16 & $<$ $-$0.50 & $-$0.22 & -- & $<$ $-$0.47 & $-$0.02 & -- \\ 
2MASSI J1217$-$0311 &  {\phs}0.06$\pm$0.14 & {\phs}0.01 & {\phs}0.05 & $<$ $-$0.12 & {\phs}0.23 & -- & $<$ $-$0.06 & {\phs}0.24 & -- \\ 
2MASSI J1237+6526 &  {\phs}0.31$\pm$0.18 & {\phs}0.00 & {\phs}0.31 & $<$ $-$0.18 & $-$0.41 & -- & $<$ {\phs}0.13 & $-$0.41 & -- \\ 
SDSSp J1346$-$0031 &  $-$0.19$\pm$0.22 & {\phs}0.02 & $-$0.21 & $<$ {\phs}0.30 & {\phs}0.15 & -- & $<$ {\phs}0.11 & {\phs}0.17 & -- \\ 
Gliese 570D &  {\phs}0.05$\pm$0.10 & {\phs}0.09 & $-$0.04 & {\phs}0.01$\pm$0.19 & $-$0.01 & {\phs}0.02 & {\phs}0.06$\pm$0.18 & {\phs}0.07 & $-$0.01 \\ 
2MASSI J1553+1532 &  $-$0.11$\pm$0.19 & {\phs}0.14 & $-$0.25 & {\phs}0.41$\pm$0.25 & {\phs}0.05 & {\phs}0.36 & {\phs}0.30$\pm$0.21 & {\phs}0.19 & {\phs}0.11 \\ 
2MASSI J2254+3123 &  {\phs}0.24$\pm$0.10 & {\phs}0.20 & {\phs}0.04 & {\phs}0.21$\pm$0.17 & $-$0.09 & {\phs}0.30 & {\phs}0.45$\pm$0.15 & {\phs}0.11 & {\phs}0.34 \\ 
2MASSI J2339+1352 &  $-$0.16$\pm$0.19 & {\phs}0.12 & $-$0.28 & {\phs}0.34$\pm$0.33 & $-$0.08 & {\phs}0.42 & {\phs}0.18$\pm$0.29 & {\phs}0.04 & {\phs}0.14 \\ 
2MASSI J2356$-$1553 &  {\phs}0.16$\pm$0.12 & {\phs}0.18 & $-$0.02 & $-$0.19$\pm$0.21 & {\phs}0.12 & $-$0.31 & $-$0.03$\pm$0.20 & {\phs}0.30 & $-$0.33 \\ 
\tablenotetext{a}{$\delta$ $\equiv$ photometry minus spectrophotometry.}
\tablenotetext{b}{Separate observations obtained on 2000 January 23-24 (UT).}
\enddata

\end{deluxetable}

\clearpage

\begin{deluxetable}{lcccc}
\tabletypesize{\scriptsize}
\tablenum{6}
\tablewidth{0pt}
\tablecaption{New T Dwarfs Identified in the 2MASS Catalog.}
\tablehead{
\colhead{2MASS Designation} &
\colhead{Primary Sample\tablenotemark{a}} &
\colhead{2MASS J} &
\colhead{2MASS H} &
\colhead{2MASS K$_s$}
}
\startdata
2MASSI J0243137$-$245329 & wdb0699 & 15.40$\pm$0.05 & 15.15$\pm$0.10 & 15.21$\pm$0.17 \\ 
2MASSI J0415195$-$093506 & wdb0699 & 15.71$\pm$0.06 & 15.57$\pm$0.12 & 15.45$\pm$0.20 \\ 
2MASSI J0727182+171001 & wdb0699 & 15.55$\pm$0.07 & 15.82$\pm$0.18 & 15.56$\pm$0.21 \\ 
2MASSI J0755480+221218 & wdb0699 & 15.72$\pm$0.07 & 15.66$\pm$0.14 & 15.77$\pm$0.21 \\
2MASSI J0937347+293142 & wdb0699 & 14.65$\pm$0.04 & 14.68$\pm$0.07 & 15.54$\pm$0.24 \\ 
2MASSI J1534498$-$295227 & wdb0699 & 14.90$\pm$0.04 & 14.89$\pm$0.09 & 14.86$\pm$0.11 \\ 
2MASSI J1546271$-$332511 & wdb0699 & 15.60$\pm$0.05 & 15.44$\pm$0.09 & 15.42$\pm$0.17 \\ 
2MASSI J1553022+153236 & rdb0400 & 15.81$\pm$0.08 & 15.92$\pm$0.17 & 15.51$\pm$0.19 \\ 
2MASSI J2254188+312349 & wdb0699 & 15.28$\pm$0.05 & 15.04$\pm$0.09 & 14.83$\pm$0.14 \\ 
2MASSI J2339101+135230 & rdb0400 & 15.88$\pm$0.08 & 16.04$\pm$0.17 & 15.70$\pm$0.28 \\ 
2MASSI J2356547$-$155310 & wdb0699 & 15.80$\pm$0.06 & 15.64$\pm$0.10 & 15.83$\pm$0.19 \\
\tablenotetext{a}{Some objects found in multiple samples due to overlapping
search criteria.}
\enddata
\end{deluxetable}

\clearpage

\begin{deluxetable}{llccccl}
\tabletypesize{\scriptsize}
\tablenum{7}
\tablewidth{0pt}
\tablecaption{Known T Dwarfs.}
\tablehead{
\colhead{Object} &
\colhead{Type} & 
\colhead{J\tablenotemark{a}} &
\colhead{J$-$H\tablenotemark{a}} &
\colhead{H$-$K$_s$\tablenotemark{a}} & 
\colhead{J$-$K$_s$\tablenotemark{a}} & 
\colhead{Ref.} \\
\colhead{(1)} &
\colhead{(2)} &
\colhead{(3)} &
\colhead{(4)} &
\colhead{(5)} &
\colhead{(6)} &
\colhead{(7)} 
}
\startdata
2MASSI J0243137$-$245329 & T6 V & 15.40$\pm$0.05 & {\phs}0.25$\pm$0.11 & $-$0.06$\pm$0.20 & {\phs}0.19$\pm$0.18 & 1 \\
2MASSI J0415195$-$093506 & T8 V & 15.71$\pm$0.06 & {\phs}0.14$\pm$0.13 & {\phs}0.12$\pm$0.23 & {\phs}0.26$\pm$0.21 & 1 \\
2MASSI J0559191$-$140448 & T5 V & 13.82$\pm$0.03 & {\phs}0.14$\pm$0.05 & {\phs}0.07$\pm$0.06 & {\phs}0.21$\pm$0.06 & 2 \\
Gliese 229B & T6.5 V & 14.32$\pm$0.05\tablenotemark{b} & $-$0.03$\pm$0.07\tablenotemark{b} & $-$0.10$\pm$0.07\tablenotemark{b} & $-$0.13$\pm$0.07\tablenotemark{b} & 3 \\
2MASSI J0727182+171001 & T7 V & 15.55$\pm$0.07 & $-$0.27$\pm$0.19 & {\phs}0.26$\pm$0.28 &  $-$0.01$\pm$0.22 & 1 \\
2MASSI J0755480+221218 & T5 V: & 15.72$\pm$0.07 & {\phs}0.06$\pm$0.16 & $-$0.11$\pm$0.25 &  $-$0.05$\pm$0.22 & 1 \\
SDSSp J083717.22$-$000018.3 & T1 V & 17.23$\pm$0.25 & {\phs}1.25$\pm$0.30 & $<$ {\phs}0.33 &  $<$ {\phs}1.58 & 4 \\
2MASSI J0937347+293142 & T6 Vp & 14.65$\pm$0.04 & $-$0.03$\pm$0.08 & $-$0.86$\pm$0.25 & $-$0.89$\pm$0.24 & 1 \\
SDSSp J102109.69$-$030420.1 & T3 V & 16.26$\pm$0.10 & {\phs}0.93$\pm$0.15 & {\phs}0.23$\pm$0.20 &  {\phs}1.16$\pm$0.20 & 4 \\
2MASSI J1047538+212423 & T6.5 V & 15.82$\pm$0.06 & {\phs}0.03$\pm$0.13 & $<$ $-$0.50 &  $<$ $-$0.47 & 5 \\
NTTDF 1205-0744 & T6 V: & 20.15\tablenotemark{c} & -- & -- & $-$0.15\tablenotemark{c} & 6 \\
2MASSI J1217110$-$031113 & T7.5 V & 15.85$\pm$0.07 & {\phs}0.06$\pm$0.14 & $<$ $-$0.12 &  $<$ $-$0.06 & 5 \\
2MASSI J1225543$-$273946 & T6 V & 15.23$\pm$0.05 & {\phs}0.13$\pm$0.09 & {\phs}0.04$\pm$0.17 & {\phs}0.17$\pm$0.16 & 5 \\
2MASSI J1237392+652615 & T6.5 V & 16.03$\pm$0.09 & {\phs}0.31$\pm$0.18 & $<$ $-$0.18 &  $<$ {\phs}0.13 & 5 \\
SDSSp J125453.90$-$012247.4 & T2 V & 14.88$\pm$0.03 & {\phs}0.84$\pm$0.05 & {\phs}0.21$\pm$0.07 & {\phs}1.05$\pm$0.07 & 4 \\
SDSSp J134646.45$-$003150.4 & T6 V & 15.86$\pm$0.08 & $-$0.19$\pm$0.22 & $<$ {\phs}0.30 & $<$ {\phs}0.11 & 7 \\
Gliese 570D & T8 V & 15.33$\pm$0.05 & {\phs}0.05$\pm$0.10 & {\phs}0.01$\pm$0.19 & {\phs}0.06$\pm$0.18 & 8 \\
2MASSI J1534498$-$295227 & T5.5 V & 14.90$\pm$0.04 & {\phs}0.01$\pm$0.10 & {\phs}0.03$\pm$0.14 & {\phs}0.04$\pm$0.12 & 1 \\
2MASSI J1546271$-$332511 & T5.5 V & 15.60$\pm$0.05 & {\phs}0.16$\pm$0.10 & {\phs}0.02$\pm$0.19 & {\phs}0.18$\pm$0.18 & 1 \\
2MASSI J1553022+153236 & T7 V & 15.81$\pm$0.08 & $-$0.11$\pm$0.19 & {\phs}0.41$\pm$0.25 & {\phs}0.30$\pm$0.21 & 1 \\
SDSSp J162414.37+002915.6 & T6 V & 15.49$\pm$0.06 & $-$0.03$\pm$0.12 & $<$ {\phs}0.08 & $<$ {\phs}0.05 & 9 \\
2MASSI J2254188+312349 & T5 V & 15.28$\pm$0.05 & {\phs}0.24$\pm$0.10 & {\phs}0.21$\pm$0.17 & {\phs}0.45$\pm$0.15 & 1 \\
2MASSI J2339101+135230 & T5.5 V & 15.88$\pm$0.08 & $-$0.16$\pm$0.19 & {\phs}0.34$\pm$0.33 & {\phs}0.18$\pm$0.29 & 1 \\
2MASSI J2356547$-$155310 & T6 V & 15.80$\pm$0.06 & {\phs}0.16$\pm$0.12 & $-$0.19$\pm$0.21 & $-$0.03$\pm$0.20 & 1 
\tablenotetext{a}{2MASS photometry unless otherwise noted.}
\tablenotetext{b}{UKIRT JHK photometry from Leggett et al.\ (1999).}
\tablenotetext{c}{JK photometry from Cuby et al.\ (1999).}
\tablerefs{
(1) This paper; (2) Burgasser et al.\ (2000b); (3) Nakajima et al.\ (1995);
(4) Leggett et al.\ (2000); (5) Burgasser et al.\ (1999); (6) Cuby et al.\ (1999);
(7) Tsvetanov et al.\ (2000); (8) Burgasser et al.\ (2000a); 
(9) Strauss et al.\ (1999).}
\enddata
\end{deluxetable}

\clearpage

\begin{deluxetable}{lll}
\tabletypesize{\scriptsize}
\tablenum{8}
\tablewidth{0pt}
\tablecaption{Near-Infrared Spectral Properties of T Dwarf Subtypes.}
\tablehead{
\colhead{Type} & 
\colhead{Description} & 
\colhead{Standard} \\ 
\colhead{(1)} &
\colhead{(2)} &
\colhead{(3)} 
}
\startdata
T1 V & Weak CH$_4$ bands seen at 1.3, 1.6, and 2.2 $\micron$ & SDSS 0837-0000 \\
 & Distinct 1.07 and 1.27 $\micron$ peaks separated by 1.15 $\micron$ H$_2$O/CH$_4$ feature & \\
 & K-band peak noticeably depressed relative to J and H & \\
 & CH$_4$ and CO bands at K equal in strength & \\
 & K I lines at 1.25 $\micron$ strong & \\
T2 V & CH$_4$ bands strengthening & SDSS 1254-0122 \\
 & Flux at 1.15 $\micron$ feature roughly 50\% J-band peak & \\
 & K-band CH$_4$ stronger than CO & \\
 & K-band peak rounded & \\
T3 V & Flux at 1.15 $\micron$ feature roughly 40\% J-band peak & SDSS 1021-0304 \\
 & Flux at 1.6 $\micron$ feature roughly 60\% H-band peak & \\
 & CO barely visible at K-band & \\
T5 V & Flux at 1.6 $\micron$ trough roughly 50\% H-band peak & 2MASS 0559-1404 \\
 & No CO present & \\
 & H-band suppressed relative to J and K & \\
 & K I lines at 1.25 $\micron$ peak in strength & \\
T6 V & Flux at 1.15 $\micron$ feature roughly 20\% J-band peak & 2MASS 0243-2453 \\
 & Flux at 1.6 $\micron$ feature roughly 30\% H-band peak & \\
 & 1.25 $\micron$ K I lines beginning to weaken & \\
 & 1.3 $\micron$ CH$_4$ band blended with 1.4 $\micron$ H$_2$O & \\
 & 2.2 $\micron$ CH$_4$ absorption nearly saturated & \\
 & K-band beginning to flatten, asymmetric peak centered at 2.11 $\micron$ & \\
T7 V & Flux at 1.15 $\micron$ feature roughly 10\% J-band peak & 2MASS 0727+1710 \\
 & Flux at 1.6 $\micron$ feature roughly 10\% H-band peak & \\
 & 1.25 $\micron$ K I lines barely discernible & \\
 & H- and K-band peaks maximally suppressed relative to J & \\
 & J-band peak increasingly narrow & \\
T8 V & Flux at 1.15 $\micron$ feature nearly saturated & 2MASS 0415-0935 \\
 & Flux at 1.6 $\micron$ feature nearly saturated & \\
 & No 1.25 $\micron$ K I lines present & \\
 & Slight increase in H- and K-band peaks relative to J & \\
 & K-band peak more sharply peaked and symmetric about 2.07 $\micron$ & \\
\enddata
\end{deluxetable}

\clearpage

\begin{deluxetable}{lcccc}
\tabletypesize{\scriptsize}
\tablenum{9}
\tablewidth{0pt}
\tablecaption{Ratios Used as T Dwarf Spectral Diagnostics.}
\tablehead{
\colhead{Diagnostic} &
\colhead{Numerator ($\micron$)} &
\colhead{Denominator ($\micron$)} &
\colhead{Feature Measured} \\
\colhead{(1)} &
\colhead{(2)} &
\colhead{(3)} &
\colhead{(4)} 
}
\startdata
H$_2$O-A\tablenotemark{a} & ${\langle}F_{1.12-1.17}{\rangle}$ & ${\langle}F_{1.25-1.28}{\rangle}$ & 1.15 $\micron$ H$_2$O/CH$_4$ \\
H$_2$O-B\tablenotemark{a} & ${\langle}F_{1.505-1.525}{\rangle}$ & ${\langle}F_{1.575-1.595}{\rangle}$ & 1.4 $\micron$ H$_2$O \\
H$_2$O-C & ${\langle}F_{2.00-2.04}{\rangle}$ & ${\langle}F_{2.09-2.13}{\rangle}$ & 1.9 $\micron$ H$_2$O \\
CH$_4$-A\tablenotemark{a} & ${\langle}F_{1.295-1.325}{\rangle}$ & ${\langle}F_{1.25-1.28}{\rangle}$ & 1.3 $\micron$ CH$_4$ \\
CH$_4$-B\tablenotemark{a} & ${\langle}F_{1.64-1.70}{\rangle}$ & ${\langle}F_{1.575-1.595}{\rangle}$ & 1.6 $\micron$ CH$_4$ \\
CH$_4$-C\tablenotemark{a} & ${\langle}F_{2.225-2.275}{\rangle}$ & ${\langle}F_{2.09-2.13}{\rangle}$ & 2.2 $\micron$ CH$_4$ \\
H/J\tablenotemark{a} & ${\langle}F_{1.50-1.75}{\rangle}$ & ${\langle}F_{1.20-1.325}{\rangle}$ & NIR color \\
K/J\tablenotemark{a} & ${\langle}F_{2.00-2.30}{\rangle}$ & ${\langle}F_{1.20-1.325}{\rangle}$ & NIR color \\
K/H & ${\langle}F_{2.00-2.30}{\rangle}$ & ${\langle}F_{1.50-1.75}{\rangle}$ & NIR color \\
CO & ${\langle}F_{2.325-2.375}{\rangle}$ & ${\langle}F_{2.09-2.13}{\rangle}$ & 2.3 $\micron$ CO \\
2.11/2.07\tablenotemark{a} & ${\langle}F_{2.10-2.12}{\rangle}$ & ${\langle}F_{2.06-2.08}{\rangle}$ & K-band shape/CIA H$_2$ \\
K shape & ${\langle}F_{2.12-2.13}{\rangle}$ - ${\langle}F_{2.15-2.16}{\rangle}$ & ${\langle}F_{2.17-2.18}{\rangle}$ - ${\langle}F_{2.19-2.20}{\rangle}$ & K-band shape/CIA H$_2$ \\
\tablenotetext{a}{Spectral ratio used as a classification diagnostic.}
\enddata
\end{deluxetable}

\clearpage

\begin{deluxetable}{llcccccccc}
\tabletypesize{\scriptsize}
\tablenum{10}
\tablewidth{0pt}
\tablecaption{Spectral Diagnostics for L and T Dwarf Standards.}
\tablehead{
\colhead{Object} &
\colhead{Type} &
\colhead{H$_2$O-A} & 
\colhead{H$_2$O-B} & 
\colhead{CH$_4$-A} & 
\colhead{CH$_4$-B} & 
\colhead{CH$_4$-C} & 
\colhead{H/J} & 
\colhead{K/J} & 
\colhead{2.11/2.07} \\
\colhead{(1)} & 
\colhead{(2)} & 
\colhead{(3)} & 
\colhead{(4)} & 
\colhead{(5)} & 
\colhead{(6)} & 
\colhead{(7)} & 
\colhead{(8)} & 
\colhead{(9)} &
\colhead{(10)} 
}
\startdata
DENIS 1228$-$1547 & L5 V & 0.765 & 0.771 & 1.044 & 1.035 & 0.938 & 0.891 & 0.575 & 1.102 \\
DENIS 0205$-$1159 & L7 V & 0.764 & 0.760 & 1.026 & 1.010 & 0.841 & 0.837 & 0.524 & 1.079 \\
SDSS 0837$-$0000 & T1 V & 0.669 & 0.637 & 0.991 & 0.920 & 0.662 & 0.746 & 0.326 & 1.061 \\
SDSS 1254$-$0122 & T2 V & 0.501 & 0.569 & 0.910 & 0.832 & 0.527 & 0.637 & 0.299 & 1.081 \\
SDSS 1021$-$0304 & T3 V & 0.393 & 0.518 & 0.864 & 0.657 & 0.498 & 0.581 & 0.250 & 1.068 \\
2MASS 0559$-$1404\tablenotemark{a} & T5 V & 0.351 & 0.450 & 0.786 & 0.464 & 0.210 & 0.380 & 0.185 & 0.999 \\
2MASS 0243$-$2453 & T6 V & 0.174 & 0.353 & 0.555 & 0.292 & 0.139 & 0.310 & 0.136 & 0.902 \\
2MASS 0727+1710 & T7 V & 0.117 & 0.296 & 0.498 & 0.176 & 0.087 & 0.241 & 0.097 & 0.864 \\
2MASS 0415$-$0935\tablenotemark{a} & T8 V & 0.062 & 0.228 & 0.407 & 0.102 & 0.071 & 0.301 & 0.127 & 0.825 \\
\tablenotetext{a}{Indices averaged from multiple spectra.}
\enddata
\end{deluxetable}

\clearpage

\begin{deluxetable}{llllllllll}
\rotate
\tabletypesize{\scriptsize}
\tablenum{11a}
\tablewidth{0pt}
\tablecaption{T Dwarf Spectral Indices: NIRC and D78 Data.}
\tablehead{
\colhead{Object} &
\colhead{H$_2$O-A} & 
\colhead{H$_2$O-B} & 
\colhead{CH$_4$-A} & 
\colhead{CH$_4$-B} & 
\colhead{CH$_4$-C\tablenotemark{a}} & 
\colhead{H/J\tablenotemark{a}} & 
\colhead{K/J\tablenotemark{a}} & 
\colhead{2.11/2.07} & 
\colhead{Type} \\
\colhead{(1)} & 
\colhead{(2)} & 
\colhead{(3)} & 
\colhead{(4)} & 
\colhead{(5)} & 
\colhead{(6)} & 
\colhead{(7)} & 
\colhead{(8)} & 
\colhead{(9)} & 
\colhead{(10)} 
}
\startdata
2MASS 2254+3123 & 0.442(2--3) & 0.496(3--5) & 0.721(5) & 0.513(5) & 0.243(5) & 0.378(5) & 0.133(6) & 0.965(5) & 4.8$\pm$0.4 \\
2MASS 0559$-$1404\tablenotemark{b} & 0.351(5) & 0.450(5) & 0.786(5) & 0.464(5) & 0.210(5) & 0.380(5) & 0.185(5) & 0.999(5) & 5.0$\pm$0.0 \\
2MASS 2339+1352 & 0.300(5) & 0.418(5) & 0.547(6) & 0.388(5--6) & 0.182(5) & 0.337(6) & 0.119(6--7) & 0.990(5) & 5.4$\pm$0.5 \\
2MASS 0755+2212 & -- & 0.548(2--3) & 0.758(5) & 0.450(5) & -- & 0.327(6) & 0.159(5--6) & -- & 5.2:\tablenotemark{c} \\
2MASS 2356$-$1553 & 0.261(5--6) & 0.389(6) & 0.720(5) & 0.322(6) & 0.140(6) & 0.351(5--6) & 0.150(6) & 0.946(6) & 5.8$\pm$0.4 \\
2MASS 0937+2931\tablenotemark{b} & 0.185(6) & 0.410(5--6) & 0.602(6) & 0.287(6) & 0.199(5) & 0.314(6) & 0.082(7) & 0.920(6) & 5.9$\pm$0.2 \\
2MASS 0243$-$2453 & 0.174(6) & 0.353(6) & 0.555(6) & 0.292(6) & 0.139(6) & 0.310(6) & 0.136(6) & 0.902(6) & 6.0$\pm$0.0 \\
SDSS 1346$-$0031 & 0.157(6) & 0.354(6) & 0.561(6) & 0.240(6--7) & 0.145(6) & 0.286(6) & 0.125(6) & 0.900(6) & 6.0$\pm$0.0 \\
2MASS 1047+2124\tablenotemark{b} & 0.166(6) & 0.345(6) & 0.524(7) & 0.235(6--7) & 0.170(6) & 0.339(6) & 0.105(7) & 0.884(6--7) & 6.3$\pm$0.4 \\
2MASS 1237+6526 & 0.127(7) & 0.328(6--7) & 0.511(7) & 0.221(7) & 0.144(6) & 0.284(6) & 0.073(7) & 0.902(6) & 6.6$\pm$0.5 \\
2MASS 0727+1710 & 0.117(7) & 0.296(7) & 0.498(7) & 0.176(7) & 0.087($\geq$ 7) & 0.241($\geq$ 7) & 0.097($\geq$ 7) & 0.864(7) & 7.0$\pm$0.0 \\
2MASS 1553+1532 & 0.130(7) & 0.311(7) & 0.444(8) & 0.175(7) & 0.078($\geq$ 7) & 0.317($\geq$ 7) & 0.128($\geq$ 7) & 0.864(7) & 7.0$\pm$0.0 \\
2MASS 1217$-$0311 & 0.091(7--8) & 0.292(7) & 0.412(8) & 0.138(7--8) & 0.074($\geq$ 7) & 0.276($\geq$ 7) & 0.132($\geq$ 7) & 0.817(8) & 7.7$\pm$0.2 \\
Gliese 570D & 0.080(8) & 0.242(8) & 0.441(8) & 0.135(7--8) & 0.063($\geq$ 7) & 0.293($\geq$ 7) & 0.110($\geq$ 7) & 0.938(6) & 7.8$\pm$0.3 \\
2MASS 0415$-$0935\tablenotemark{b} & 0.062(8) & 0.228(8) & 0.407(8) & 0.102(8) & 0.071($\geq$ 7) & 0.301($\geq$ 7) & 0.127($\geq$ 7) & 0.825(8) & 8.0$\pm$0.0 \\
\tablenotetext{a}{Index ambiguous for subtypes T7 V through T8 V.}
\tablenotetext{b}{Indices averaged from multiple spectra.}
\tablenotetext{c}{Spectral type uncertain due to poor quality of D78 spectrum.}
\enddata
\end{deluxetable}

\clearpage

\begin{deluxetable}{lllllll}
\tabletypesize{\scriptsize}
\tablenum{11b}
\tablewidth{0pt}
\tablecaption{T Dwarf Spectral Indices: OSIRIS Data.}
\tablehead{
\colhead{Object} &
\colhead{H$_2$O-B} & 
\colhead{CH$_4$-A} & 
\colhead{CH$_4$-B} & 
\colhead{H/J\tablenotemark{a}} & 
\colhead{K/J\tablenotemark{a}} & 
\colhead{Type} \\
\colhead{(1)} & 
\colhead{(2)} & 
\colhead{(3)} & 
\colhead{(4)} & 
\colhead{(5)} & 
\colhead{(6)} & 
\colhead{(7)}  
}
\startdata
SDSS 1254$-$0122 & 0.558(2) & 0.943(2) & 0.825(2) & 0.680(2) & 0.344(1) &  2.0$\pm$0.0  \\
2MASS 2254+3123 & 0.469(5) & 0.886(2--3) & 0.484(5) & 0.410(5) & 0.184(5) &  5.0$\pm$0.0  \\
2MASS 0559$-$1404 & 0.456(5) & 0.811(5) & 0.383(5--6) & 0.337(5--6) & 0.153(5--6) &  5.3$\pm$0.3  \\
2MASS 2339+1352 & 0.456(5) & 0.756(5) & 0.344(6) & 0.266(7) & 0.176(5) &  5.3:\tablenotemark{b} \\
2MASS 1534$-$2952 & 0.441(5) & 0.824(3--5) & 0.380(5--6) & 0.305(6) & 0.103(7) &  5.5$\pm$0.4  \\
2MASS 1546$-$3325 & 0.418(5) & 0.788(5) & 0.283(6) & 0.297(6) & 0.134(6) &  5.7$\pm$0.6  \\
2MASS 2356$-$1553 & 0.409(5--6) & 0.795(5) & 0.254(6) & 0.297(6) & 0.153(6) &  5.8$\pm$0.3  \\
2MASS 0937+2931 & 0.388(6) & 0.710(5) & 0.246(6) & 0.282(6) & 0.147(6) &  6.0$\pm$0.0  \\
2MASS 0243$-$2453 & 0.328(6) & 0.605(6) & 0.225(7) & 0.277(6) & 0.205(5) &  6.0$\pm$0.0  \\
2MASS 1225$-$2739 & 0.403(5--6) & 0.745(5) & 0.262(6) & 0.264(7) & 0.110(7) &  6.2$\pm$0.8  \\
SDSS 1624+0029 & 0.230(8) & 0.739(5) & 0.188(7) & 0.235(7) & 0.057(7) &  7.0:\tablenotemark{b}  \\
2MASS 0727+1710 & 0.332(6) & 0.511(7) & 0.133(8) & 0.183(7) & 0.137($\geq$ 7) &  7.0$\pm$0.0 \\
2MASS 1553+1532\tablenotemark{c} & 0.300(7) & 0.601(6) & 0.114(8) & 0.250(7) & 0.109($\geq$ 7) &  7.0$\pm$0.0 \\
Gliese 570D & 0.294(7) & 0.537(6) & 0.106(8) & 0.261(7) & 0.111($\geq$ 7) &  7.0$\pm$0.0  \\
\tablenotetext{a}{Index ambiguous for subtypes T7 V through T8 V.}
\tablenotetext{b}{Spectral type uncertain due to poor quality spectrum.}
\tablenotetext{c}{Indices averaged from multiple spectra.}
\enddata
\end{deluxetable}

\clearpage

\begin{deluxetable}{lllllllllll}
\tabletypesize{\scriptsize}
\rotate
\tablenum{11c}
\tablewidth{0pt}
\tablecaption{T Dwarf Spectral Indices: Data from Literature.}
\tablehead{
\colhead{Object} &
\colhead{Ref.} &
\colhead{H$_2$O-A} & 
\colhead{H$_2$O-B} & 
\colhead{CH$_4$-A} & 
\colhead{CH$_4$-B} & 
\colhead{CH$_4$-C} & 
\colhead{H/J} & 
\colhead{K/J} & 
\colhead{2.11/2.07\tablenotemark{a}} & 
\colhead{Type} \\
\colhead{(1)} & 
\colhead{(2)} & 
\colhead{(3)} & 
\colhead{(4)} & 
\colhead{(5)} & 
\colhead{(6)} & 
\colhead{(7)} & 
\colhead{(8)} & 
\colhead{(9)} &
\colhead{(10)} &
\colhead{(11)} 
}
\startdata
SDSS 0837$-$0000 & 1 & 0.669(1) & 0.637(1) & 0.991(1) & 0.920(1) & 0.662(1) & 0.746(1) & 0.326(1) & 1.061($\leq$ 3) & 1.0$\pm$0.0 \\
SDSS 1254$-$0122 & 1 & 0.501(2) & 0.569(2) & 0.910(2) & 0.832(2) & 0.527(2) & 0.637(2) & 0.299(2) & 1.081($\leq$ 3) & 2.0$\pm$0.0  \\
SDSS 1021$-$0304 & 1 & 0.393(3) & 0.518(3) & 0.864(3) & 0.657(3) & 0.498(3) & 0.581(3) & 0.250(3) & 1.068($\leq$ 3) & 3.0$\pm$0.0  \\
2MASS 0559$-$1404 & 2 & 0.350(5) & 0.494(3--5) & 0.862(3) & 0.470(5) & 0.232(6--7) & 0.377(6) & 0.127(7) & 1.001(5) & 5.3$\pm$0.9  \\
SDSS 1624+0029 & 3 & 0.167(6) & 0.371(6) & 0.703(5) & 0.280(6) & 0.150(6) & 0.279(6--7) & 0.099(7) & 0.937(6) & 6.1$\pm$0.2 \\
SDSS 1346$-$0031 & 5 & 0.175(6) & 0.352(6) & 0.635(6) & 0.247(6) & 0.094(7) & 0.261(7) & 0.113(6--7) & 0.937(6) & 6.3$\pm$0.4  \\
NTTDF 1205$-$0744 & 4 & 0.087(7--8) & 0.289(7) & 0.723(5) & 0.229(6--7) & 0.261(5) & 0.279(6--7) & 0.117(6--7) & 0.913(6) & 6.3:\tablenotemark{b}  \\
Gliese 229B\tablenotemark{c} & 6,7 & 0.114(7) & 0.425(5--6) & 0.485(7) & 0.200(7) & 0.195(5) & 0.277(6--7) & 0.100(7) & 0.926(6) & 6.5$\pm$0.6  \\
\tablenotetext{a}{Index ambiguous for subtypes T1 V through T3 V.}
\tablenotetext{b}{Spectral type uncertain due to poor quality spectrum.}
\tablenotetext{c}{Indices averaged from multiple spectra.}
\tablerefs{
(1) \citet{leg00}; 
(2) \citet{me00c}; (3)  \citet{str99};
(4) \citet{cub99}; (5) \citet{tsv00}; 
(6) \citet{geb96}; (7) \citet{opp98}.}
\enddata
\end{deluxetable}

\clearpage

\begin{deluxetable}{lcccc}
\tabletypesize{\scriptsize}
\tablenum{12a}
\tablewidth{0pt}
\tablecaption{Linear Fits to Spectral Indices: T Dwarfs\tablenotemark{a}}
\tablehead{
\colhead{} &
\colhead{} &
\colhead{} &
\colhead{} &
\colhead{RMS Error} \\
\colhead{Index} &
\colhead{c0} &
\colhead{c1} &
\colhead{Range} &
\colhead{(SpT)\tablenotemark{b}} \\
\colhead{(1)} & 
\colhead{(2)} & 
\colhead{(3)} & 
\colhead{(4)} & 
\colhead{(5)} 
}
\startdata
H$_2$O-A & {\phn}8.5$\pm$2.8{\phn} & -11.9$\pm$0.9 & T1 V -- T8 V & 0.6 \\
H$_2$O-B & 12.9$\pm$3.1{\phn} & -18.2$\pm$1.4 & T1 V -- T8 V & 0.7 \\
CH$_4$-A & 13.2$\pm$3.9{\phn} & -11.3$\pm$1.1 & T1 V -- T8 V & 0.9 \\
CH$_4$-B & {\phn}8.3$\pm$1.5{\phn} & {\phn}-8.0$\pm$0.3 & T1 V -- T8 V & 0.4 \\
CH$_4$-C & {\phn}8.2$\pm$2.4{\phn} & -12.0$\pm$0.8 & T1 V -- T8 V & 0.5 \\
H/J & {\phn}9.7$\pm$2.3{\phn} & -12.0$\pm$0.7 & T1 V -- T7 V & 0.5 \\
K/J & {\phn}9.0$\pm$3.4{\phn} & -23.9$\pm$2.2 & T1 V -- T7 V & 0.7 \\
2.11/2.07 & 24.0$\pm$10.0 & -19.1$\pm$3.8 & T3 V -- T8 V & 1.2 \\
\tablenotetext{a}{Coefficients are for linear fit SpT = c0 + c1$\times$Index,
where SpT(T0) = 0, SpT(T5) = 5, SpT(L5) = -4, etc.}
\tablenotetext{b}{RMS of SpT minus adopted spectral type, the latter quantity 
from the literature or listed in Table 7.}
\enddata
\end{deluxetable}

\clearpage

\begin{deluxetable}{lcccc}
\tabletypesize{\scriptsize}
\tablenum{12b}
\tablewidth{0pt}
\tablecaption{Linear Fits to Spectral Indices: M and L Dwarfs\tablenotemark{a}}
\tablehead{
\colhead{} &
\colhead{} &
\colhead{} &
\colhead{} &
\colhead{RMS Error} \\
\colhead{Index} &
\colhead{c0} &
\colhead{c1} &
\colhead{Range} &
\colhead{(SpT)\tablenotemark{b}} \\
\colhead{(1)} & 
\colhead{(2)} & 
\colhead{(3)} & 
\colhead{(4)} & 
\colhead{(5)} 
}
\startdata
H$_2$O-A & {\phs}37.7$\pm$7.0 & $-$41.0$\pm$5.8{\phn} & M5 V -- L7 V & 2.5 \\
H$_2$O-B & {\phs}47.7$\pm$4.6 & $-$53.9$\pm$4.2{\phn} & M5 V -- L7 V & 1.7 \\
H$_2$O-C & {\phs}57.8$\pm$8.1 & $-$65.5$\pm$7.7{\phn} & M5 V -- L7 V & 2.2 \\
CH$_4$-C & {\phs}20.8$\pm$2.7 & $-$17.1$\pm$1.3{\phn} & L3 V -- T3 V & 0.8 \\
H/J & $-$32.7$\pm$4.6 & {\phs}43.9$\pm$7.8{\phn} & M5 V -- L7 V & 3.9 \\
K/J & $-$18.3$\pm$2.5 & {\phs}40.4$\pm$6.8{\phn} & M5 V -- L7 V & 3.6 \\
K/H & $-$30.7$\pm$8.0 & {\phs}68.2$\pm$16.4 & M5 V -- L7 V & 5.3 \\
\tablenotetext{a}{Coefficients are for linear fit SpT = c0 + c1$\times$Index,
where SpT(L0) = 0, SpT(L5) = 5, SpT(M5) = -5, etc.}
\tablenotetext{b}{RMS of SpT minus adopted spectral type, the latter quantity
from the literature.}
\enddata
\end{deluxetable}

\clearpage

\begin{deluxetable}{llcccc}
\tabletypesize{\scriptsize}
\tablenum{13a}
\tablewidth{0pt}
\tablecaption{K I Pseudo-equivalent Widths - OSIRIS Data}
\tablehead{
\colhead{} &
\colhead{} &
\multicolumn{2}{c}{1.2432 $\micron$} &
\multicolumn{2}{c}{1.2522 $\micron$} \\
\cline{3-4} 
\cline{5-6} \\
\colhead{Object} &
\colhead{Type} &
\colhead{$\lambda$$_c$ ($\micron$)} &
\colhead{PEW ({\AA})} &
\colhead{$\lambda$$_c$ ($\micron$)} &
\colhead{PEW ({\AA})} \\
\colhead{(1)} & 
\colhead{(2)} & 
\colhead{(3)} & 
\colhead{(4)} & 
\colhead{(5)} &
\colhead{(6)} 
}
\startdata
LHS 511AB & M4.5 V & 1.2425 & 1.50$\pm$0.15 & 1.2514 & 1.50$\pm$0.12 \\
HB 2115$-$4518 & M7.5 V & 1.2433 & {\phn}8.1$\pm$0.7{\phn} & 1.2518 & {\phn}4.3$\pm$0.7{\phn} \\
HB 2124$-$4228 & M8.5 V & 1.2429 & {\phn}9.2$\pm$0.7{\phn} & 1.2518 & {\phn}5.6$\pm$0.8{\phn} \\
2MASS 2224$-$0158 & L4.5 V & 1.2425 & {\phn}8.6$\pm$0.9{\phn} & 1.2513 & {\phn}6.9$\pm$0.9{\phn} \\
DENIS 0205$-$1159 & L7 V & 1.2433 & {\phn}3.3$\pm$0.8{\phn} & 1.2522 & {\phn}6.2$\pm$0.7{\phn} \\
SDSS 1254$-$0122 & T2 V & 1.2433 & {\phn}5.5$\pm$0.9{\phn} & 1.2521 & {\phn}9.1$\pm$0.9{\phn} \\
2MASS 2254+3123 & T5 V & 1.2434 & {\phn}7.5$\pm$1.4{\phn} & 1.2527 & 10.6$\pm$1.3{\phn} \\
2MASS 0559$-$1404 & T5 V & 1.2423 & {\phn}6.4$\pm$1.0{\phn} & 1.2515 & {\phn}9.3$\pm$0.9{\phn} \\
2MASS 1534$-$2952 & T5.5 V & 1.2432 & {\phn}5.9$\pm$0.6{\phn} & 1.2521 & {\phn}7.7$\pm$0.6{\phn} \\
2MASS 2339+1352 & T5.5 V & 1.2405 & $<$ 3.2 & 1.2516 & {\phn}7.4$\pm$3.5{\phn} \\
2MASS 1546$-$3325 & T5.5 V & 1.2427 & {\phn}5.9$\pm$1.3{\phn} & 1.2528 & 11.1$\pm$1.2{\phn} \\
2MASS 2356$-$1553 & T6 V & 1.2427 & {\phn}4.1$\pm$1.4{\phn} & 1.2506 & 11.7$\pm$1.1{\phn} \\
2MASS 0243$-$2453 & T6 V & 1.2430 & {\phn}4.2$\pm$1.0{\phn} & 1.2523 & {\phn}8.9$\pm$1.2{\phn} \\
2MASS 0937+2931 & T6 V & 1.2407 & $<$ 1.3 & 1.2508 & {\phn}2.5$\pm$1.2{\phn} \\
2MASS 1225$-$2739 & T6 V & 1.2436 & {\phn}6.6$\pm$1.1{\phn} & 1.2524 & {\phn}7.8$\pm$0.6{\phn} \\
SDSS 1624+0029 & T6 V & 1.2429 & {\phn}3.4$\pm$1.8{\phn} & 1.2517 & {\phn}9.8$\pm$1.8{\phn} \\
2MASS 0727+1710 & T7 V & 1.2428 & {\phn}1.9$\pm$1.2{\phn} & 1.2516 & {\phn}6.8$\pm$1.0{\phn} \\
2MASS 1553+1532 & T7 V & 1.2430 & {\phn}2.9$\pm$1.0{\phn} & 1.2519 & {\phn}5.8$\pm$0.9{\phn} \\
2MASS 1553+1532 & T7 V & 1.2425 & {\phn}1.5$\pm$1.1{\phn} & 1.2522 & {\phn}5.7$\pm$1.1{\phn} \\
Gliese 570D & T8 V & 1.2411 & {\phn}3.7$\pm$0.7{\phn} & 1.2491 & {\phn}3.0$\pm$0.5{\phn} \\
\enddata
\end{deluxetable}

\clearpage

\begin{deluxetable}{llccccc}
\tabletypesize{\scriptsize}
\tablenum{13b}
\tablewidth{0pt}
\tablecaption{K I Pseudo-equivalent Widths - Data from the Literature}
\tablehead{
\colhead{} &
\colhead{} &
\multicolumn{2}{c}{1.2432 $\micron$} &
\multicolumn{2}{c}{1.2522 $\micron$} \\
\cline{3-4} 
\cline{5-6} &
\colhead{} \\
\colhead{Object} &
\colhead{Type} &
\colhead{$\lambda$$_c$ ($\micron$)} &
\colhead{PEW ({\AA})} &
\colhead{$\lambda$$_c$ ($\micron$)} &
\colhead{PEW ({\AA})} &
\colhead{Ref.} \\
\colhead{(1)} & 
\colhead{(2)} & 
\colhead{(3)} & 
\colhead{(4)} & 
\colhead{(5)} &
\colhead{(6)} &
}
\startdata
LHS 3406 & M5.5 V & 1.2416 & {\phn}5.1$\pm$0.9 & 1.2503 & 4.2$\pm$0.2 & 1 \\
LHS 2930 & M6.5 V & 1.2404 & {\phn}4.5$\pm$1.1 & 1.2503 & 2.9$\pm$0.5 & 1 \\
LHS 3003 & M7 V & 1.2404 & {\phn}4.2$\pm$0.9 & 1.2503 & 2.0$\pm$0.4 & 1 \\
LHS 429 & M7 V & 1.2404 & {\phn}4.0$\pm$0.7 & 1.2503 & 1.8$\pm$0.4 & 1 \\
LHD 94 & M7 V & 1.2434 & {\phn}5.4$\pm$0.6 & 1.2519 & 4.3$\pm$0.7 & 2 \\
2MASS 0320+1854 & M8 V & 1.2438 & {\phn}7.2$\pm$0.3 & 1.2524 & 4.3$\pm$0.4 & 2 \\
TVLM 513$-$46546 & M8.5 V & 1.2436 & {\phn}9.0$\pm$0.4 & 1.2528 & 6.2$\pm$0.3 & 1 \\
BRI 0021$-$0214 & M9.5 V & 1.2432 & {\phn}9.0$\pm$0.6 & 1.2525 & 5.2$\pm$0.5 & 1 \\
2MASS 0345+2540 & L0 V & 1.2435 & {\phn}8.0$\pm$0.5 & 1.2522 & 6.1$\pm$0.3 & 1 \\
2MASS 0345+2540 & L0 V & 1.2438 & {\phn}8.0$\pm$0.6 & 1.2525 & 6.2$\pm$0.5 & 2 \\
2MASS 0746+2000 & L0.5 V & 1.2436 & {\phn}9.3$\pm$0.4 & 1.2522 & 6.8$\pm$0.3 & 2 \\
2MASS 0829+1456 & L2 V & 1.2431 & 11.2$\pm$1.8 & 1.2518 & 6.9$\pm$0.6 & 2 \\
Kelu 1 & L2 V & 1.2435 & {\phn}9.8$\pm$0.5 & 1.2527 & 6.6$\pm$0.3 & 1 \\
2MASS 1029+1626 & L2.5 V & 1.2422 & {\phn}8.2$\pm$1.5 & 1.2513 & 5.0$\pm$0.8 & 2 \\
DENIS 1058$-$1548 & L3 V & 1.2434 & 11.6$\pm$0.5 & 1.2521 & 8.4$\pm$0.6 & 1 \\
2MASS 0036+1821 & L3.5 V & 1.2435 & 10.1$\pm$0.5 & 1.2522 & 9.4$\pm$0.4 & 2 \\
GD 165B  & L4 V & 1.2436 & 12.5$\pm$0.7 & 1.2522 & 6.3$\pm$0.4 & 1 \\
2MASS 1112+3548 & L4.5 V & 1.2439 & {\phn}9.2$\pm$1.0 & 1.2518 & 5.2$\pm$1.2 & 2 \\
DENIS 1228$-$1547 & L5 V & 1.2431 & {\phn}9.4$\pm$0.9 & 1.2525 & 7.8$\pm$0.8 & 1 \\
SDSS 0539$-$0059 & L5 V & 1.2440 & {\phn}8.1$\pm$0.6 & 1.2527 & 7.3$\pm$0.5 & 3 \\
DENIS 0205$-$1159 & L7 V & 1.2435 & {\phn}4.9$\pm$1.0 & 1.2529 & 5.3$\pm$0.4 & 1 \\
DENIS 0205$-$1159 & L7 V & 1.2435 & $<$ 1.0 & 1.2516 & 6.2$\pm$1.7 & 2 \\
2MASS 0825+2115 & L7.5 V & 1.2424 & {\phn}4.1$\pm$1.9 & 1.2522 & 4.6$\pm$1.0 & 2 \\
2MASS 0310+1648 & L8 V & 1.2439 & $<$ 1.8 & 1.2532 & 7.1$\pm$1.5 & 2 \\
SDSS 0837$-$0000 & T1 V & 1.2430 & {\phn}3.4$\pm$1.2 & 1.2517 & 3.2$\pm$1.4 & 3 \\
SDSS 1254$-$0122 & T2 V & 1.2460 & {\phn}3.5$\pm$0.5 & 1.2517 & 7.2$\pm$0.6 & 3 \\
SDSS 1021$-$0304 & T3 V & 1.2430 & {\phn}4.7$\pm$0.5 & 1.2517 & 7.8$\pm$0.4 &  3 \\
SDSS 1346$-$0031 & T6 V & 1.2429 & {\phn}2.9$\pm$1.1 & 1.2516 & 9.1$\pm$0.9 & 4 \\
SDSS 1624+0029 & T6 V & 1.2425 & {\phn}1.9$\pm$0.4 & 1.2513 & 5.1$\pm$0.7 & 5 \\
Gliese 229B  & T6.5 V & 1.2410 & {\phn}1.3$\pm$0.6 & 1.2520 & 4.3$\pm$0.5 & 6 \\
\tablerefs{
(1) Leggett et al.\ (2001); (2) Reid et al.\ (2001); (3) Leggett et al.\ (2000);
(4) Tsvetanov et al.\ (2000); (5) Strauss et al.\ 1999;
(6) Geballe et al.\ (1996).}
\enddata
\end{deluxetable}

\clearpage

\begin{deluxetable}{lcccccc}
\tabletypesize{\scriptsize}
\tablenum{14}
\tablewidth{0pt}
\tablecaption{Average 2MASS Colors for T dwarfs}
\tablehead{
\colhead{Type} &
\colhead{$\langle$2MASS J-H$\rangle$} &
\colhead{n\tablenotemark{a}} &
\colhead{$\langle$2MASS H-K$_s$$\rangle$} &
\colhead{n\tablenotemark{a}} &
\colhead{$\langle$2MASS J-K$_s$$\rangle$} &
\colhead{n\tablenotemark{a}} \\
\colhead{(1)} & 
\colhead{(2)} & 
\colhead{(3)} & 
\colhead{(4)} & 
\colhead{(5)} &
\colhead{(6)} & 
\colhead{(7)} 
}
\startdata
T2 V & {\phs}0.84$\pm$0.06 & 1 & {\phs}0.22$\pm$0.07 & 1 & {\phs}1.05$\pm$0.07 & 1 \\
T3 V & {\phs}0.93$\pm$0.15 & 1 & {\phs}0.23$\pm$0.21 & 1 & {\phs}1.16$\pm$0.21 & 1 \\
T5 V & {\phs}0.17$\pm$0.04 & 2 & {\phs}0.09$\pm$0.06 & 2 & {\phs}0.25$\pm$0.05 & 2 \\
T5.5 V & {\phs}0.05$\pm$0.07 & 3 & {\phs}0.05$\pm$0.11 & 3 & {\phs}0.08$\pm$0.09 & 3 \\
T6 V & {\phs}0.07$\pm$0.05 & 6 & $-$0.19$\pm$0.10 & 4 & $-$0.03$\pm$0.09 & 4 \\
T6.5 V & {\phs}0.14$\pm$0.11 & 2 & ... & 0 & ... & 0 \\
T7 V & $-$0.18$\pm$0.13 & 2 & {\phs}0.33$\pm$0.19 & 2 & {\phs}0.15$\pm$0.15 & 2 \\
T7.5 V & {\phs}0.06$\pm$0.14 & 1 & ... & 0 & ... & 0 \\
T8 V & {\phs}0.09$\pm$0.08 & 2 & {\phs}0.05$\pm$0.15 & 2 & {\phs}0.14$\pm$0.13 & 2 \\
\tablenotetext{a}{Number of objects used in average.}
\enddata
\end{deluxetable}

\clearpage

\begin{figure}
\epsscale{0.9}
\plotone{f1.eps}
\end{figure}
\clearpage


\begin{figure}
\epsscale{0.9}
\plotone{f3.eps}
\end{figure}
\clearpage

\begin{figure}
\epsscale{0.9}
\plotone{f4.eps}
\end{figure}
\clearpage

\begin{figure}
\epsscale{0.9}
\plotone{f5.eps}
\end{figure}
\clearpage

\begin{figure}
\epsscale{0.9}
\plotone{f6a.eps}
\end{figure}
\clearpage

\begin{figure}
\epsscale{0.9}
\plotone{f6b.eps}
\end{figure}
\clearpage

\begin{figure}
\epsscale{0.9}
\plotone{f7.eps}
\end{figure}
\clearpage

\begin{figure}
\epsscale{0.9}
\plotone{f8.eps}
\end{figure}
\clearpage

\begin{figure}
\epsscale{0.9}
\plotone{f9.eps}
\end{figure}
\clearpage





\begin{figure}
\epsscale{0.9}
\plotone{f11.eps}
\end{figure}
\clearpage

\begin{figure}
\epsscale{1.0}
\plottwo{f12a.eps}{f12b.eps}
\end{figure}

\begin{figure}
\epsscale{1.0}
\plottwo{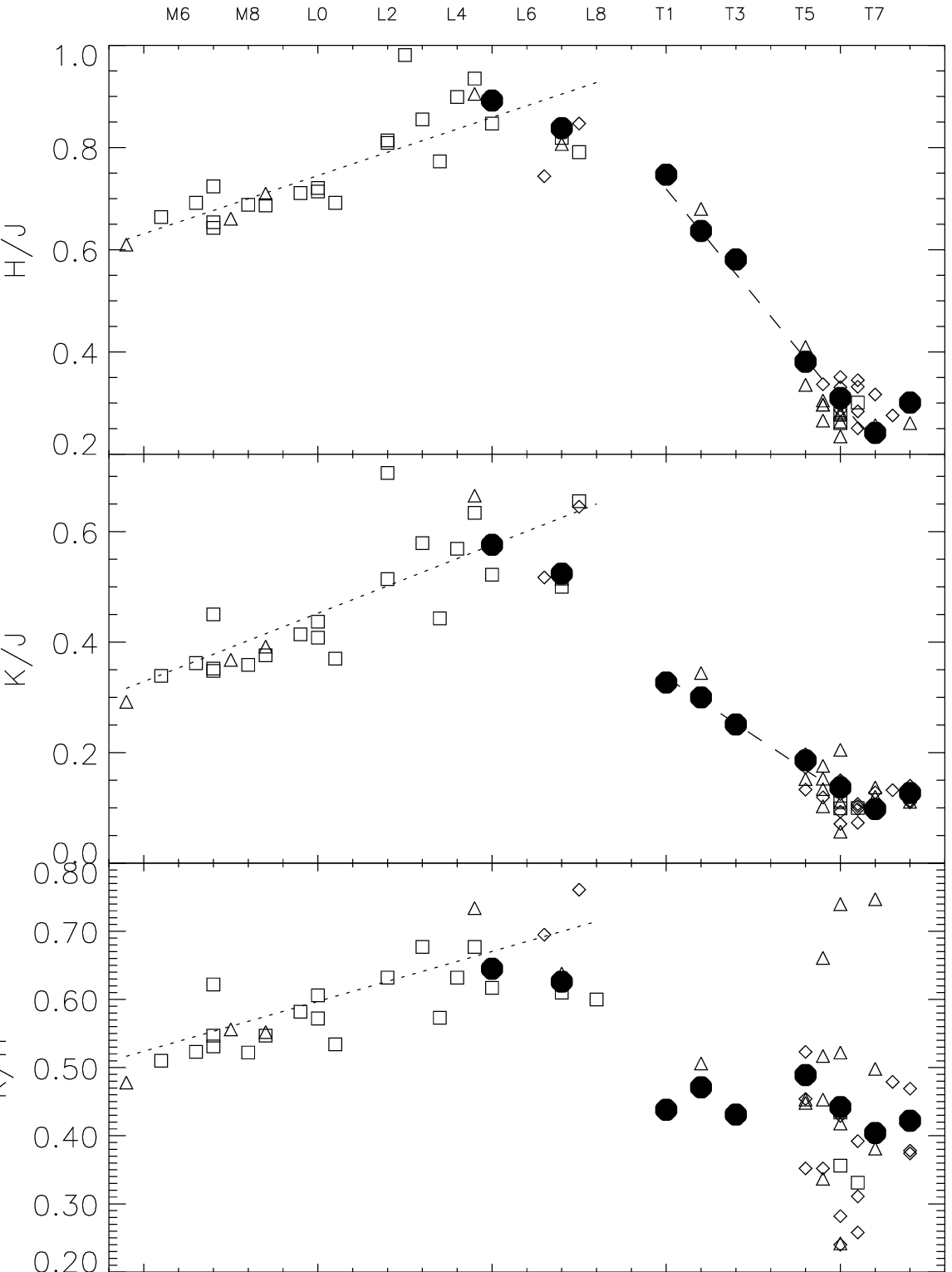}{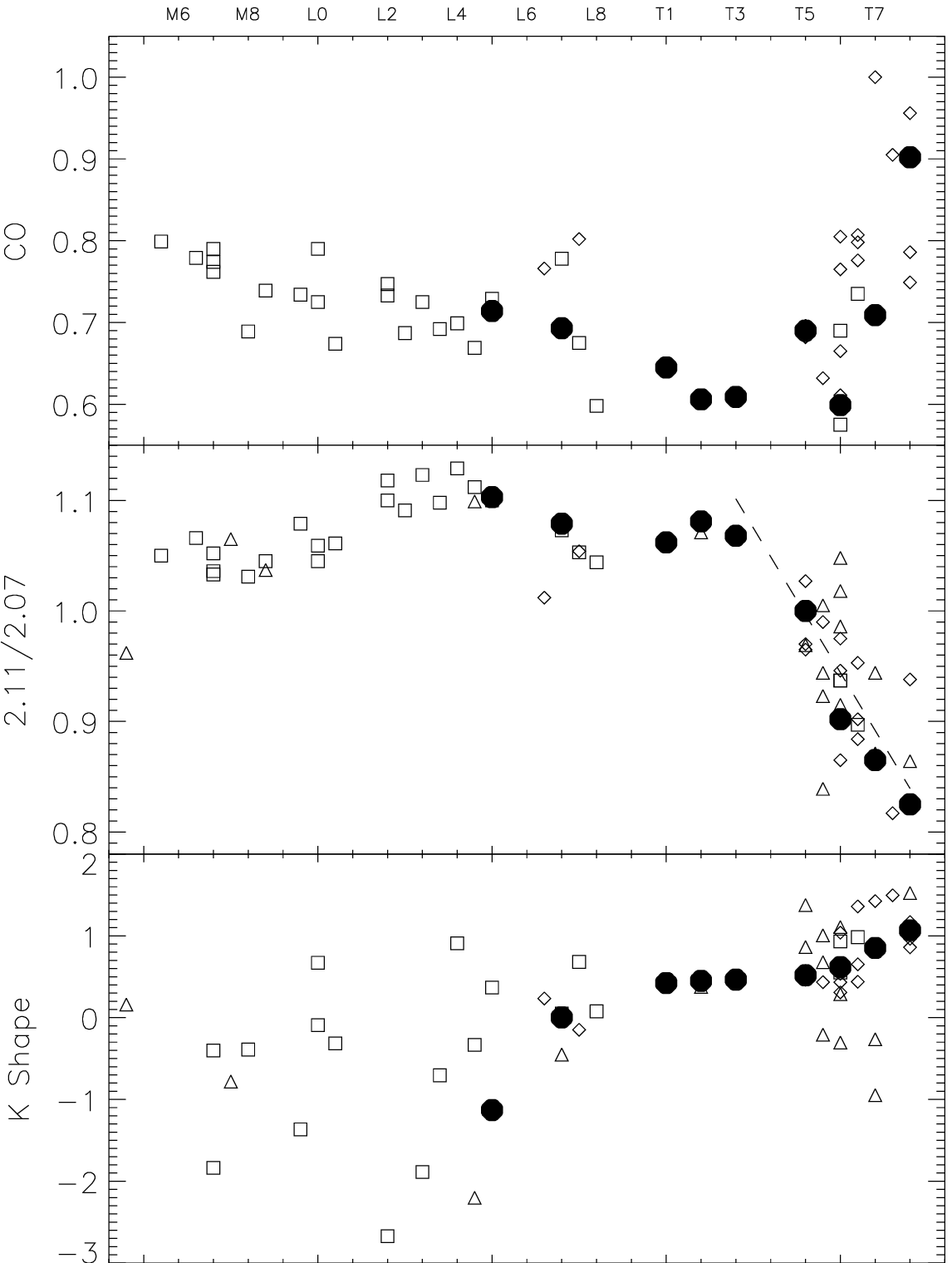}
\end{figure}
\clearpage

\begin{figure}
\epsscale{0.9}
\plotone{f13.eps}
\end{figure}
\clearpage

\begin{figure}
\epsscale{0.9}
\plotone{f14.eps}
\end{figure}
\clearpage

\begin{figure}
\epsscale{0.9}
\plotone{f15.eps}
\end{figure}
\clearpage

\begin{figure}
\epsscale{0.9}
\plotone{f16.eps}
\end{figure}
\clearpage

\begin{figure}
\epsscale{0.9}
\plotone{f17.eps}
\end{figure}
\clearpage

\begin{figure}
\epsscale{0.9}
\plotone{f18.eps}
\end{figure}
\clearpage

\begin{figure}
\epsscale{0.9}
\plotone{f19.eps}
\end{figure}
\clearpage

\end{document}